\documentclass[a4paper,11pt]{article}
\pdfoutput=1 % if your are submitting a pdflatex (i.e. if you have
\usepackage{jheppub} % for details on the use of the package, please
\usepackage{subfig} % to use subfloat 
\usepackage{multirow} % to use multirow
\usepackage{hyperref} % hyperlink for reference

\usepackage[T1]{fontenc} % if needed

\title{Pixel data real-time processing as a next step for HL-LHC upgrades and beyond}

\author[a]{Junho Kim,}
\author[b]{Jongho Lee,}
\author[b]{Chang-Seong Moon,}
\author[c]{Aurore Savoy-Navarro}
\author[a]{and Un-Ki Yang}

\affiliation[a]{Department of Physics \& Astronomy, Seoul National University,Seoul, 08826, Republic of Korea}
\affiliation[b]{Department of Physics, Kyungpook National University, Daegu, 41566, Republic of Korea}
\affiliation[c]{IRFU-CEA, University Paris Saclay, Department of Particle Physics, and CNRS/IN2P3, \\Gif-Sur-Yvette, 91190, France}

\emailAdd{jhkim@cern.ch}
\emailAdd{jongho.lee@cern.ch}
\emailAdd{csmoon@cern.ch}
\emailAdd{asavoy@cern.ch}
\emailAdd{ukyang@snu.ac.kr}

\abstract{The experiments at LHC are implementing novel and challenging detector upgrades for the High Luminosity LHC, among which the tracking systems. 
This paper reports on performance studies,
illustrated by an electron trigger, using a simplified pixel tracker.
To achieve a real-time trigger (e.g. processing HL-LHC collision events at 40 MHz), simple algorithms are developed for reconstructing pixel-based tracks and track isolation, utilizing look-up tables based on pixel detector information.
Significant gains in electron trigger performance are seen when pixel detector information is included.
In particular, a rate reduction up to a factor of 20 is obtained with a signal selection efficiency of more than 95\% over the whole $\eta$ coverage of this detector. 
Furthermore, it reconstructs p-p collision points in the beam axis (z) direction, with a high precision of 20 $\mu$m resolution in the very central region ($|\eta| < 0.8$), and, up to 380 $\mu$m in the forward region (2.7 $< |\eta| <$ 3.0). 
This study as well as the results can easily be adapted to the muon case and to the different tracking systems at LHC and other machines beyond the HL-LHC. The feasibility of such real-time processing of the pixel information is mainly constrained by the Level-1 trigger latency of the experiment. How this might be overcome by the Front-End ASIC design, new processors, and embedded Artificial Intelligence algorithms is briefly tackled as well.}
\keywords{High Luminosity LHC upgrade, First-level trigger upgrade, Real-time track trigger, Pixel detector based track reconstruction algorithm, Pixel-based track isolation, Pixel-based track momentum determination, Pixel-based precise vertex resolution}

\begin{document} 
\maketitle
\flushbottom

\section{Introduction}
\label{sec:intro}

The exploration of the overall Higgs sector will enter a new era with the High-Luminosity LHC (HL-LHC). 
It will focus on processes with much smaller cross-sections around the femtobarn level and even below. 
It will request at the same time very high precision for handling huge backgrounds. 
The HL-LHC will run at a higher luminosity (up to 10 times) and with higher pileup rates (3 times), compared to the present values in Run 3. This will lead to a huge increase in the data rate to be recorded, as well as a more sophisticated selection requested by the Physics and the machine running conditions. It will impose stringent and challenging detector performances, especially at the Level-1 (L1) trigger selection level. 
%This is the main focus of this paper.

%At the high-luminosity upgrade of the LHC (HL-LHC), the proton-proton collision rate will be increased by almost a factor of three with respect to the LHC Run 3 operation. 
%It is thus essential to develop the most powerful tools to face this to search for new physics and to study physics at the electroweak scale.
The L1 trigger using the pixel detector can serve as one of the powerful triggers at the earliest hardware-based data selection stage. It can indeed further strengthen the selection of leptons; it can achieve a higher rejection factor in a high-rate collision environment while allowing lower $p_T$ selection on leptons and extended coverage of the forward region, even facing higher pileup rates. It provides a precise primary vertex determination, of particular importance for high pileups. Its role is crucial for the precise determination of secondary vertices (e.g. b-tagging).

There is a large variety of important electroweak and Beyond Standard Model Physics motivations for selecting one or several leptons (electrons or muons) in real-time at LHC. The Higgs pair production, the overall Top-Higgs sector (e.g. 4 tops, $t\bar{t}H$ and beyond), the Lepton Flavour violation, some Dark Matter cases, and long-lived particles are among the many essential topics that will benefit from such a trigger.
%This is the main focus of this paper.

%The physics motivations of a real-time pixel-based trigger cover generic leptons which involve electroweak physics and new physics phenomena with lepton, and a large spectrum of interesting physics cases essential for the HL-LHC upgrade. namely the Double Higgs production, the Lepton Flavour violation, and the long-lived particles that cover various appealing scenarios beyond the Standard Model.

All the four main experiments (ALICE, ATLAS, CMS, and LHCb) that will run at the HL-LHC (also denoted as Phase-2) are preparing very new, important, and impressive upgrades of all the key parts of their experiment in order to confront at best these new conditions. 
The triggering systems of ATLAS, CMS, and LHCb have been the object of innovative redefinition and developments, where the tracking system plays an even more essential role. They all now start addressing upgrades for the second stage of HL-LHC towards 2035.

This work concentrates on the tracking trigger system and especially the added value of including the innermost part, i.e. the microvertex or pixel detector. The two general purpose experiments ATLAS and CMS are the considered cases, also as showcases for future high energy colliders.
Both experiments are undergoing a complete reconstruction of their tracking system based on Silicon technology, already pioneered by the CMS experiment in Phase-1~\cite{CMS-trackPhase1, CMS-trackPhase1-Add}.

The ATLAS trigger system is drastically upgraded for Phase-2 ~\cite{atlasTDR-TDAQ}. 
A single Level-0 (L0) hardware trigger, processes the data from the calorimeters and muon detectors at 40 MHz, identifying the physics objects and computing event-physics quantities within a total latency budget of 10 $\mu$s. The resulting L0 trigger decision is transmitted to all the detectors with, for the first time in the ATLAS experiment, the overall inner tracking (ITk) detector. Moreover, the innermost pixel-based tracker in the ITk, extending the tracking capability up to $\eta$ of 4.0,  is integrated into this trigger strategy \cite{Filimonov_2020}.
The resulting trigger data and detector data are transmitted through the Data Acquisition (DAQ) system at 1 MHz. The last decision step, i.e. the Event Filter system (EF) is a heterogenous processor farm (including FPGAs and GPUs); it reduces the event rate to 10 KHz. Apart from L0 which is made purely by hardware, the following selection stages in the DAQ chain are based on commercial processing units and the use of advanced algorithms for instance, Graph Neural Networks (GNN) for finding the right track candidates.
An important amendment~\cite{atlasTDR-TDAQ-Add} has been added to the original ATLAS TDR ~\cite{atlasTDR-TDAQ}. 
It is related to the Event Filter design which relies on the industrial progress on new CPUs as well as FPGAs and GPUs and on the Artificial Intelligence (AI) field.
%The L1 trigger (i.e. indeed the second stage in this trigger strategy) with an input rate at 1 MHz and latency of 25 $\mu$s (at least for the first period of the HL-LHC), includes for the first time in this experiment the overall ATLAS tracking system. 
%This hardware-based Track Trigger (HTT) uses both strip and pixel (microvertex) parts of the tracking information. 
%Important also to note the two stages in the ATLAS HTT, with rHTT (regional HTT) based on the Regions of Interest (RoI) defined by L0 and the second stage gHTT (global HTT) that further improves the track trigger reconstruction with the overall tracking information.

CMS pursues as well at HL-LHC with the original CMS trigger strategy i.e. a 2 stages trigger: the hardware-based L1 at 40 MHz followed in, a second stage, by the software-based High-Level Trigger (HLT). 
In addition, CMS includes the new outer tracker within the new L1 trigger at HL-LHC. 
The sophisticated and innovative L1 trigger decisions of CMS for HL-LHC, including the overall calorimeter, the muon detectors, and the outer tracker system, will have to be achieved within the total L1 latency of 12.5 $\mu$s. 
This is presented in the Phase-2 Upgrade of the CMS Level-1 Trigger TDR by CMS~\cite{CMS-phase2L1TDR}. 
% The CMS-like challenging scenario is chosen here as a showcase scenario.

Over several years, feasibility and performance studies on an L1 pixel-based trigger have been developed within the framework of the CMS Phase-2 upgrade studies. 
They focused on the possible benefits of including the pixel information for improving the electron selection and including the b tagging, both at L1~\cite{Moon_2015, Moon_2016}. 
This option is not currently retained by CMS for Phase-2~\cite{CMS-phase2L1TDR}. 
However, the goal here is to further develop these studies, to keep open an eventual beyond-baseline option in Phase-2 for including the pixel information in a beneficial and realistic way in the L1 trigger.  
Unlike what is presently done by ATLAS at the second 1 MHz step in the trigger chain or CMS HLT\footnote{As in Phase-1, the processing of the pixel detector information for Phase-2 in CMS, is currently performed at HLT \cite{CMS-HLTPhase2TDR, Bocci_2020}}, the purpose of this study is to explore the feasibility and interest of including part of the pixel information at 40 MHz, with, as objective, a second stage upgrade after the start of the HL-LHC.

This paper concentrates, as a first example, on a new approach for further improving the electron trigger over the overall pseudorapidity ($\eta$) coverage, with increased granularity of the calorimetry and a novel and detailed track reconstruction algorithm. This can be applied to triggering applications other than the electron case. 
%The simulation framework and generation of MC samples are described in %Section~\ref{sec:simulation} and Appendix A.
%The L1 pixel track reconstruction algorithm (PiXTRK), first introduced in~\cite{Moon_2015, Moon_2016}, is developed in great detail in this paper, and reported in Section~\ref{sec:PiXTRK}. 
%Section~\ref{sec:TrkIso} explains the added improvement of this track trigger algorithm by including the track isolation in the pixel clusters used by the PiXTRK algorithm. 
%Section~\ref{sec:result} gathers the main results of this study. 
It stresses the benefits in the L1 trigger performance by including the pixel information in the electron trigger used here as an example case. 
%Section~\ref{sec:future} summarizes the two main categories of technological challenges to be overcome to make this option feasible within the HL-LHC scenario.
%This implies the Pixel Front-End ASIC and the real-time related algorithms to perform this triggering scheme, thanks also to the novel development in the processor technology.
%Section~\ref{sec:remark} concludes by showing the perspectives for a possible beyond baseline upgrade at HL-LHC and also for application to future colliders.

%%%% The comments as follows
%-> although it is understandable to try introducing all sections in this paragraph, the inclusion of all makes the reader get lost from the main "story" of the paper. I would recommend to keep only the (introduction of) Sections 3 and 4 here, moving the sentence on the simulation framework into Section 2, and the (introduction of) Section 5 and after into the start of Section 5, to keep the main story simple. The paragraph could be, e.g.

Section 2 briefly describes the simulation framework completed for details in Appendix A. The L1 pixel track reconstruction algorithm (PiXTRK), first introduced in [5, 6], is developed in great detail in Section 3. An improvement of this track trigger algorithm by including the track isolation in the pixel clusters is introduced in Section 4.
The results and performances are summarized in Section 5. The main technological challenges are tackled in Section 6. Section 7 concludes with the perspectives.

\section{Simulation Framework}
\label{sec:simulation}

%As pointed out in Section~\ref{sec:intro}, including the pixel detector within the overall L1 trigger CMS upgrade for HL-LHC has not been endorsed by the CMS collaboration~\cite{CMS-phase2L1TDR}. 
%Therefore
%The work reported here is performed at a “generic level”, although using 
%and in some sense more general approach, but keeping 
%the overall CMS-Phase-2 upgrade as an example case. 

The simulation framework and generation of MC samples are described in this section and Appendix A. The performance studies are based on a rather generic detector design, although still well representative of the key detector components of a multipurpose experiment for HL-LHC. The simulation framework here is based on \texttt{DELPHES}~\cite{Delphes3} with a version  tuned for HL-LHC 
%For this reason we choose to use the multipurpose detector fast simulation framework \texttt{DELPHES}~\cite{Delphes3}, with a version adapted to the overall CMS-Phase-2 detector design, although simplified, as in particular, the rather simple pixel detector geometry (see Fig.~\ref{fig:Phase2Pixel}).
%The other important reason to
The use of \texttt{DELPHES}
%is that it 
allows generating very large samples of data as requested for this study. 
Parton processes generated with \texttt{PYTHIA} 8.2~\cite{PYthia8} which includes a library of models for initial- and final-state parton showers, and multiple proton-proton interactions (pileup) from inelastic collisions are parameterized by \texttt{DELPHES}. 
%The studies performed with the full simulation software in the context of the CMS HL-LHC upgrade are used here for improving the \texttt{DELPHES} modeling. This is stressed in the relevant sections of this paper.

%Three types of fully simulated MC samples are generated using \texttt{DELPHES} in this study: (i) 5 million single electron gun events without pileup to measure 3$\sigma$ boundaries of signal windows; (ii) 1 million single electron gun events with 200 pileup for measurement of the L1 trigger efficiency; (iii) 10 million of minimum bias events with 200 pileup to estimate the L1 trigger rate. (ii) and (iii) samples are also used for the pixel-based charged isolation algorithm.

%Although the \texttt{DELPHES} simulation allows producing more realistic results than a simple ``parton-level'' framework, it has, as any fast simulation tool, some limitations such as a The effects due to a simplified detector geometry,%neglectedsecondary interactions,the multiple scatterings, photon conversion, and bremsstrahlung. To overcome these aspects, 
%as stressed above, 
%a careful comparison is done with studies based on a full detector simulation, e.g. the upgraded CMS detector for HL-LHC. This is detailed in the relevant sections of this paper.
The way to handle the effects due to multiple scattering, photon conversion, and bremsstrahlung are carefully included in the simulation package and detailed in the corresponding sections.
%in the case of the upgraded CMS detector for HL-LHC is done.

%In particular, to properly reproduce the signal events for the pixel-based track isolation, another sample of bremsstrahlung events has been added to the single electron gun sample with 200 pileup. The modeling of this contribution is based on a study performed with the detailed CMSSW full simulation package~\cite{CMSSW} of the CMS detector which indeed includes this Physics process.

\begin{figure}[hbtp]
    \centering
    \includegraphics[width=\textwidth]{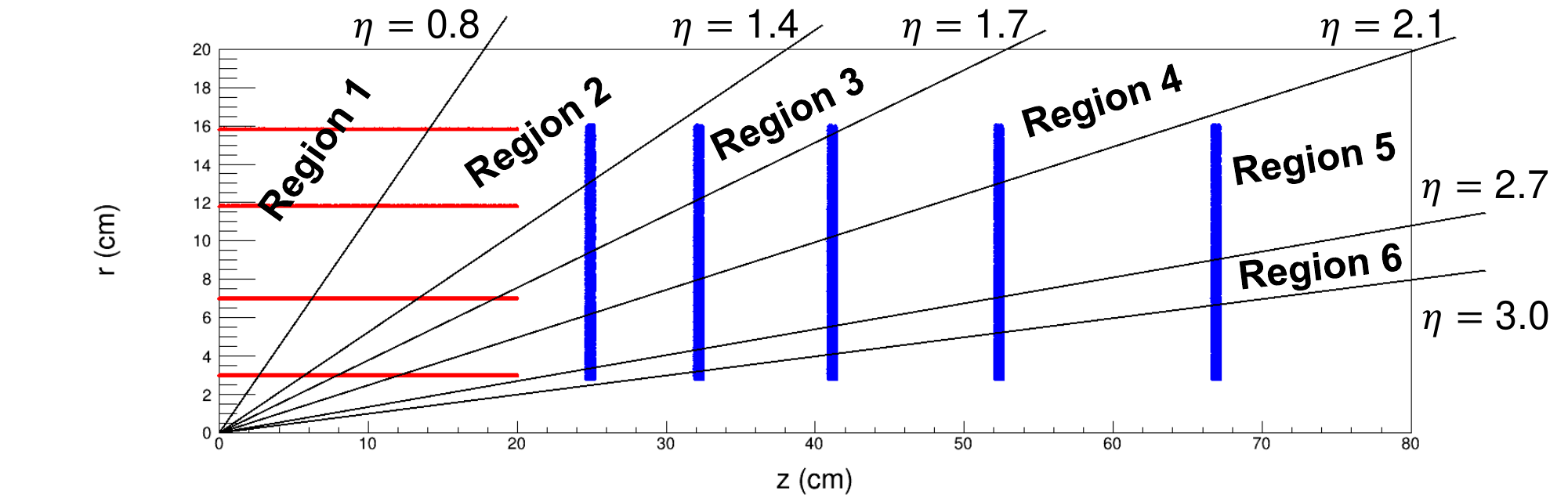}
    \caption{Schematic quadrant view of the pixel detector layout used as a showcase. It covers $|\eta|\;\leq$ 3 in pseudorapidity and shows the segmentation in $\eta$ regions used in this study.
}
    \label{fig:Phase2Pixel}
\end{figure}

A simplified pixel detector set-up is implemented with 4 barrel layers covering the pseudorapidity range, $|\eta|\;\leq$ 1.457, and 5 disk layers covering 1.457 $< |\eta|\;\leq$ 3.0. 
Both ATLAS and CMS will use for Phase-2, the same small pitch silicon pixel sensors of 100-150~$\mu$m thickness, with pixel size of 50$\times$50~$\mu$m\textsuperscript{2} for the barrel part and 25$\times$100~$\mu$m\textsuperscript{2} for the endcap and forward parts~\cite{ATLAS-phase2tracker, CMS-phase2tracker}.
The pixel detector set-up is divided into six different  $\eta$ regions. This is used in the description of the pixel-based track trigger in Section~\ref{sec:TrkIso}, and shown in Fig.~\ref{fig:Phase2Pixel}. 
These regions are defined by the 3-out-of-4 pixel clusters track reconstruction strategy (see Section~\ref{sec:PiXTRK}). 
%As already stressed this is a very simplified design if we compare with the 12 disks in CMS Phase-2 or the quite sophisticated design of the ATLAS pixel detector for Phase-2 (5 barrel layers and titled end cap disks). Therefore the results obtained will be kind of slightly underestimated as discussed later.  
The detector description in this simulation framework also includes a fine-grained barrel calorimeter and an endcap highly granular calorimeter a-la-HGCAL CMS Calorimeter.

In this simulation environment, as well as in any simulation at the first-level trigger, the beam spot (B0) is reduced to a point ($\phi$=0, $\eta$=0, z=0) and z0 is the collision position along the beam axis (z=0), in the rest frame of the experiment.

More details on the DELPHES simulation used here are in Appendix A.

\section{Real-time Pixel track reconstruction algorithm with vertexing capability}
\label{sec:PiXTRK}

This section presents the PiXTRK algorithm, able to be performed in real-time (i.e.~at 40 MHz) and with the determination of the vertex as a ``by-product''. 
The example case is the electron trigger. 
The used simulation samples for this study are minimum bias data for the background, and single electron gun plus pileup (200) events for the signal. 
The signal sample consists only of single prompt electrons ($e^-$) with a $p_T$ range ($p_T$ > 10 GeV) corresponding to electron events produced in p-p collisions at 14 TeV c.m energy. The signal windows are defined with the electron sample. 
For positrons, the $\Delta \phi$ signal windows are directly derived from the ones of the electrons, taking into account their opposite curvature in the transverse plane. 
Both electrons and positrons have the same $\Delta \eta$ signal windows.

\subsection{Real-time Pixel-based track reconstruction algorithm: the strategy}
\label{PiXTRKStrategy}

Unlike the tracks reconstructed in real-time (i.e. at 40 MHz) with the outer tracker information in the case of the CMS upgrade for HL-LHC~\cite{CMS-phase2tracker}, the tracks that are based on the pixel information request to be ``seeded'' i.e. to be searched for within a Region of Interest (RoI) (``L1 clusters/tracks'') that is provided either by the electromagnetic (EM) calorimeter for the electrons, by the muon detectors for the muons, or the outer tracker in the case of the b-tagging. 
This is due to the very high information rate provided by the extremely high granularity of the microvertex detectors, further increased for the HL-LHC (see Section~\ref{sec:future})
%
%
%Unlike the tracks reconstructed in real-time (i.e. at 40 MHz) with the outer tracker information (``L1 tracks'') in the case of the CMS upgrade for HL-LHC~\cite{CMS-phase2tracker}, the tracks based on the pixel information request to be ``seeded'' i.e. to be searched for within a Region of Interest (RoI). 
%This is due to the very high information rate provided by the extremely high granularity of the microvertex detectors, still increased for the HL-LHC (see Section~\ref{sec:future}).
%
%Therefore the global strategy is based on a {\it seed} that can be provided either by the electromagnetic (EM) calorimeter for the electrons, or the muon detectors for the muons, or the outer tracker in the case of the b-tagging. 

What is proposed here for the electron case, is easily applicable as well to the muon case. Another interesting goal for a Level-1 trigger is for real-time b-tagging. This has been addressed in the case of the CMS experiment for HL-LHC~\cite{Moon_2016}. Because of the increasing importance of b-tagging, this objective will be revisited in another dedicated study, including the latest technological developments, and applied to all the experiments at the HL-LHC and future HEP machines.
%is addressed in~\cite{Moon_2016} with the specific CMSSW framework; 
%it will be the object of a new paper, because of the increasing importance of the b-tagging for all the experiments at HL-LHC and at future HEP machines.

The pixel-based reconstruction follows the concept of the ``PiXTRK'' algorithm, first introduced in ~\cite{Moon_2015, Moon_2016}. 
Its detailed strategy as well as its realistic implementation is developed for the first time in this study.

In order to optimize the efficiency of the fast reconstruction efficiency, PiXTRK allows for one missing pixel cluster (e.g. some detector inefficiency) in the pixel track segment. 
This means, for the detector design considered here, that PiXTRK uses only 3 pixel clusters within 4 selected pixel layers or disks, to reconstruct the pixel tracks in the barrel and/or the endcap or the forward regions of the microvertex. 
We label it as the ``3-out-of-4 pixel clusters'' reconstruction strategy.\\

The PiXTRK algorithm works in 3 steps:
\begin{itemize}
    \item {\it Step 1: (RoI) Identifying the pixel clusters in each RoI}.
    
    The first step in pattern recognition for reconstructing the tracks in the microvertex implies identifying the relevant clusters. 
    This procedure is {\it{Region of Interest, RoI, based}}. 
    One RoI is assigned to each L1 e/$\gamma$\footnote{L1 e/$\gamma$ designates the physics object based on the EM calorimeter cluster created by an electron or a photon as reconstructed by the level-1 trigger.} candidate that corresponds to an L1 EM  trigger cluster with a measured transverse energy, $E_{\textrm{\scriptsize T}}$\footnote{L1 e/$\gamma$ $E_{\textrm{\scriptsize T}}$ is the transverse energy of the e/$\gamma$ object, measured by the level-1 trigger in the EM calorimeter}. 
    The RoI is defined, by linking the L1 EM cluster (EM) to the {\it{beam origin}}, B0, defined by the coordinates ($\phi$ (azimuthal angle)=0, $\eta$ (pseudorapidity)=0, z (beam direction)=0). 
    The RoI is chosen to have an opening of 0.1 rad in $\phi$ and covers the whole $|\eta|$ region, i.e. 3.0. This definition of the RoI is thus compatible with a $\it {wedge-shaped~RoI}$. The $|\eta|$ range is defined by the present detector design coverage and could be further segmented into two, three, or four parts in $|\eta|$ if needed i.e. if too large for the trigger rate. This allows proper handling of the spread of the primary z vertex.
    %as discussed in the last section (see Section~\ref{sec:result}).
    
    For each RoI, the selected pixel clusters in each layer or disk of the pixel detector, are therefore those which are comprised within the $\Delta\phi$ window (Fig.~\ref{fig:step-1}):
    
    \begin{equation} \label{eq:Step1Define}
        \centering
        \Delta\phi = \phi(\textrm{B0}, L_{i})-\phi(\textrm{B0}, \textrm{EM}) < 0.1
    \end{equation}

    %The size of the RoI define by 0.1 in $\phi$, and the full $\eta$ coverage of the EM calorimeter and pixel detector, thus %by $|\eta|$=3 corresponds to a reasonable trigger rate (see Section~\ref{sec:result}).
    The size of the RoI we choose corresponds to a reasonable trigger rate (see Section~\ref{sec:result}).
        
    \begin{figure}[htbp]
        \centering
        \includegraphics[width=0.65\textwidth]{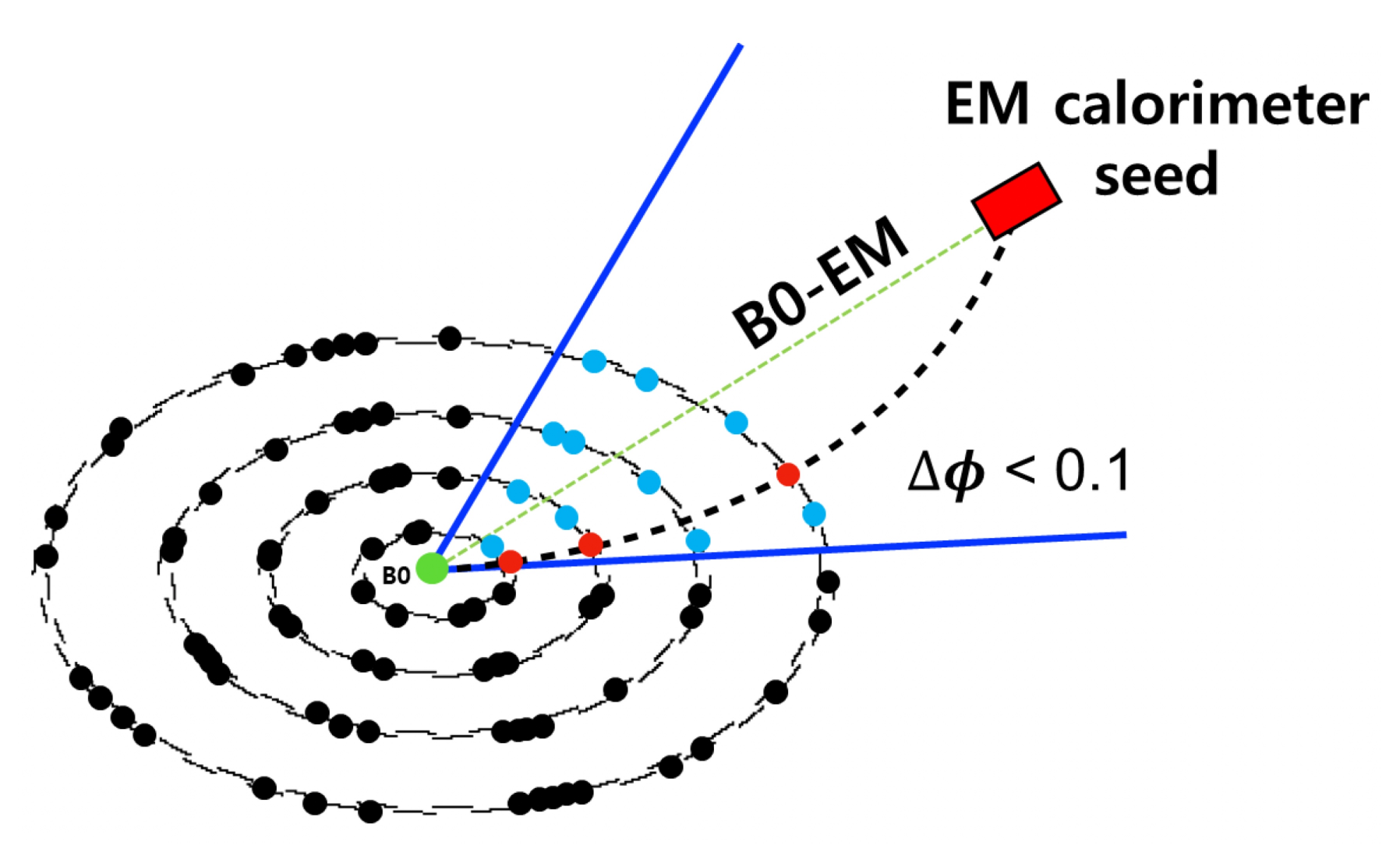}
        \caption{Step 1 of the PiXTRK algorithm for reconstructing track segments based on pixel only information: definition of the Region of Interest.}
        \label{fig:step-1}
    \end{figure}
    
    \item {\it Step 2: (Vector Segment Search) Refined pattern recognition seeded by the L1 EM cluster}.
    
    The next step in the pattern recognition consists in identifying the pixel clusters gathered by pairs, within defined $\Delta\phi$ and $\Delta\eta$ 3$\sigma$ boundaries w.r.t. the segment (B0, EM), as sketched in (Fig.~\ref{fig:step2}).
    The $\Delta\phi$ and $\Delta\eta$ signal windows are computed as a function of the measured EM E$_{T}$. 
    In operation, they will be provided by the analysis of real L1 single electron data allowing the determination of the $\Delta\phi$ and $\Delta\eta$ 3$\sigma$ boundaries.

    For all the pixel clusters selected by the equation~\ref{eq:Step1Define}, for each layer/disk, the algorithm considers each combination of a pair of pixel layers or disks in order to form all the possible [$L_{i}, L_{j}$] track segments. 
    It then compares the matching in ($\phi$, $\eta$) of each of these [$L_{i}, L_{j}$] segments with the segment [B0, EM] joining the beam origin, B0, with the EM cluster. 
    This matching is defined by the following set of boundary conditions: 

    \begin{equation} \label{eq:step2-2Define}
        \centering  
        \Delta\eta_{i, j} = \eta(L_{i}, L_{j})-\eta(\textrm{B0}, \textrm{EM}) < 3\sigma
    \end{equation}
    \begin{equation} \label{eq:step2-1Define}
        \centering
        \Delta\phi_{i, j} = \phi(L_{i}, L_{j})-\phi(\textrm{B0}, \textrm{EM}) < 3\sigma
    \end{equation}

    where {\it i, j} $=$1…4 or 5 (if disks only) and {\it i} $\neq$ {\it j}. 
    The pixel cluster in each layer/disk is then only selected if it passes the 3$\sigma$ requirements as specified in Equation~\ref{eq:step2-2Define} and Equation~\ref{eq:step2-1Define}.
    
    \begin{figure}[htbp]
        \centering
        \includegraphics[width=0.85\textwidth]{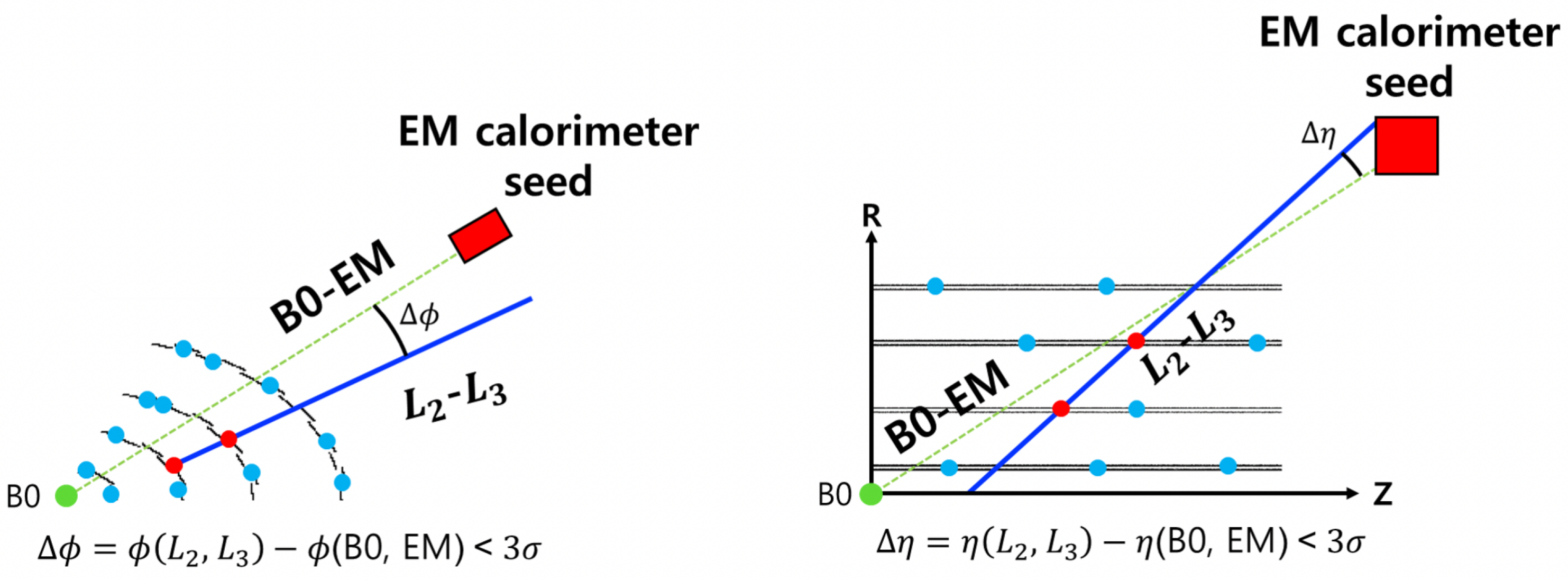}
        \caption{Step 2 of the PiXTRK algorithm for reconstructing track segments based on pixel only information: refining the coupling in $\eta$ and $\phi$ of the relevant Pixel clusters.}
        \label{fig:step2}
    \end{figure}
    
    \item {\it Step 3: (Bending Correction) The standalone pattern recognition}
    
    This step aims to further reduce the number of combinations with fake clusters. 
    To do so, the algorithm considers all the possible 3-layer combinations with the surviving 2-layer (disk) vector selection in each of the 4 barrel layers or in each of the barrel layers plus disks combinations or in each of the 5 disks combinations, depending on where the RoI is located in $\eta$.
    The beam origin (B0) is also included. 
    This is sketched in Fig.~\ref{fig:step3}.
    
    \begin{figure}[htbp]
        \centering
        \includegraphics[width=0.85\textwidth]{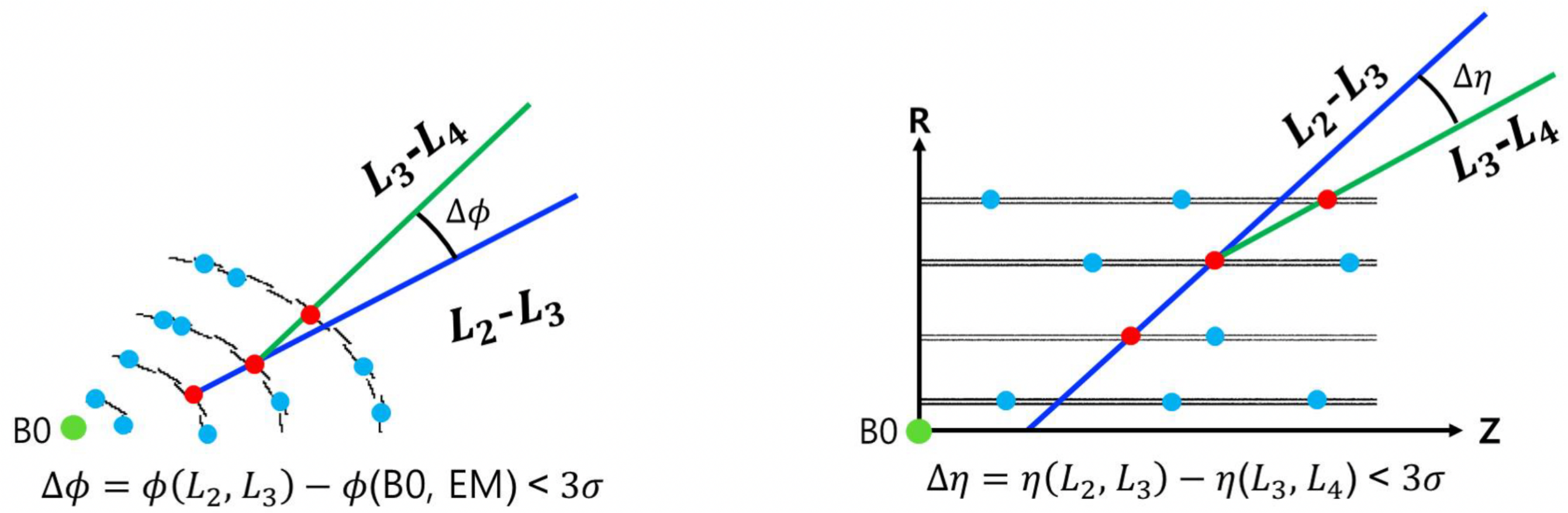}
        \caption{Step 3 of the PiXTRK algorithm for reconstructing track segments based on pixel only information: The standalone reconstruction is based on 3 out of 4 pixel clusters.}
        \label{fig:step3}
    \end{figure}
    
    The pixel clusters must satisfy  all the signal windows requirements within 3 standard deviations :

    \begin{equation} \label{eq:step3-2Define}
        \centering  
        \Delta\eta_{i, j, k} = \eta(L_{i}, L_{j})-\eta(L_{j}, L_{k}) < 3\sigma
    \end{equation}
    \begin{equation} \label{eq:step3-1Define}
        \centering
        \Delta\phi_{i, j, k} = \phi(L_{i}, L_{j})-\phi(L_{j}, L_{k}) < 3\sigma
    \end{equation} 

    where {\it i, j, k} $=$1…4 or 5 (if disks only) and {\it i} $\neq$ {\it j} $\neq$ {\it k}. 
    The pixel cluster in each layer/disk is then only selected if it passes the 3$\sigma$ requirements as specified by Conditions~\ref{eq:step3-2Define} and ~\ref{eq:step3-1Define}.

    This 3-step procedure thus performs progressively a good pattern recognition. 
    This pattern is then used for a simple track fitting.
\end{itemize}

The optimized combinations of pixel barrel layers and end-cap disks based on the $\eta$ range they cover in one of the six $\eta$ regions of the microvertex (Fig.~\ref{fig:Phase2Pixel}) are listed in Table~\ref{tab:pixelCombi}. This optimization can be used to further decrease the overall data rate when applying the L1 Pixel trigger. It is also a basic tool for the detailed computation of the PiXTRK algorithm explained in Section~\ref{sec:PixTRKLUT}.

\begin{table}[htbp]
    \centering
    \scalebox{1}{
        \begin{tabular}{|c|c|c|}
            \hline
            Region & $\eta$ Range & Pixel Combination \\
            \hline
            Region 1 & $|\eta| < $ 0.8 & 1-4 layer \\
            \hline
            Region 2 & 0.8 $ < |\eta| < $ 1.4 & 1-3 layer, 1 disk \\
            \hline
            Region 3 & 1.4 $ < |\eta| < $ 1.7 & 1-2 layer, 1-2 disk \\
            \hline
            Region 4 & 1.7 $ < |\eta| < $ 2.1 & 1 layer, 1-3 disk \\
            \hline
            Region 5 & 2.1 $ < |\eta| < $ 2.7 & 1-4 disk \\
            \hline
            Region 6 & 2.7 $ < |\eta| < $ 3.0 & 2-5 disk \\
            \hline
        \end{tabular}
    }
    \caption{The combinations of pixel layers and disks for the different regions in the $\eta$ coverage of the pixel detector.}
    \label{tab:pixelCombi}
\end{table}

\subsection{Real-time Pixel-based track reconstruction algorithm: the Look-Up-Tables as computing basis}
\label{sec:PixTRKLUT}

%In order to efficiently perform the 3 steps of the PiXTRK algorithm, the
Look-up tables (LUT) are used as a computing basis for efficiently performing the three steps of the PiXTRK algorithm
They include the information needed for applying the requirements of Step 2 and Step 3 defined in Section~\ref{PiXTRKStrategy}. 
%Indeed, as shown in Section~\ref{PiXTRKStrategy}, both Step 2 and Step 3 
These Steps impose constraints on both the $\eta$ and $\phi$ coordinates of the sets of pixels remaining after each of these steps. These constraints (e.g. Conditions~\ref{eq:step2-2Define} and~\ref{eq:step2-1Define} for Step 2 and Conditions~\ref{eq:step3-2Define} and~\ref{eq:step3-1Define} for Step 3) are defined by the size of the 3$\sigma$ windows in $\eta$ and in $\phi$, within which the pixel coordinates must be included. In addition, the size of these windows also varies with the $E_{\textrm{\scriptsize T}}$ of the electron candidate provided by the EM calorimeter at L1.

A detailed study of the dependence in $E_{\textrm{\scriptsize T}}$ of the 3$\sigma$ windows in $\eta$ and in $\phi$, for step 2 and step 3 is thus performed over the complete acceptance of the pixel detector$\footnote{Nota Bene: the results of the detailed studies performed with \texttt{DELPHES}, reported here below, were carefully cross-checked and shown to well agree with the results obtained with a full LHC detector simulation}$.
%the full simulation package of the CMS experiment, taken as a showcase.
%underlining again that CMS is }.

\subsubsection{Look-Up Tables describing the dependence in \texorpdfstring{$\eta$}{Lg}}
\label{sec:PixTRKLUTpseudo}
The variation in $\eta$ of the 3$\sigma$ window is studied over the  $E_{\textrm{\scriptsize T}}$ range, 10 to 100 GeV, and the full $\eta$ coverage of the pixel detector.
%Over this full $p_{\textrm{\scriptsize T}}$ range, the 3$\sigma$  window in $\eta$ is found to be constant in size.This window is found to be constant both in $\eta$ and $\phi$ for Step 2, with a value of $\pm{0.01}$ around zero for the $\Delta\eta$-window.
%For step 2, this window size is even constant over the full $\eta$ coverage of the pixel detector. This window size is thus given by a single value, namely: $\pm{0.01}$ around zero. 
This gives a quick and simple decision about whether or not the considered couple of pixel clusters
%are valid or not and thus if they 
must be kept at this early decision step.  
%If not verified, these couple of pixel clusters are disregarded, without having even to verify the condition~\ref{eq:step2-1Define} on the $\Delta\phi$ window.

%For step 3, 
%the detailed study shows that
%three values are found the be sufficient to define the size of the window in $\eta$. It depends on the $\eta$ region where the pixel clusters are located
%this is provided by the L1 EM calorimeter seed to which they are assigned, but also by and then the Front End ASIC with even more precision.

Three $\Delta\eta$-windows are needed for step 3. They correspond each one to different sectors in $\eta$, namely: $\pm{0.002}$ around zero for $0<\eta<1.7$, $\pm{0.004 }$ around zero for $1.7<\eta<2.7$ and $\pm{0.01}$ around zero for $2.7<\eta<3$. 
%Thus again as in step 2,
If the pixel clusters under consideration do not fulfill the Condition~\ref{eq:step3-2Define} corresponding to their $\eta$ location$\footnote{As briefly described in Section~\ref{sec:future}, this location in ($\eta$, $\phi$) is transmitted with even a high precision, by the corresponding Front-End ASIC, triggered by the EM calorimeter seed}$ and as defined here above, they will be discarded without even needing to verify Condition~\ref{eq:step3-1Define}. 
This is a quick and simple cross-check.

Table~\ref{tab:etaLUT} represents the content of the simple LUT that summarizes the size in $\eta$ of the 3$\sigma$ windows for both Step 2 and Step 3, for applying the Condition~\ref{eq:step2-2Define} to~\ref{eq:step3-2Define} in PiXTRK.

\begin{figure}
    \centering
     \subfloat[step 2]{
        \includegraphics[width=0.32\textwidth]{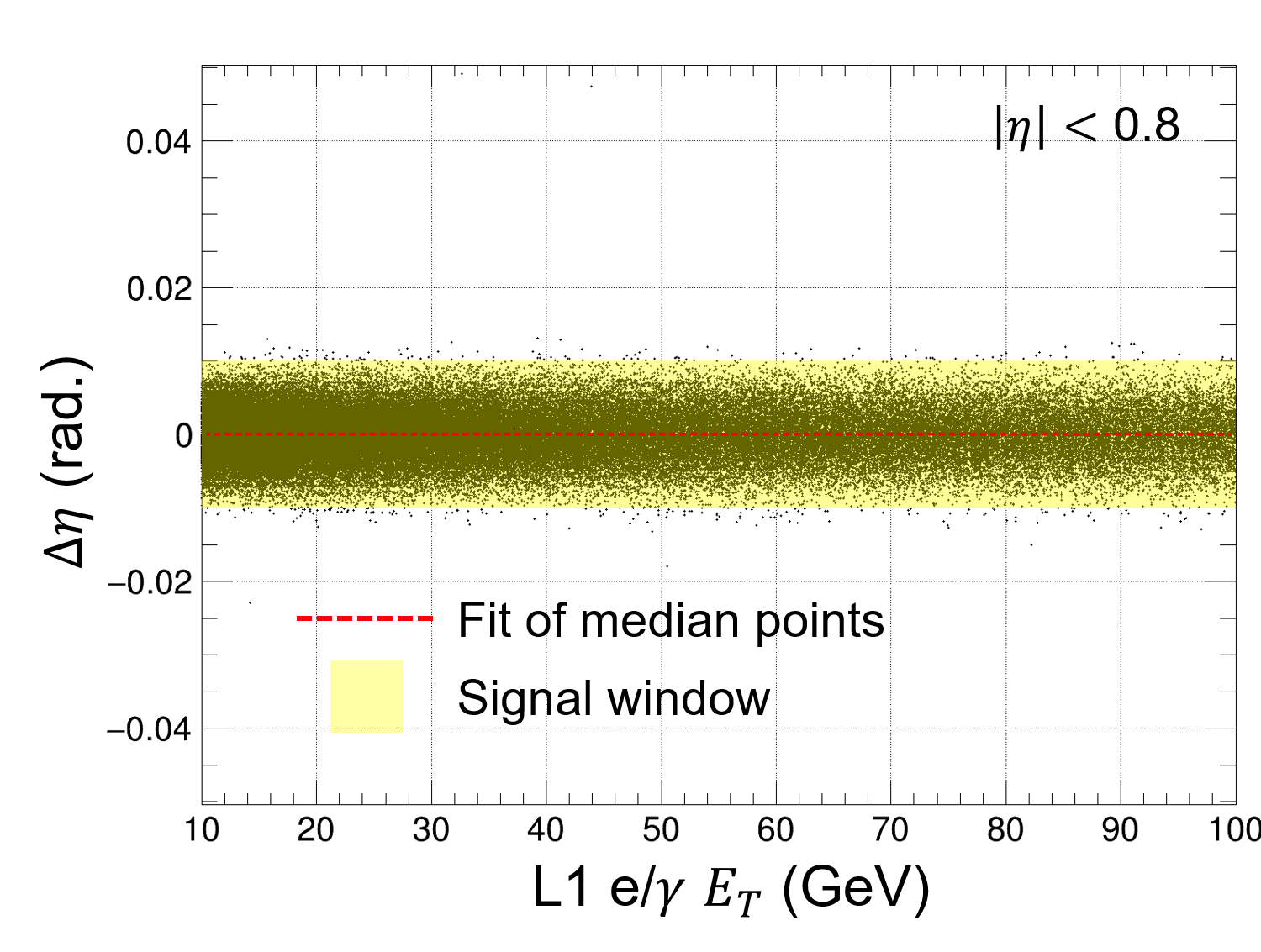}
        \includegraphics[width=0.32\textwidth]{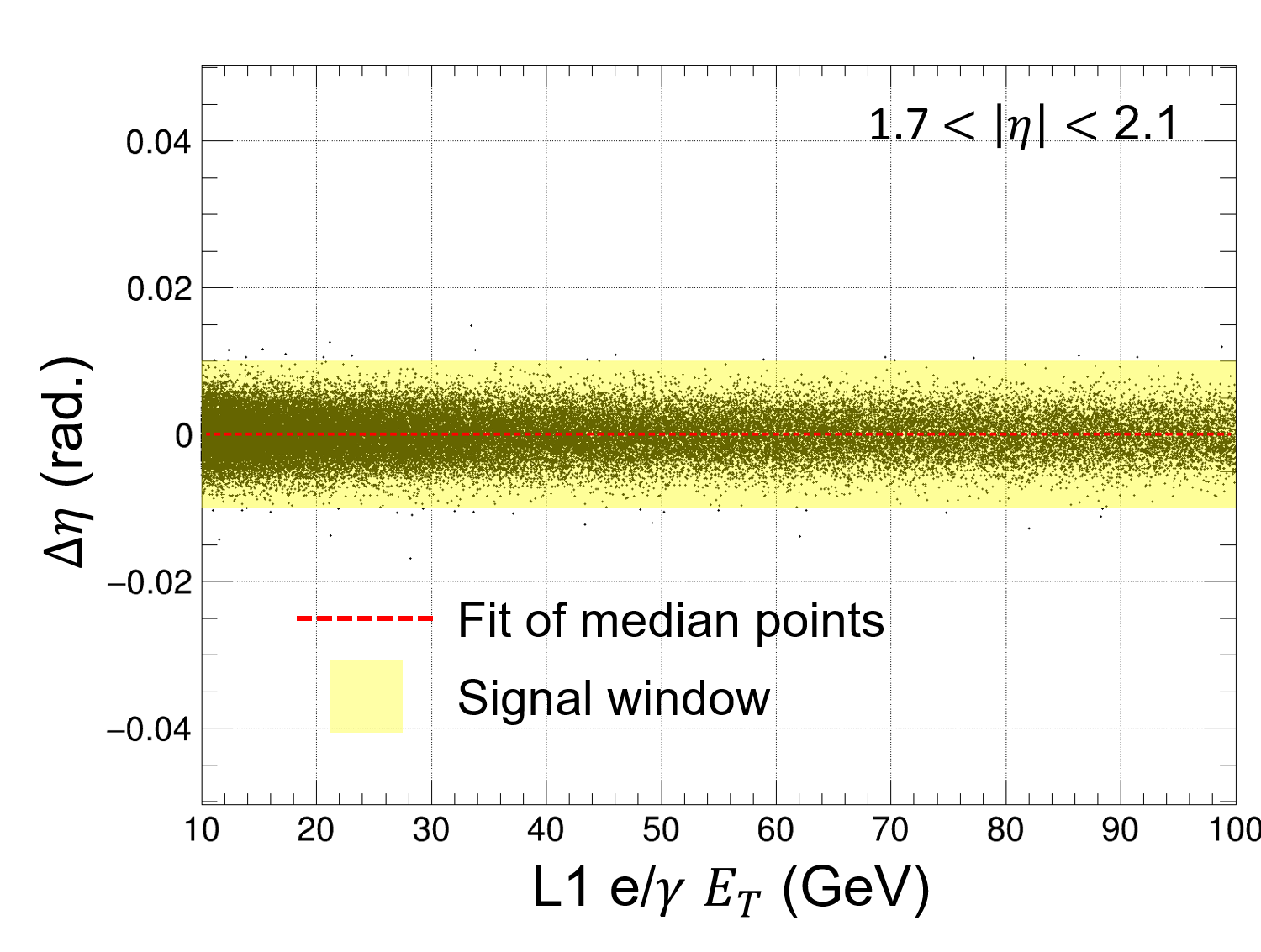}
        \includegraphics[width=0.32\textwidth]{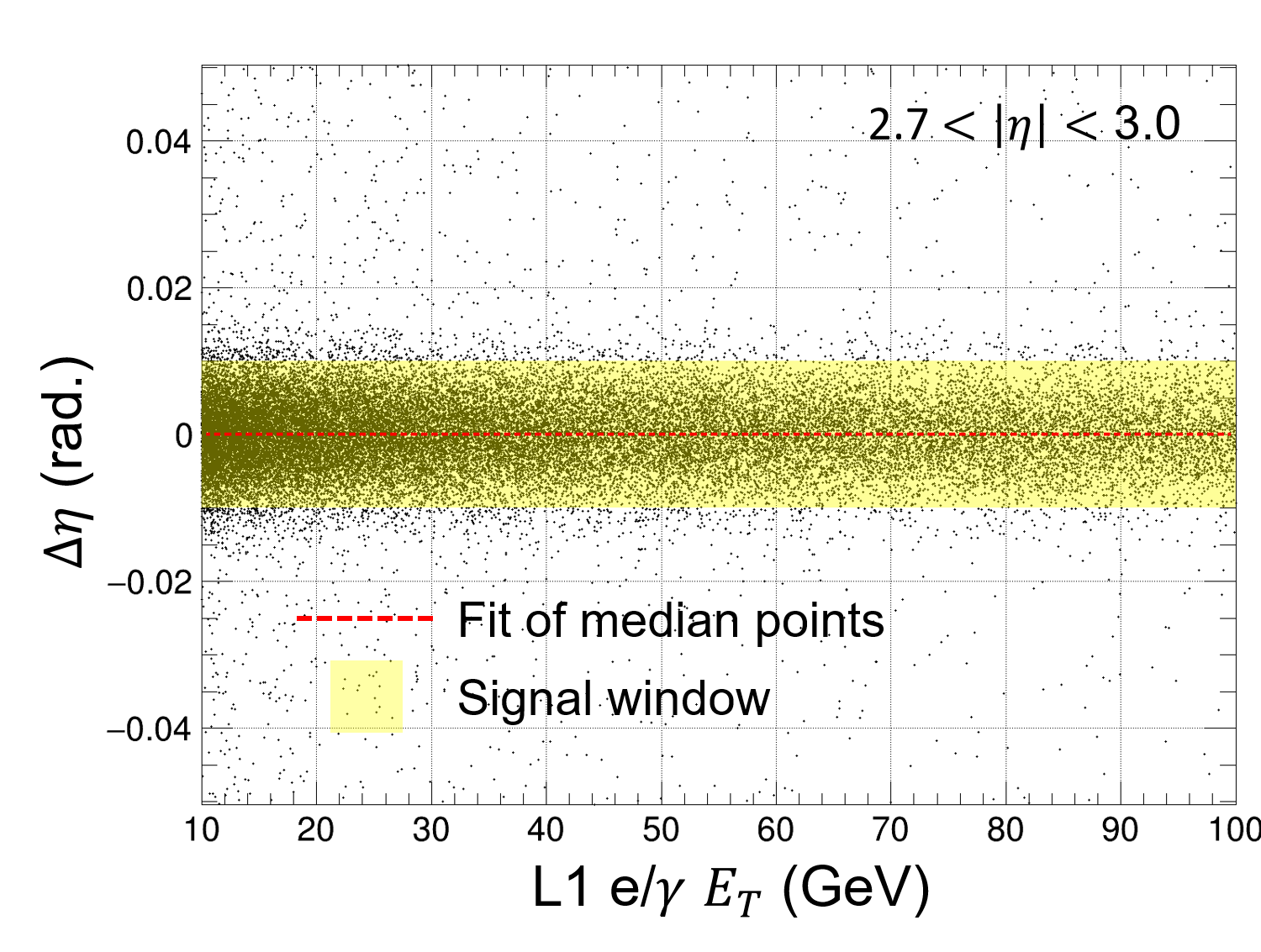}
    }
    
    \subfloat[step 3]{
        \includegraphics[width=0.32\textwidth]{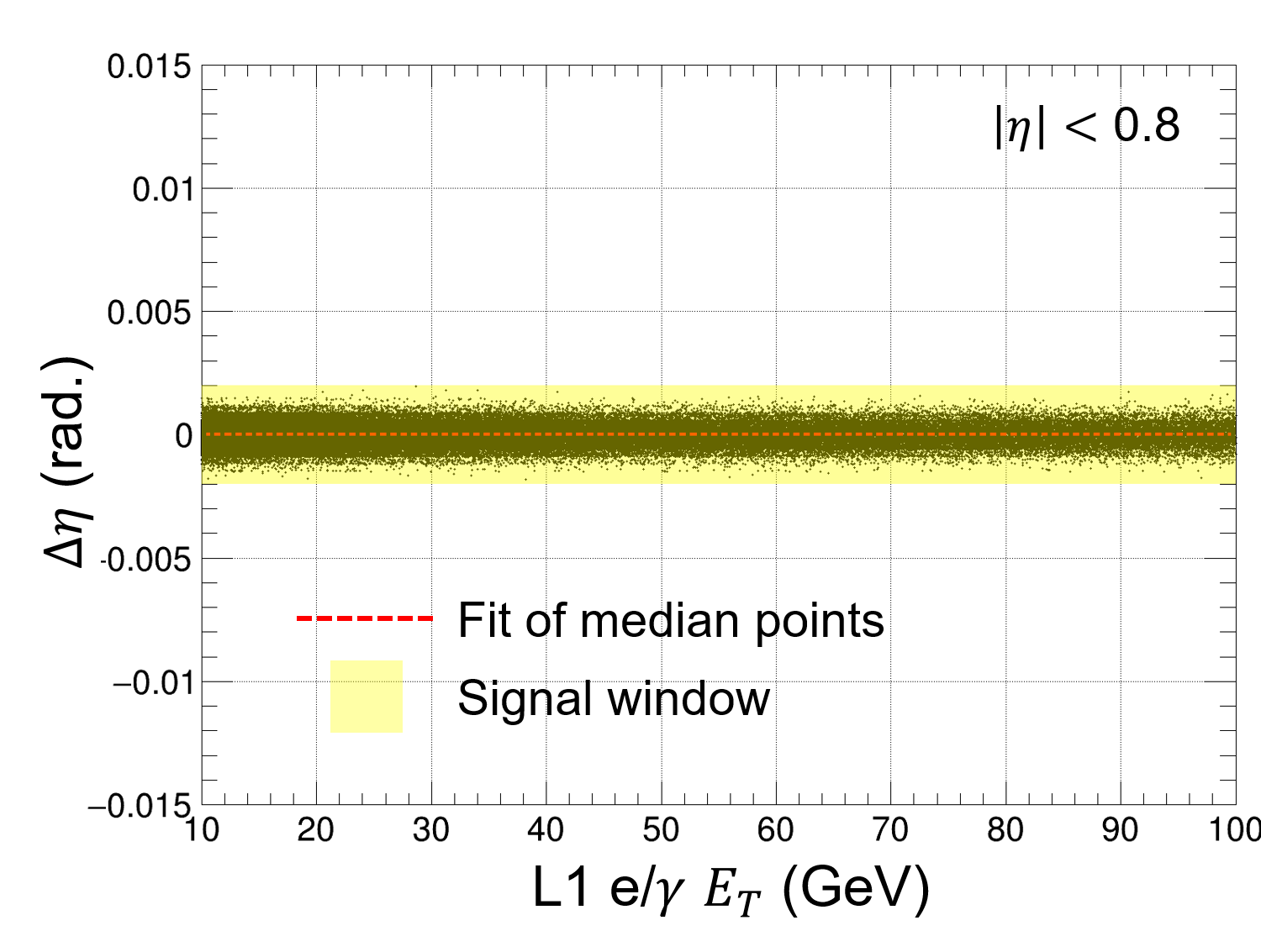}
        \includegraphics[width=0.32\textwidth]{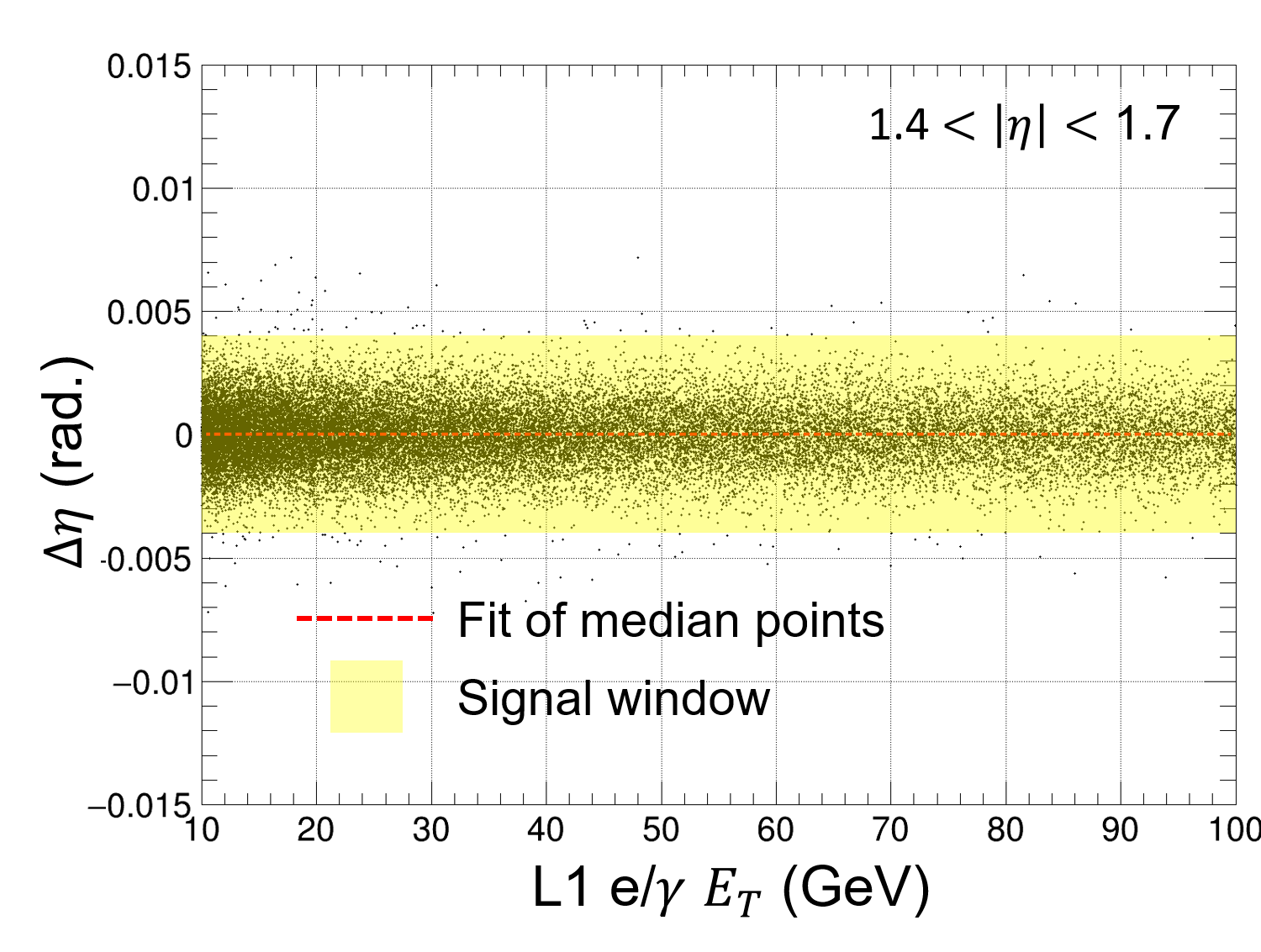}
        \includegraphics[width=0.32\textwidth]{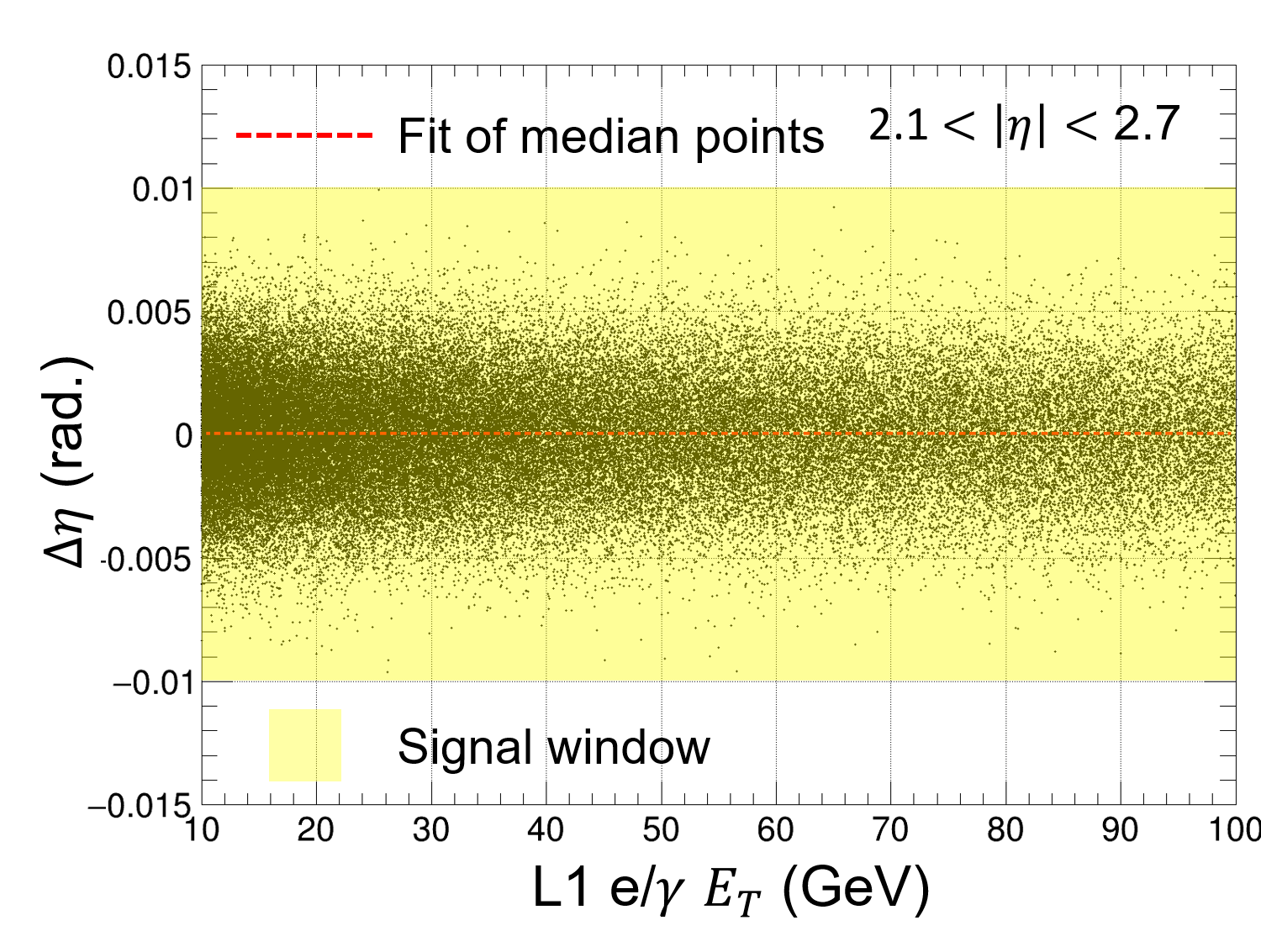}
    }
    \caption{(a) Step 2: Variation, in function of the L1 e/$\gamma$ $E_{\textrm{\scriptsize T}}$, of the size of the 3$\sigma$ window in $\eta$ (yellow band) and of the residuals in $\eta$ (black points) from Equation~\ref{eq:step2-2Define}, for central barrel (left), endcap (middle) and forward (right). (b) Step 3: Equivalent distributions as above, but for step 3, thus corresponding to Equation~\ref{eq:step3-2Define}. These curves correspond to the electron case. Both electrons and positrons have the same $\Delta\eta$ distributions.}   
    \label{fig:etaSW}
\end{figure}

The plots in Fig.~\ref{fig:etaSW}(a) show the variation as a function of the L1 e/$\gamma$ $E_{\textrm{\scriptsize T}}$ of the size of the 3$\sigma$ window in $\eta$ for three different regions taken as examples: namely for Region 1, Region 4 and Region 6, for Step 2, thus stressing how well this unique value represents all the cases for this step.

For Step 3, the three different values characterizing the size of the 3$\sigma$ window in $\eta$, in the function of L1 e/$\gamma$ $E_{\textrm{\scriptsize T}}$ are presented in Fig.~\ref{fig:etaSW}(b). The three plots show the variation in function of L1 e/$\gamma$ $E_{\textrm{\scriptsize T}}$ of the size of the 3$\sigma$ window in $\eta$ for Region 1, Region 3, and Region 5 respectively.

\begin{table}[htbp]
    \centering
    \begin{tabular}{|c|c|c|}
         \hline
         Step 2 & $\eta$ Region 1-6 & 3$\sigma$ $\Delta\eta_{i, j}$ = 0.01 \\
         \hline
         \multirow{3}{*}{Step 3} & $\eta$ Region 1-2 & 3$\sigma$ $\Delta\eta_{i, j, k}$ = 0.002 \\
         \cline{2-3}
         & $\eta$ Region 3 & 3$\sigma$ $\Delta\eta_{i, j, k}$ = 0.004 \\ 
         \cline{2-3}
         & $\eta$ Region 4-6 & 3$\sigma$ $\Delta\eta_{i, j, k}$ = 0.01 \\ 
         \hline
    \end{tabular}
    \caption{$\eta$-LUT table providing the 3$\sigma$ window size in $|\eta|$ to be applied for Step 2 and Step 3 of the PiXTRK algorithm.}
    \label{tab:etaLUT}
\end{table}

\subsubsection{Look-Up Tables describing the dependence in \texorpdfstring{$\phi$}{Lg}}
\label{sec:PixTRKLUTPhi}
The variation of the size of the 3$\sigma$ windows in azimuthal angle, over the full $E_{\textrm{\scriptsize T}}$ range is more complex for both Step 2 and Step 3 of the PiXTRK algorithm. 

This is so because the typical dependence in $E_{\textrm{\scriptsize T}}$ of the upper and lower bounds of these 3$\sigma$ windows, for Step 2 and Step 3, is not at all constant; it varies quite fast especially in the low $E_{\textrm{\scriptsize T}}$ range between 10 and 20 GeV and in different ways over the full $\eta$ range. Moreover, in Step 2, the $\Delta\phi$ resolution is dominated by the calorimeter granularity and the distance of the calorimeter to the beam axis. The chosen calorimeter design, in this showcase study, is made of a barrel calorimeter with a coarser granularity but a shorter distance (by about a factor of 2) from the beam axis than the endcap calorimeter. This is reflected in the Step 2 results in Fig.~\ref{fig:phiSWstep2} as well as in Appendix B.1.  Step 3 instead corresponds to the ``standalone''-based track reconstruction, thus the $\Delta\phi$ resolution is dominated by the pixel high granularity (see results in Appendix B and Fig.~\ref{fig:phiSWstep2}). It is typically here, a factor of 10 better.

The variation of the size of the $\Delta\phi$-windows has been carefully studied. The results summarized here below show an impressive agreement between the \texttt{DELPHES} results and those with a full detector upgrade simulation for HL-LHC 
%$\footnote{Here the comparison has been done with the results from a similar study performed with a full HL-LHC detector simulation study, within CMS}$.

\begin{itemize}

    \item {\it{The $\Delta\phi$-window LUT for PiXTRK Step 2}} 
    
    The variation of the 3$\sigma$ windows in $\phi$ boundaries are studied as a function of L1 e/$\gamma$ $E_{\textrm{\scriptsize T}}$ and in all the 6 regions in $\eta$ as defined in Table~\ref{tab:pixelCombi}.
    
    As a result, a LUT is defined with compressed values, reducing the dimensions of this Table, but still preserving a high precision. 
    It is made of 20 steps in transverse energy ($E_{\textrm{\scriptsize T}}$), where the first 10 steps of 1 GeV each represents the variation from 10 to 20 GeV, then by steps of 2 GeV from 20 to 30 GeV and steps of 5 GeV from 30 to 50 GeV.
    The value at 50 GeV is kept as a constant step, for $E_{\textrm{\scriptsize T}}$ larger than 50 GeV.
    %with the value at 50 GeV extended to $E_{\textrm{\scriptsize T}}$ larger than 50 GeV.
    The other parameter to include in this LUT represents the three combinatorial possibilities between the 4 barrel layers and the 5 end cap disks, over the full $\eta$ coverage. It can be split into 4 sectors, where the first one extends from 0 to 2.1 in $\eta$ (thus merging in one sector the 3 first regions) and the other sectors correspond each to regions 4, 5, and 6 as defined in Table~\ref{tab:pixelCombi}. 
    The total number of $\Delta\phi$ values included in this $\Delta\phi$-LUT is 160 and their detailed values are in Appendix B.1. Note that, unlike the $\Delta\eta$-LUT, the $\Delta\phi$-LUT does not define the size of the 3$\sigma$ window in $\Delta\phi$ by just one $\Delta$-like value centered around 0, but indeed by two values which correspond to the lower and upper boundaries for each L1 $E_{\textrm{\scriptsize T}}$ of the defined windows.
    
    One may note that the $E_{\textrm{\scriptsize T}}$ dependence of the $\Delta\phi$-windows is sharper in the first 3 regions in $\eta$ (i.e Regions 1, 2 and 3) than for the 3 other Regions 4, 5 and 6. Indeed the curve flattens when corresponding to larger $\eta$ regions and thus, especially, when getting to Region 6.
    
    As examples we show some plots of the 3$\sigma$ windows in Fig.~\ref{fig:phiSWstep2}; they correspond, from left to right, to Region 1, Region 4, and Region 5, respectively.

\begin{figure}
    \centering
    \subfloat[Step 2]{
        \includegraphics[width=0.32\textwidth]{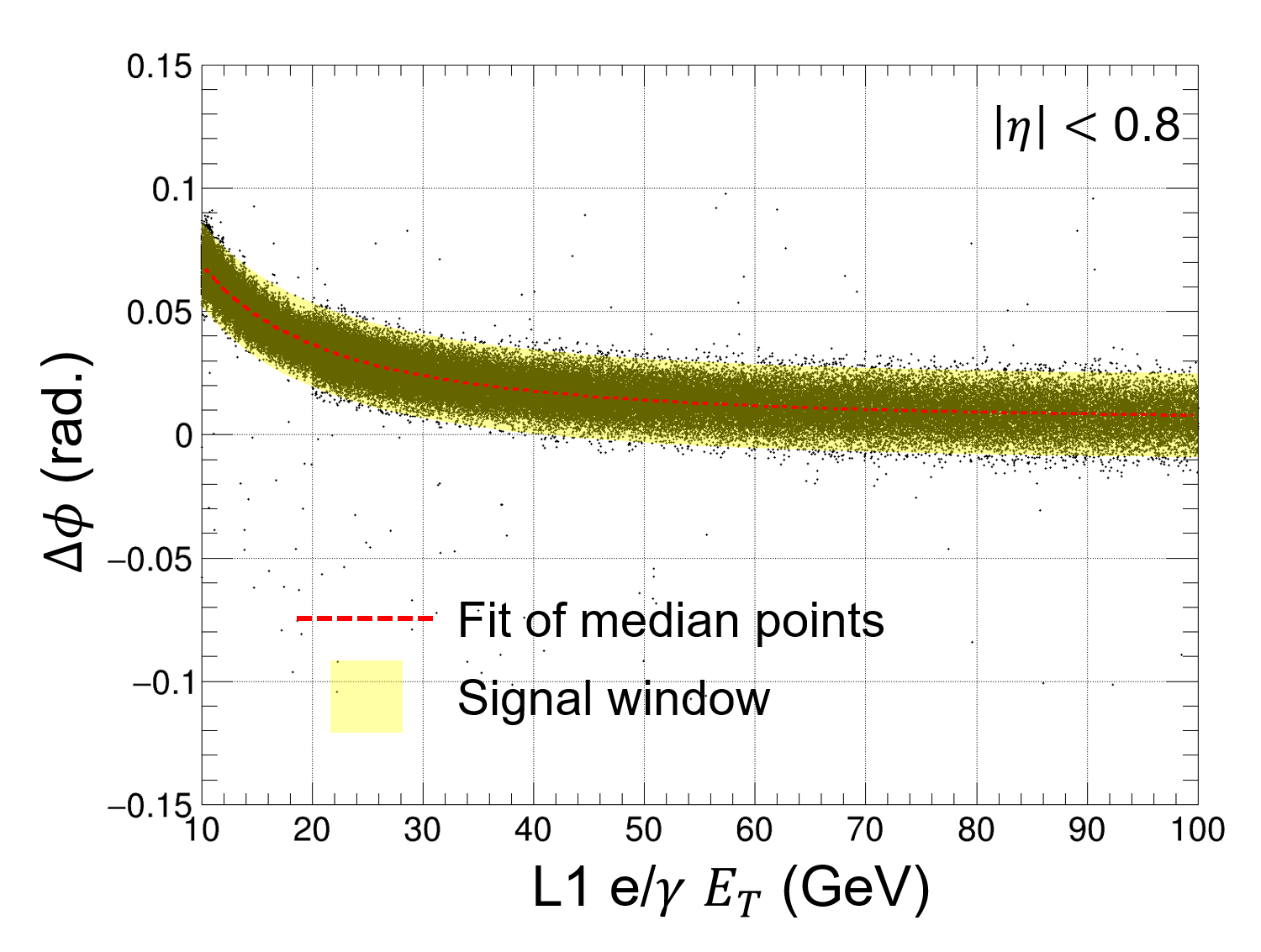}
        \includegraphics[width=0.32\textwidth]{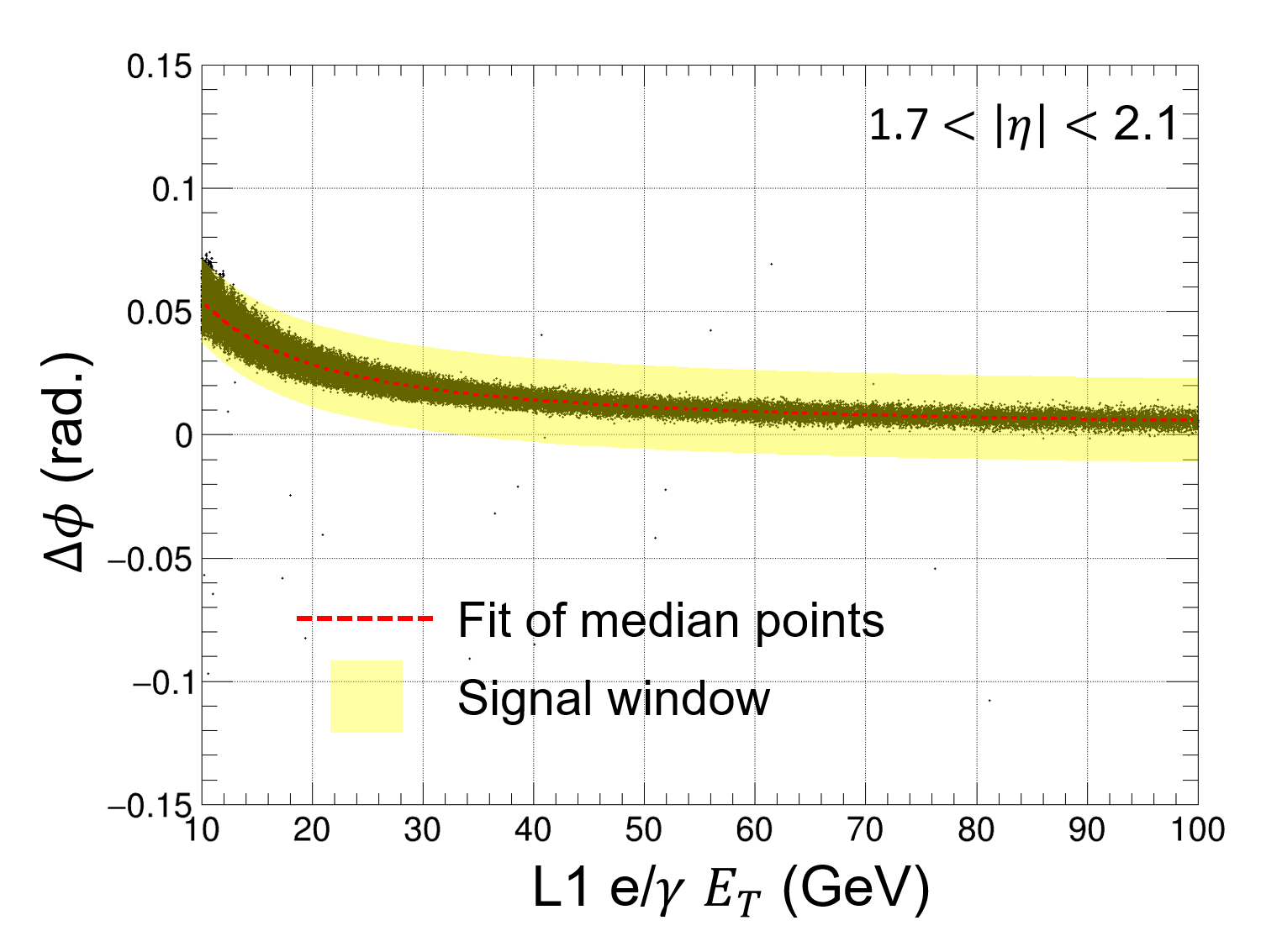}
        \includegraphics[width=0.32\textwidth]{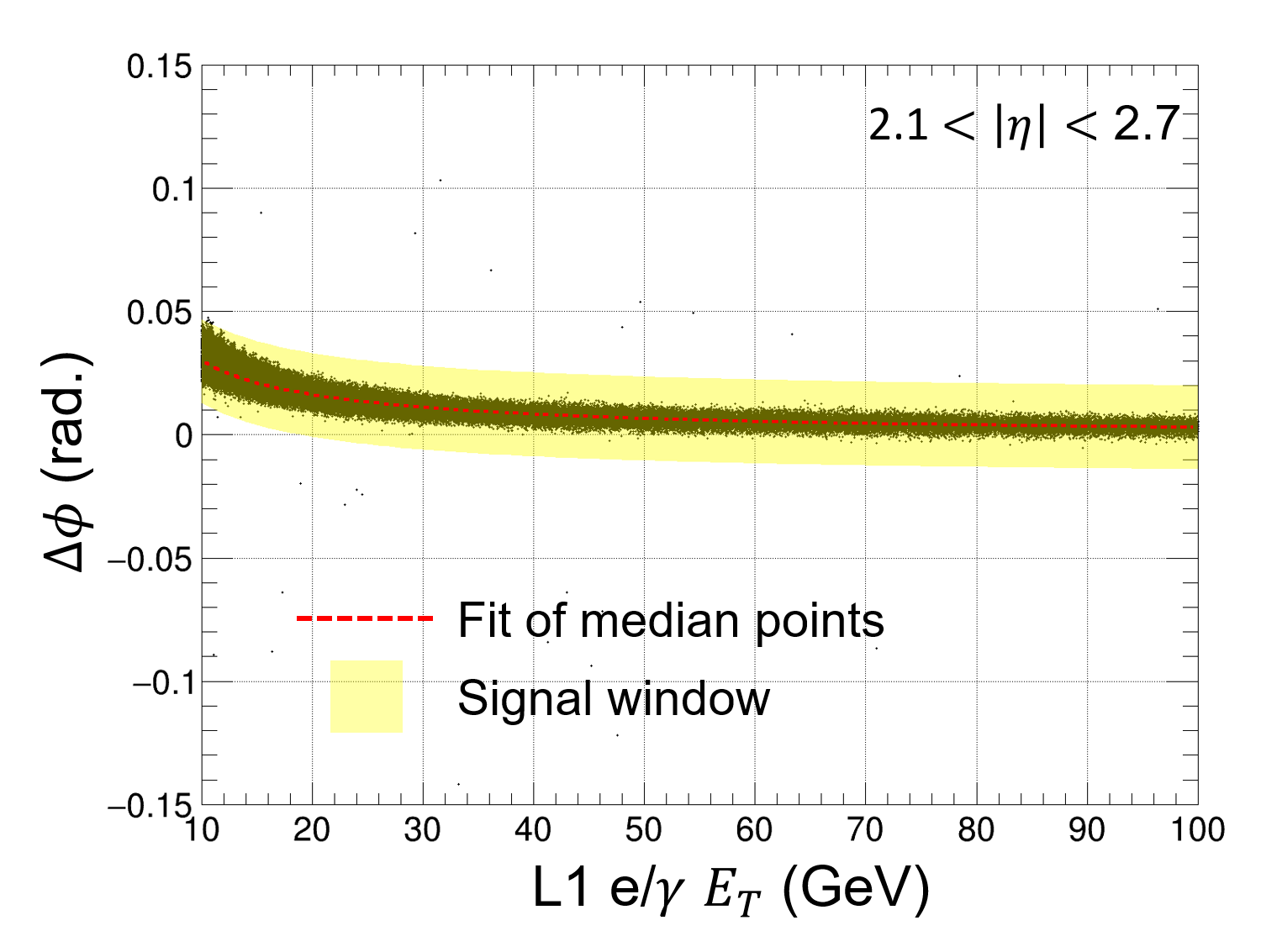}
    }
    
    \subfloat[Step 3]{
        \includegraphics[width=0.32\textwidth]{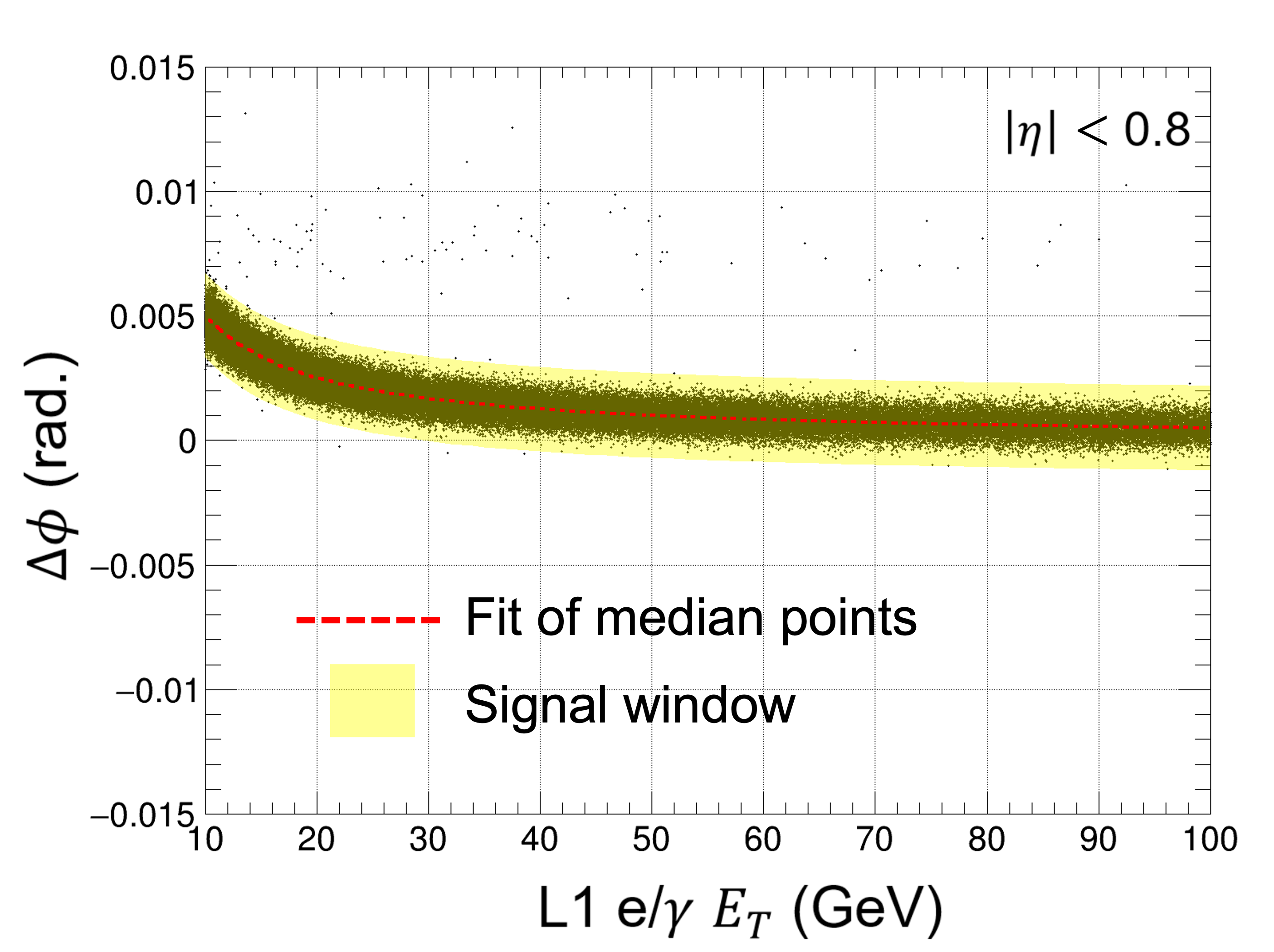}
        \includegraphics[width=0.32\textwidth]{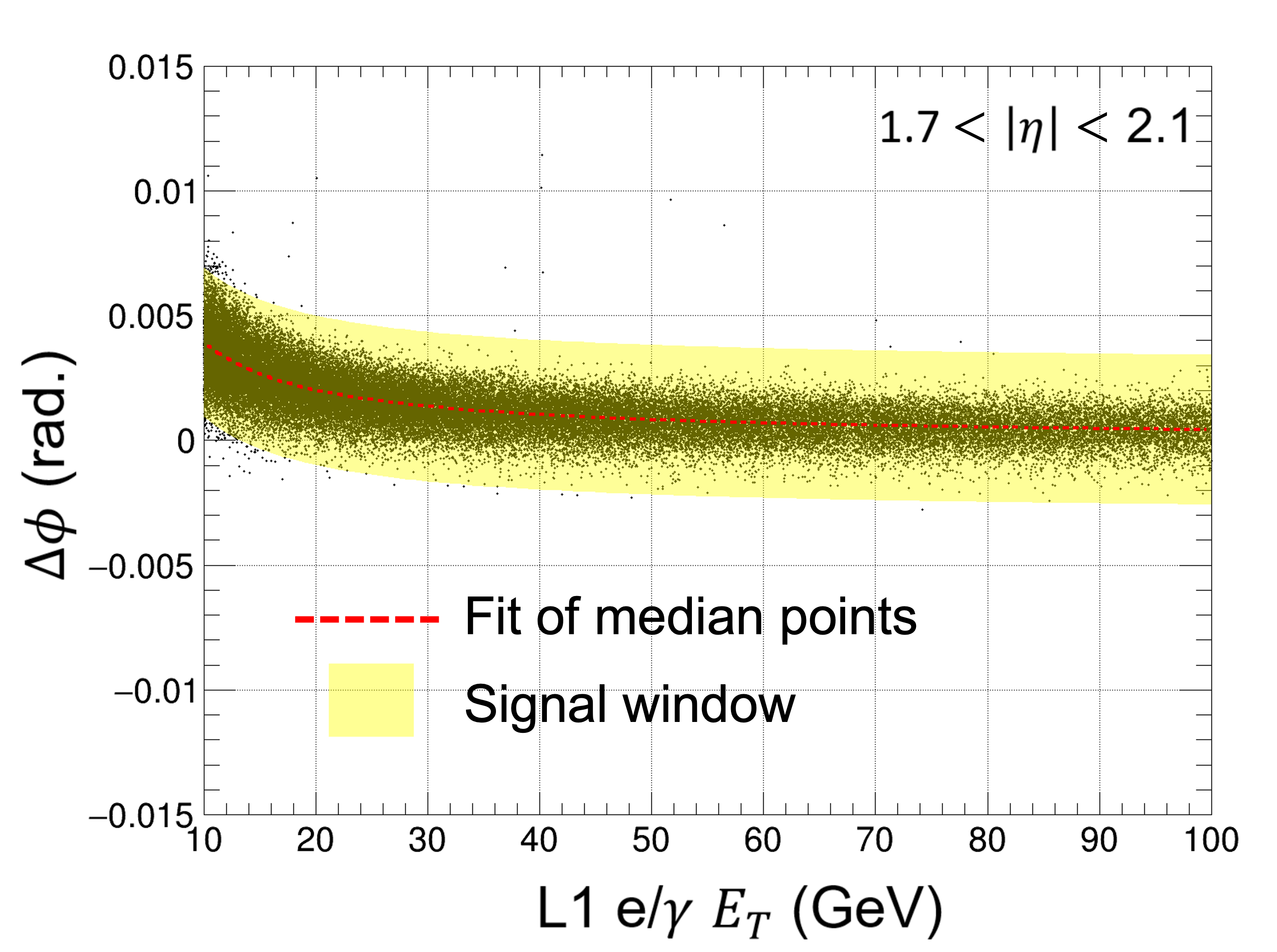}
        \includegraphics[width=0.32\textwidth]{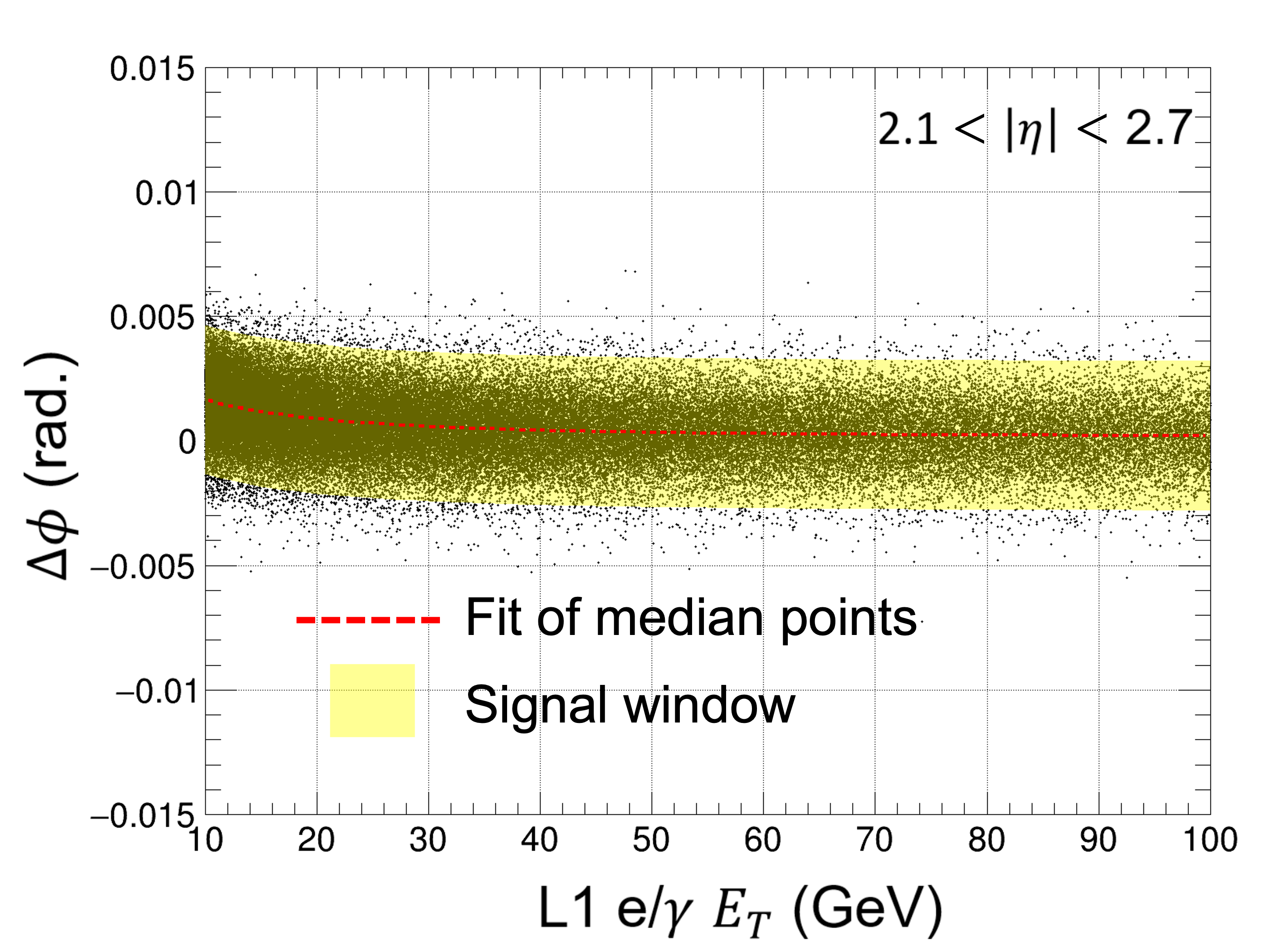}
    }
    \caption{(a) Step 2: Variation, in the function of the L1 e/$\gamma$ $E_{\textrm{\scriptsize T}}$, of the size of the 3$\sigma$ window in $\Delta\phi$ (yellow band) and of the residuals in $\phi$ (black points) from  Equation~\ref{eq:step2-1Define}, for central barrel (left), endcap (middle) and forward (right). (b) Step 3: Equivalent distributions as above, but for step 3, thus corresponding to Equation~\ref{eq:step3-1Define}. These curves correspond to the electron case. The corresponding $\Delta\phi$ distributions for the positrons are symmetrical w.r.t. $\Delta\phi$ = 0 axis, to the ones of the electrons presented in this Figure. }    
    \label{fig:phiSWstep2}
\end{figure}
    
    At this stage in Step 2, the PiXTRK algorithm cross-checks Condition~\ref{eq:step2-1Define} only for the pixel clusters satisfying Condition~\ref{eq:step2-2Define}. To do so, it looks for the specific $\Delta\phi$ window size as provided in Step 2 $\Delta\phi$-LUT, that corresponds to the $\Delta\eta$ region in the detector and the transverse energy of the electron candidate which is measured by the EM calorimeter at L1. Only the pixel clusters satisfying Condition~\ref{eq:step2-1Define} will go to Step 3.

    \item {\it {The $\Delta\phi$-window LUT for PiXTRK Step 3}}
    
    This step corresponds to the standalone (track bending) part of PiXTRK, defined with all the combinations between 3 pixel clusters in different layers and/or disks, remaining after the 2 first selection steps. 
    
    The study of the $\Delta\phi$-window evolution with the $\eta$ and $\phi$ positions of the remaining pixel clusters and on the $E_{\textrm{\scriptsize T}}$ of the electron candidate shows that higher precision is required on those parameters.
    Indeed, the corresponding $\Delta\phi$-LUT contains, in this case, 2400 values, namely: 400 boundary values for each of the six regions in $\eta$, to start with. 
    
    The goal is to compress this large data sample without jeopardizing the precision. This is successfully achieved by combining in each region the ranges in transverse energy with the same boundaries. An overall reduction factor of almost 5 is obtained, leading to a total of 526 values to be reported in this LUT$\footnote{It must be noted that for Regions 4, 5, and 6 the reduction factor is about 10, and for Regions 1 and 4 the reduction factor is 4 while only a bit below 2 for Region 2}$. 
    
    This study gives a precision on Conditions~\ref{eq:step2-1Define} and~\ref{eq:step3-1Define} of 1 mrad in $\Delta\phi$.
    The details of the content of these LUTs can be found in Appendix B.2.
    
    Applying Condition~\ref{eq:step3-1Define} to the remaining pixel clusters after passing Conditions~\ref{eq:Step1Define} to~\ref{eq:step3-2Define}, is the last selection step of the PiXTRK algorithm.
    
    %A simple linear fit is then performed on each remaining triplet of pixels clusters that passed conditions~\ref{eq:Step1Define} to~\ref{eq:step3-1Define}. It provides the reconstructed pixel track segments, corresponding to each L1 e/$\gamma$ object provided by the L1 EM calorimeters, as a final result of the L1 PiXTRK algorithm.

\end{itemize}

\subsection{Vertexing performance in z-direction with pixel-based information}
\label{sec:vertex}

The vertexing capability i.e. the capability to reconstruct vertex in the collision points is indeed a fundamental role of the microvertex detectors installed in the closest neighbourhood of the beam pipe.  
The microvertex information is used for vertexing the collision point, in the z-direction, at the High-Level Trigger (HLT) which is the final stage of the triggering system in both the ATLAS and CMS experiments for instance. 

The feasibility studies for using the pixel cluster information in real-time i.e. at 40 MHz have already been performed both for the electron trigger and the b-tagging case and reported in~\cite{Moon_2015, Moon_2016}. In the b-tagging case~\cite{Moon_2016} the seed is provided by the L1 track defined with the outer tracker~\cite{CMS-phase2tracker}; we then look for the pixel track segment (defined by 2 pixel clusters) compatible with the L1 track reconstructed by the L1 track trigger system~\cite{CMS-phase2L1TDR}. 
%The extension of the L1 track built with the outer tracker only in the CMS case, to the pixel detector allows an improvement in the precision of the reconstructed z-vertex position by about a factor of 10.

As this paper focuses on the electron case, we thus stress here the potential of the pixel detector for a high precision reconstruction of the z-vertex for the electron trigger(s); this would be similar to the muon trigger(s).
%as the bremsstrahlung is in the transverse plane (see for instance \cite{CMS-HLTPhase2TDR}).
%This trigger, as explained above, is provided by the pixel detector plus the EM calorimeter: the seed element, namely the  L1 e/$\gamma$ reconstructed object, is provided by the L1 EM calorimeter system. 

The data samples we use are single electron gun events produced at 14 TeV with 200 superimposed pileups. 
The reconstructed z-vertex position is obtained as a direct outcome of the PiXTRK as described in Section~\ref{PiXTRKStrategy} and~\cite{Moon_2015}. 
%We extend, with a straight line the pixel track segment reconstructed with PiXTRK, by taking the two most distant pixels out of the three selected clusters till the z-beam axis, and thus obtain the ``$z_{vtx}$'' point.
The straight line joining the two most distant pixel clusters out of the three used by PiXTRK to reconstruct the pixel track segment (i.e. the cluster in the innermost and the one in the outermost layer or disk), is further extended till the beam axis. 
The intercepting ``$z_{vtx}$'' point, fits quite closely with the z-coordinate of the true vertex.

\begin{figure}[hbtp]
    \centering
    \subfloat{\includegraphics[width=0.41\textwidth]{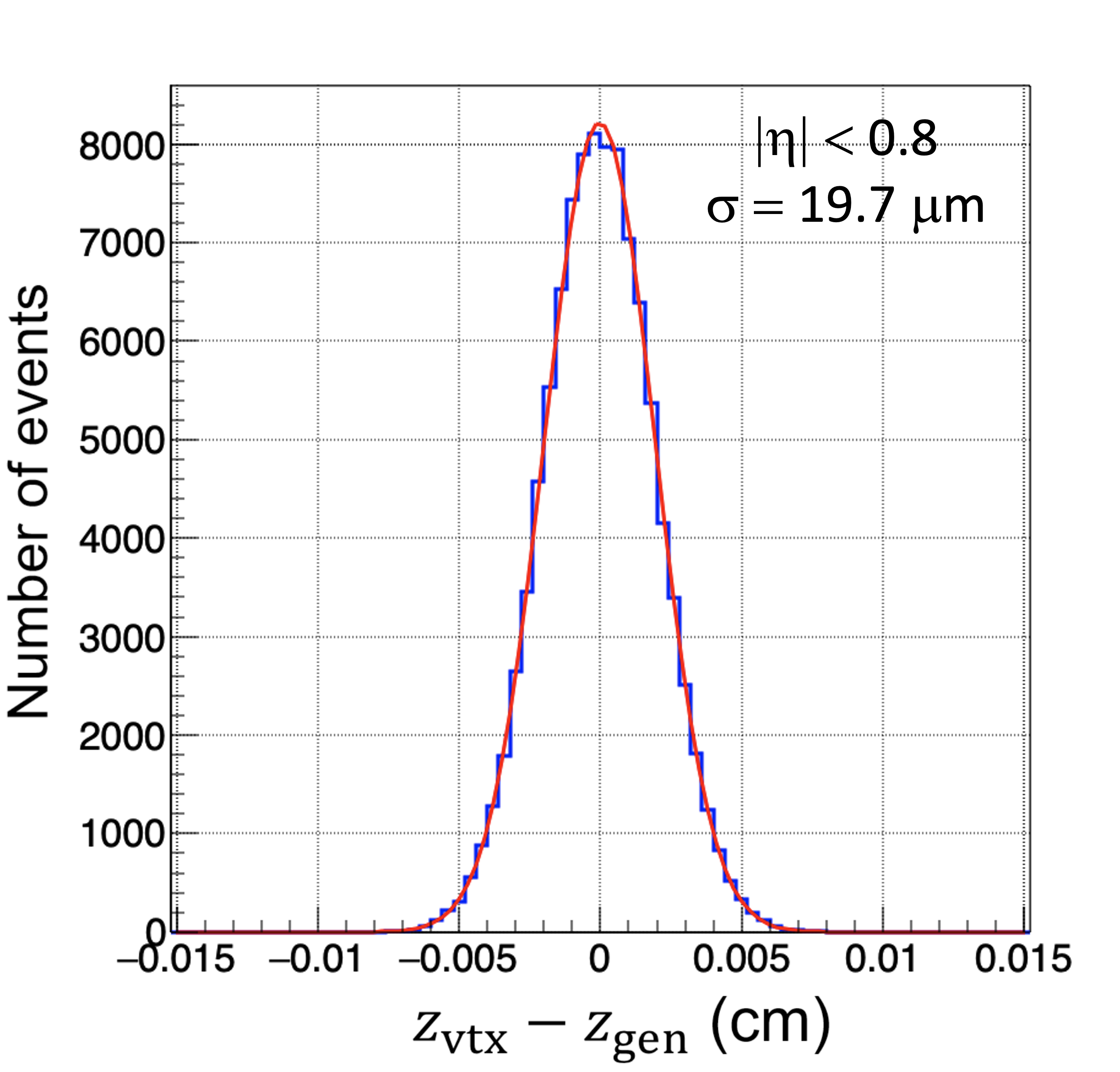}}
    \subfloat{\includegraphics[width=0.41\textwidth]{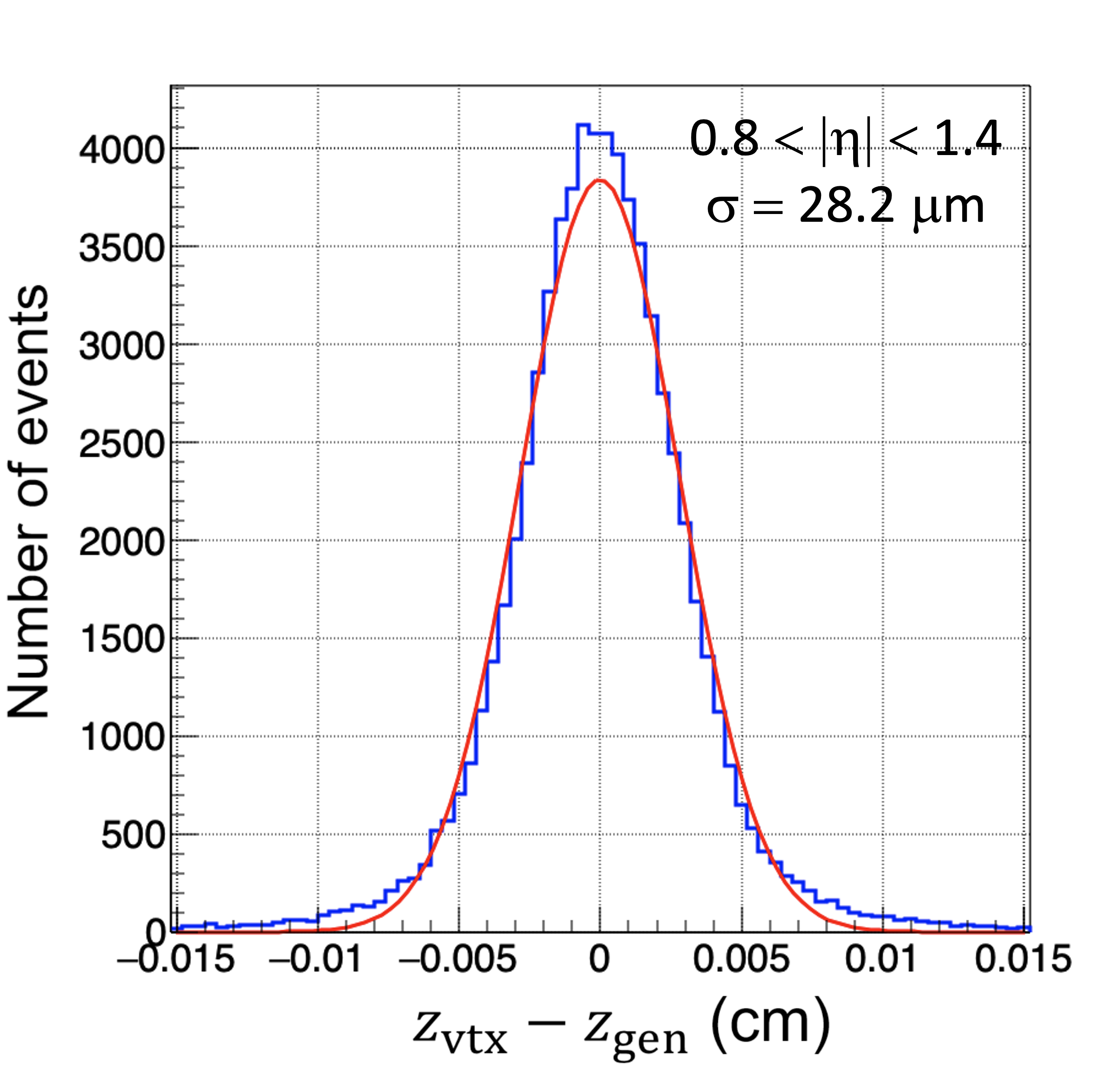}}

    \subfloat{\includegraphics[width=0.41\textwidth]{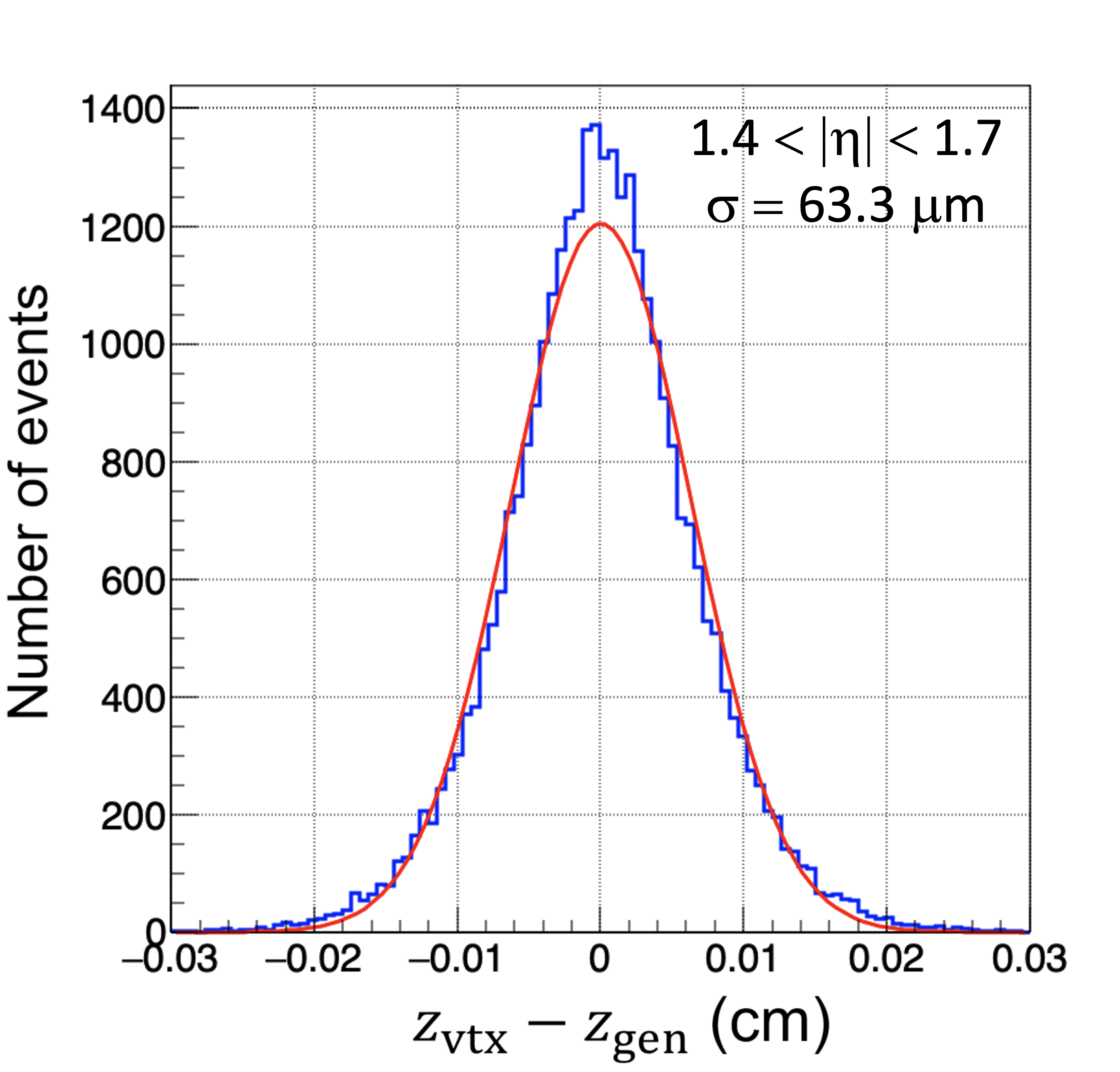}}
    \subfloat{\includegraphics[width=0.41\textwidth]{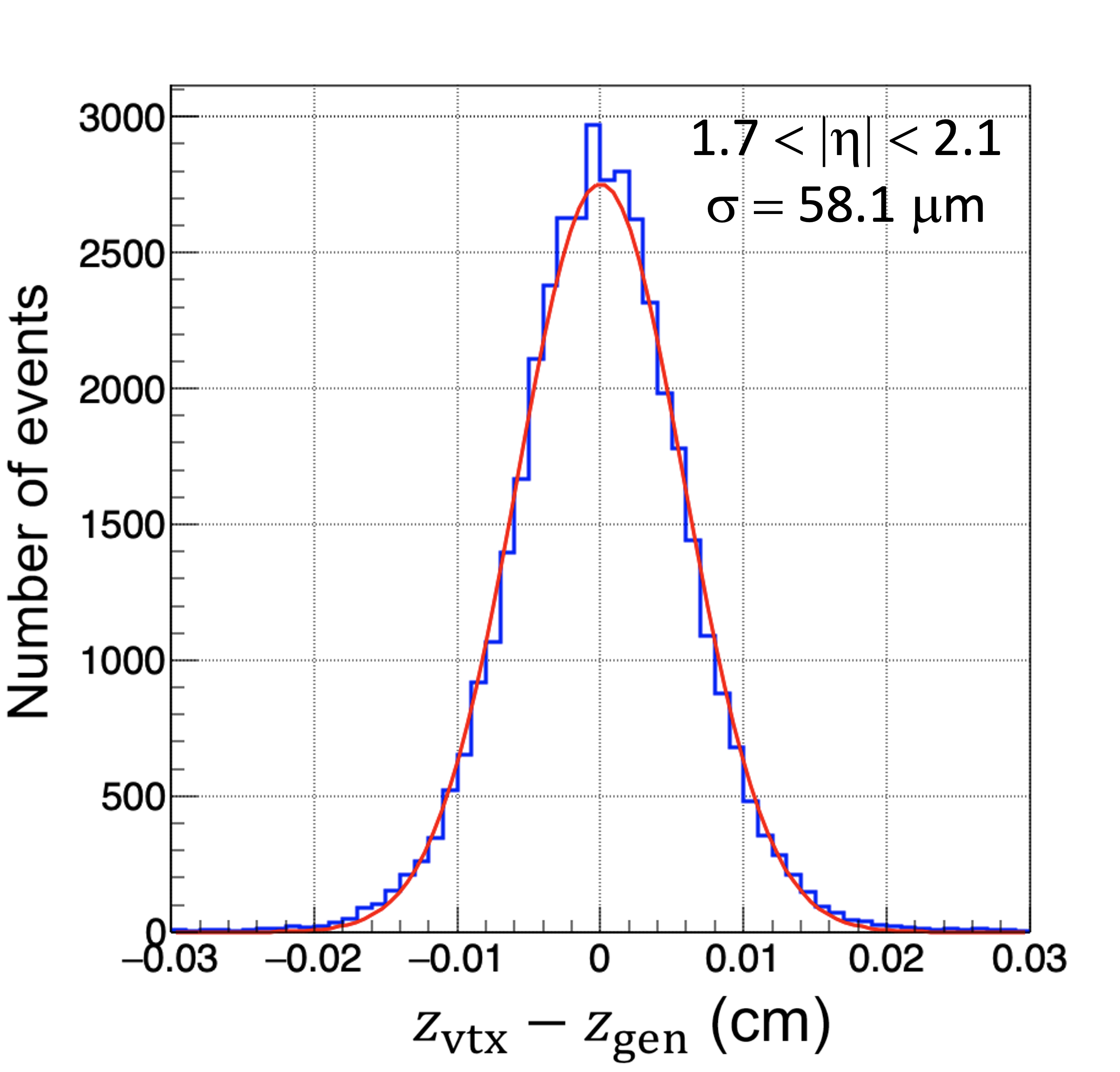}}

    \subfloat{\includegraphics[width=0.41\textwidth]{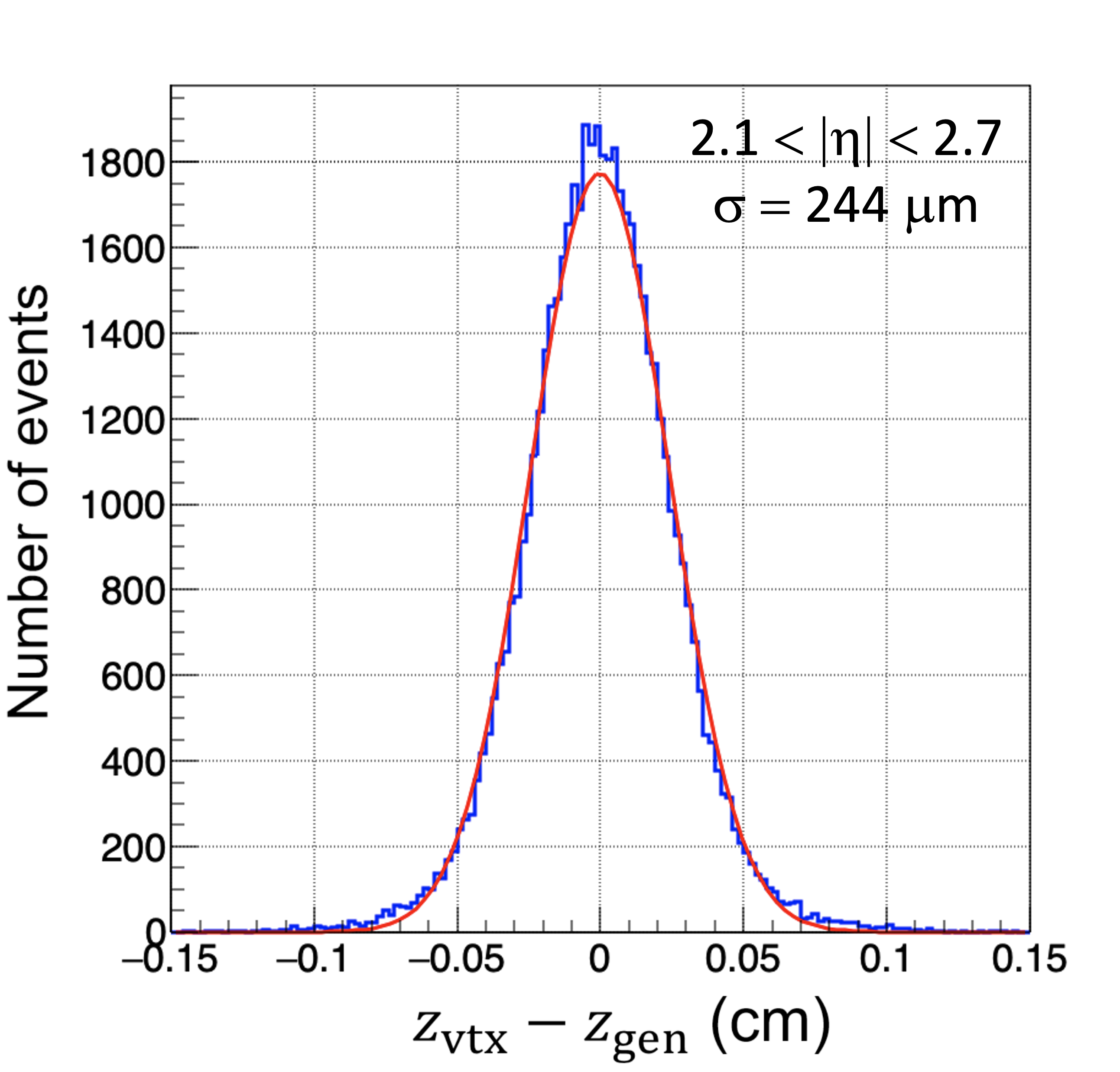}}
    \subfloat{\includegraphics[width=0.41\textwidth]{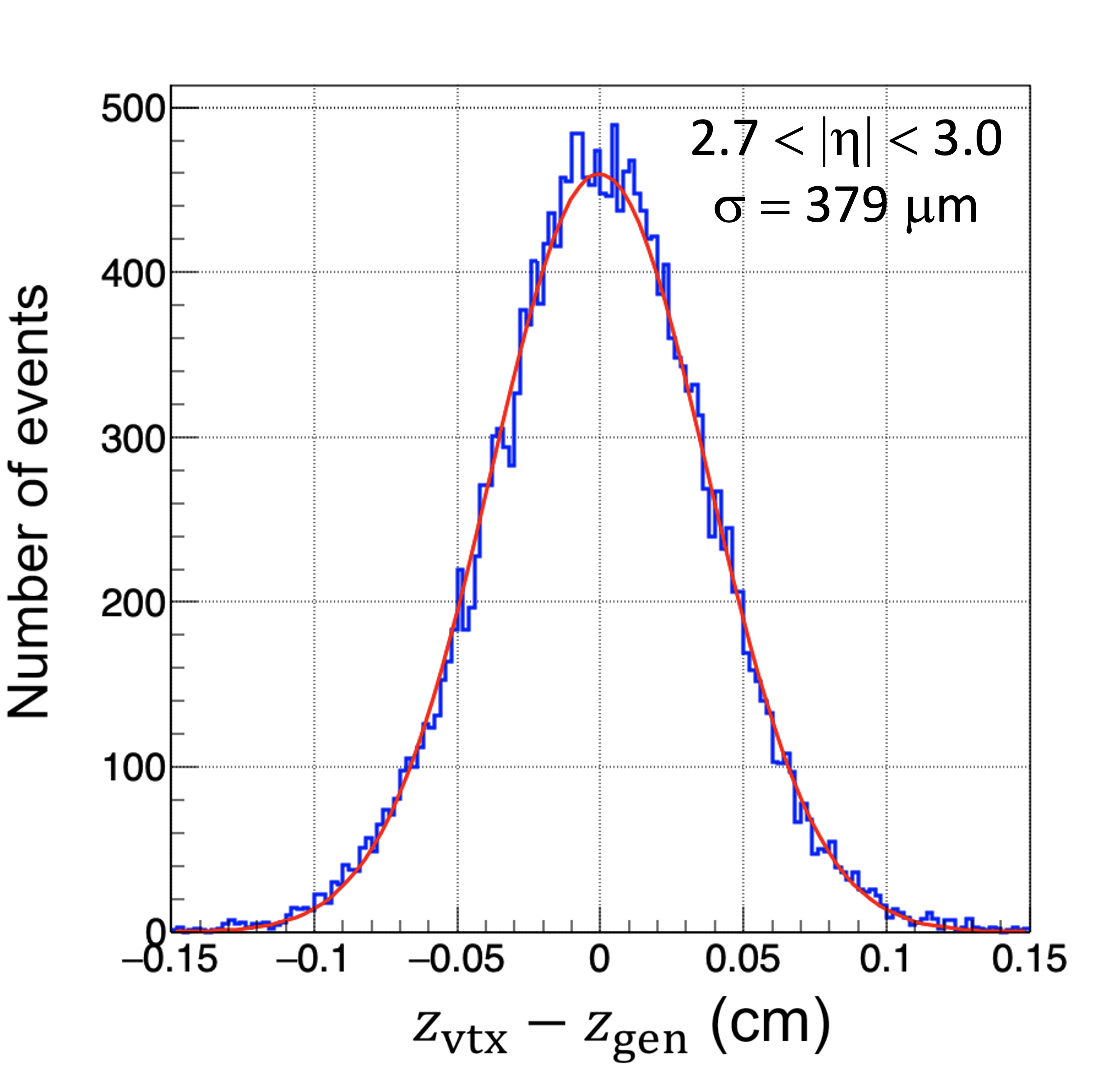}}
    \caption{Distributions of the distance between the vertex of the L1 e/$\gamma$ object determined by the simple method described above and the true vertex at generator level, in different $\eta$ regions. The fit performed with a Gaussian function is shown in red. The standard deviation ($\sigma$) of the fitted Gaussian function represents the resolution of each distribution.}
    \label{fig:dzDist}
\end{figure}

\begin{figure}[hbtp]
    \centering
    \includegraphics[width=0.6\textwidth]{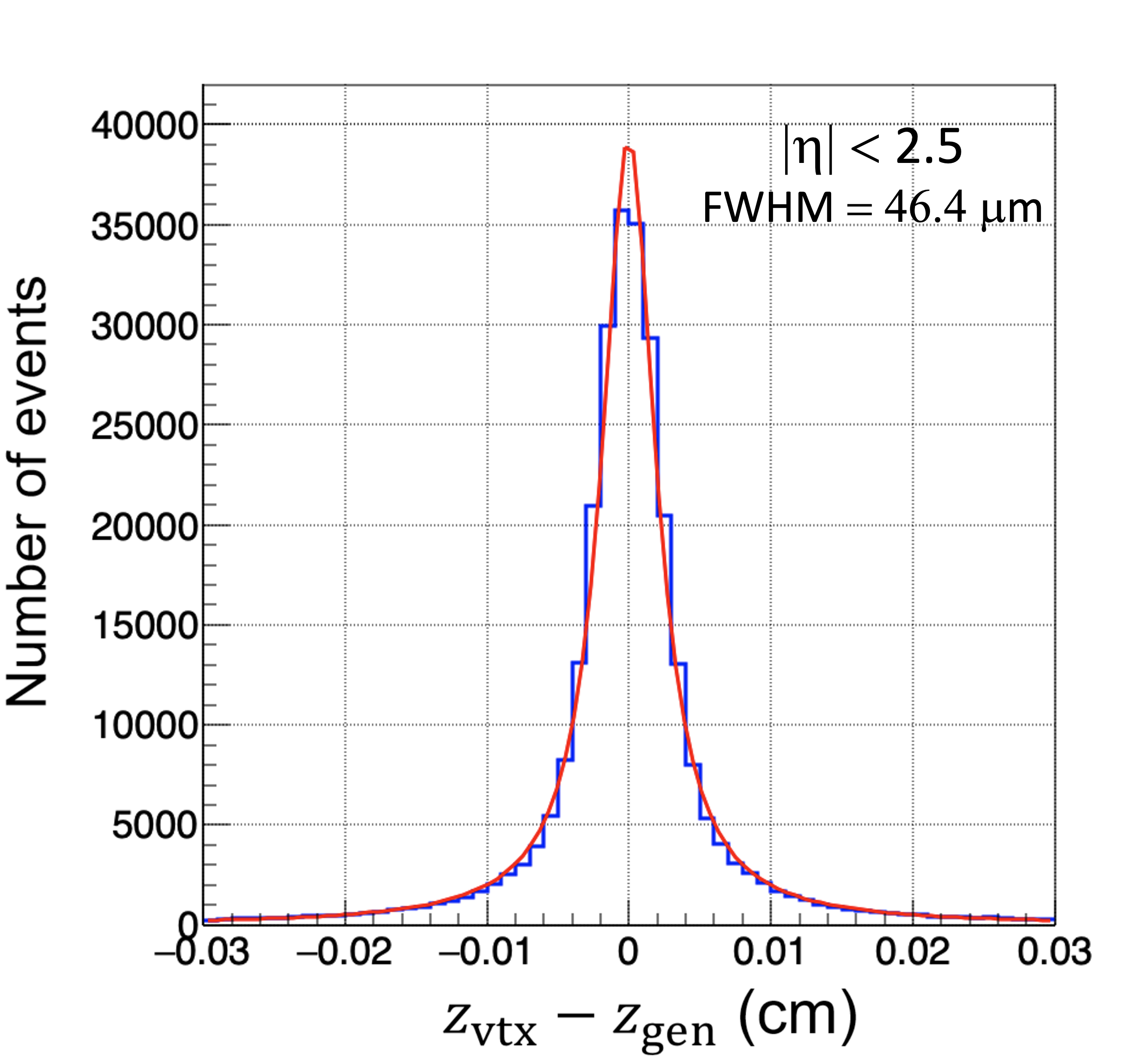}
    \caption{Distribution of the distance between the vertex of the L1 e/$\gamma$ object determined by the simple method described above and the true vertex at generator level, over the overall range in $\eta <2.5$. The fit with the Cauchy distribution function is shown in red.}
    \label{fig:dzDist_eta2p5}
\end{figure}

In order to verify the validity of this approach and to evaluate its precision, we study the difference between the reconstructed z-vertex position along the beam axis, $z_{vtx}$, and the z-vertex position of the ``true'' primary vertex as given at the generator level (gen-level), in the six different $\eta$ regions as defined in Table~\ref{tab:pixelCombi}.\\
Figure~\ref{fig:dzDist} gives the resolution in the determination of the z-vertex position as given by this simple vertex determination method. 
The resolution of each distribution is defined by a standard deviation from a fitted Gaussian function.
Excellent 1$\sigma$ resolutions are obtained: in the barrel with 20$\mu$m resolution for $\eta$ up to 0.8 and with 30$\mu$m for 0.8 <$\eta$< 1.4) as well as in the intermediate region 1.4 < $\eta$ < 2.1, with 60$\mu$m. 
The resolution increases by about a factor 4 in the end-cap (2.1 < $\eta$ < 2.7) and by a factor 6 in the very forward region ($\eta$ > 2.7 reaching a maximum value of about 360$\mu$m.\\
The 1$\sigma$ resolution of this vertex reconstruction method for $\eta$ < 2.5, is shown in Fig.~\ref{fig:dzDist_eta2p5}. 
A resolution of 46.4$\mu$m, on average, is obtained over this overall $\eta$ range. 
It should be noted here, that because the single electrons sample includes a large fraction of central electrons, this average estimate is biased towards values corresponding to $\eta$ < 1.4.\\
For completion, these results can be compared to the vertex resolution of 37$\mu$m in the central part ($\eta$ < 1.5), with the same method applied to the b-tagging case~\cite{Moon_2016} using top pair events and improving by more than an order of magnitude the resolution with the outer tracker only.  

\section{Track Isolation based on the Pixel Detector Only Information}
\label{sec:TrkIso}

The track isolation using only the pixel-based tracks is implemented to further improve the performances of the PiXTRK algorithm, for the single electron trigger as a showcase.
%It is defined here in the case of the electron trigger identification, as an example case. 
%The strategy for defining 
%The pixel-based track isolation of a single electron event with 200 pileup is thus compared to the background represented by minimum bias events. 
The isolation consists of counting the number of reconstructed pixel-based tracks included in an isolation cone. It follows the main steps here below:

%The specific and particular feature of this isolation recipe as compared to the usual definition (based primarily on the outer tracker or the overall tracker) is that only the pixel-based tracks are considered here. In particular, as described below in this section, the tracks $p_{\textrm{\scriptsize T}}$ is determined through the $\Delta\phi$ azimuthal curvature of the track, defined only with the pixel detector provided information.

%The electron pixel-based track isolation algorithm follows the main steps detailed below:

\begin{itemize}

\item {\it Reconstruction of the leading electron track and definition of the isolation cone}

\item {\it Reconstruction of all the pixel-based tracks within the isolation cone}.

\subitem {\it {1) Look for all the pixel clusters within the isolation cone}}
\subitem {\it {2) Reconstructing the track segments}}.
\subitem {\it {3) Reconstructing the pixel tracks within the cone}}
\subitem {\it {4) Measuring the azimuthal $\phi$ curvature}}

\item {\it Computation of the pixel-based tracks $p_{\textrm{\scriptsize T}}$ and application to pixel track isolation}

\item {\it Estimate of the electron track isolation}

\end{itemize}

\subsection{Reconstruction of the leading electron track and definition of the isolation cone}
\label{sec:reco_elec_trk}

%%%%%%%%%%%%%%%%%%%%%%%% step 1 %%%%%%%%%%%%%%%%%%%%%%%%%%%%%%%%%%%%%%
%%%%%%%%%%%%%%%%%%%%%%%% step 1 %%%%%%%%%%%%%%%%%%%%%%%%%%%%%%%%%%%%%%
%%%%%%%%%%%%%%%%%%%%%%%% step 1 %%%%%%%%%%%%%%%%%%%%%%%%%%%%%%%%%%%%%%
%%%%%%%%%%%%%%%%%%%%%%%% step 1 %%%%%%%%%%%%%%%%%%%%%%%%%%%%%%%%%%%%%%
%%%%%%%%%%%%%%%%%%%%%%%% step 1 %%%%%%%%%%%%%%%%%%%%%%%%%%%%%%%%%%%%%%

%The first step, i.e. 
The pixel-based track reconstruction of the considered electron, labeled as “leading or L1 electron track” is performed with PixTRK (Section~\ref{sec:PiXTRK}).
%pixel-based track reconstruction follows what is described in details in Section~\ref{sec:PiXTRK}. 
%This is the so-called “leading electron track” here. 
%The z-vertex ($z_{\textrm{\scriptsize vtx}}$) corresponding to this L1 electron track is defined as described in (Section~\ref{sec:vertex}).
Its z-vertex ($z_{\textrm{\scriptsize vtx}}$) position is defined as in (Section~\ref{sec:vertex}).
%%%%%%%%%%%%%%%%%%%%%%%% step 2 %%%%%%%%%%%%%%%%%%%%%%%%%%%%%%%%%%%%%%
%%%%%%%%%%%%%%%%%%%%%%%% step 2 %%%%%%%%%%%%%%%%%%%%%%%%%%%%%%%%%%%%%%
%%%%%%%%%%%%%%%%%%%%%%%% step 2 %%%%%%%%%%%%%%%%%%%%%%%%%%%%%%%%%%%%%%
%%%%%%%%%%%%%%%%%%%%%%%% step 2 %%%%%%%%%%%%%%%%%%%%%%%%%%%%%%%%%%%%%%
%%%%%%%%%%%%%%%%%%%%%%%% step 2 %%%%%%%%%%%%%%%%%%%%%%%%%%%%%%%%%%%%%%
%In step 2, an isolation cone is opened around the electron track. 
A cone with a typical aperture of 0.2 to 0.4, is defined around the leading electron track, originating from the z-vertex position ($z_{\textrm{\scriptsize vtx}}$), as sketched in Fig.~\ref{fig:cone}. 
%The cone aperture we tested and further optimized is typically between 0.2 and 0.4. 

%%%%%%%%%%%%%%%%%%%%%%%% step 3 %%%%%%%%%%%%%%%%%%%%%%%%%%%%%%%%%%%%%%
%%%%%%%%%%%%%%%%%%%%%%%% step 3 %%%%%%%%%%%%%%%%%%%%%%%%%%%%%%%%%%%%%%
%%%%%%%%%%%%%%%%%%%%%%%% step 3 %%%%%%%%%%%%%%%%%%%%%%%%%%%%%%%%%%%%%%
%%%%%%%%%%%%%%%%%%%%%%%% step 3 %%%%%%%%%%%%%%%%%%%%%%%%%%%%%%%%%%%%%%
%%%%%%%%%%%%%%%%%%%%%%%% step 3 %%%%%%%%%%%%%%%%%%%%%%%%%%%%%%%%%%%%%%

\subsection{Reconstruction of all the pixel-based tracks within the isolation cone}
\label{sec:reco_pixtrk}

\subsubsection{Look for all the pixel clusters within the isolation cone}

%We proceed by looking at each pixel cluster included in this cone. 
Each pixel cluster position within this cone is defined with its $\phi$, $\eta$ coordinates, and corresponding z-vertex position. The $\eta$ pixel cluster is evaluated assuming that the particle originated at the electron z vertex.
The coordinates in $\phi$, $\eta$ of the ith-cluster, for this cluster to be included in the cone, must verify that $\Delta R$ as defined in Equation~\ref{eq:cone} is within the chosen cone aperture.

\begin{equation} \label{eq:cone}
    \centering
    \Delta R = \sqrt{(\phi_{\textrm{\scriptsize L1 electron}}-\phi_{\textrm{\scriptsize pixel cluster}}^{i \textrm{\scriptsize th}})^{2} + (\eta_{\textrm{\scriptsize L1 electron}}-\eta_{\textrm{\scriptsize pixel cluster}}^{i \textrm{\scriptsize th}})^{2}}
\end{equation}

where $\phi_{\textrm{\scriptsize L1 electron}}$ and $\eta_{\textrm{\scriptsize L1 electron}}$ are the azimuthal angle and the pseudorapidity angle of the L1 electron track 
%as computed by the PiXTRK algorithm,
while $\phi_{\textrm{\scriptsize pixel cluster}}^{i \textrm{\scriptsize th}}$ and $\eta_{\textrm{\scriptsize pixel cluster}}^{i \textrm{\scriptsize th}}$ are the azimuthal angle and the pseudorapidity angle of the $i$th-cluster$\footnote{The coordinates of the pixel cluster will be provided by the signal processing on the Pixel Front-End ASIC}$.

%This step as well as the next one is sketched in 
Figure ~\ref{fig:cone} illustrates the tracks that might be fully included or just crossing the cone space (pileup tracks for instance). 

%Next, in order to fully identify the clusters within this cone that are the ones of interest, we form all the pixel track segments by combining two pixel clusters within this cone.

\subsubsection{Reconstructing the track segments within the isolation cone}

\begin{figure}[htbp]
    \centering
    \includegraphics[width=0.65\textwidth]{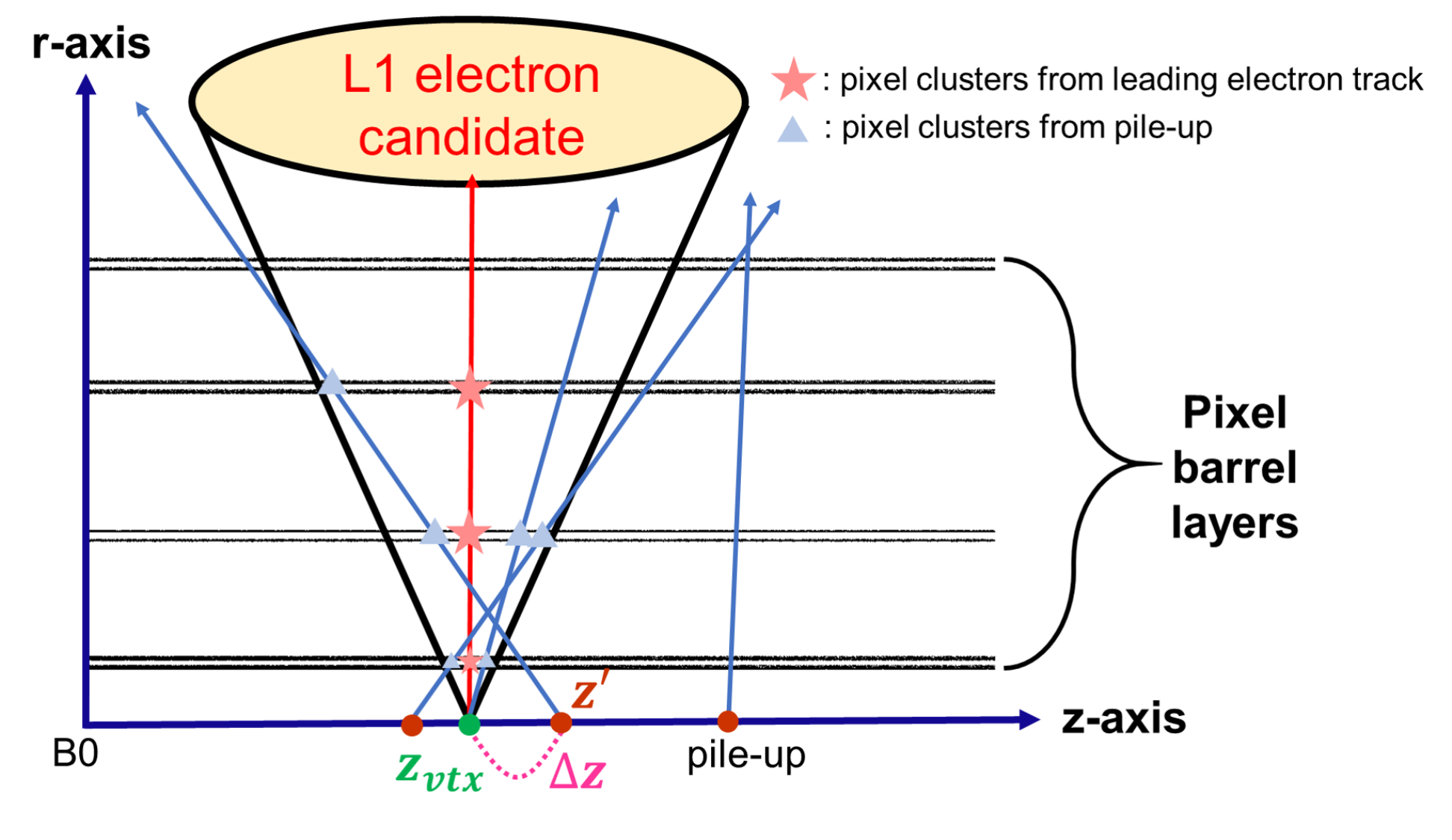} 
    \caption{A schematic of the cone from the $z_{\textrm{\scriptsize vtx}}$ to the L1 electron pixel-based track to select pixel clusters satisfying $\Delta R < $0.3. It illustrates how additional pixel track segments can be fully included or just crossed in the cone space. Here the $r$-axis is defined as $r = \sqrt{x^{2}+y^{2}}$, where $x$ and $y$ axes are perpendicular to the z-axis (i.e. in the transverse plane).}
    \label{fig:cone}
\end{figure}

Pixel track segments combining two pixel clusters within this cone are formed. Their 
%We extend these pixel track segments to the beam axis. 
intercept with the z-beam axis, defines the corresponding $z'$ value (Fig.~\ref{fig:cone}).\\ 
%This
The $z'$ value is compared to the z-vertex ($z_{\textrm{\scriptsize vtx}}$) as defined above.
%derived from the PiXTRK algorithm applied to the reconstructed leading electron track (Section~\ref{sec:vertex}). 
$\Delta z$ defined as 
% is thus obtained by computing 
the distance along the z-axis between the track segment joining these two pixel clusters and the $z_{\textrm{\scriptsize vtx}}$ from the leading electron, 
%It \label{eq:ZdistDefine}
which must satisfy Equation~\ref{eq:ZdistDefine}. 
%3$\sigma$ boundary signal window here below. 

\begin{equation} \label{eq:ZdistDefine}
    \centering
    \Delta z = |z_{\textrm{\scriptsize vtx}}-z'|< 3\sigma\footnote{This signal window in z, is represented as a standard deviation of a fitted the Gaussian function of the $\Delta z$ distribution between the $z'$ and the true $\Delta$z-vertex of the electron. The $\Delta$z-window is computed to be 0.3mm if $|\eta| \leq 0.8$ and 1mm if 2.7 $\leq |\eta| \leq$ 3.0.}
\end{equation}

%It is evaluated with a sample of simulated single electron events
%From the simulation study, a computed 3$\sigma$ window is obtained for each $\eta$ region (Fig.~\ref{fig:Phase2Pixel}). It ranges from 0.3 mm in the most central barrel region to 1 mm in the most end-cap region.\\

If Equation~\ref{eq:ZdistDefine} is not satisfied the combination of corresponding two pixel clusters is disregarded.
  
%The pixel-based track isolation also exploits the 3 out of 4 pixel cluster strategy as the PiXTRK algorithm, in order to reconstruct the pixel tracks in this cone.
In the considered pixel design, four possible combinations of layers or disks have to be taken into account. 
%depending on the considered $\eta$ region. 
For example, the possible combinations of two pixel clusters from different pixel barrel layers for measuring $\Delta z$ are shown in Table~\ref{tab:dz-combi}, for the detector region defined by $|\eta| <$ 0.8. 

\begin{table}[htpb]
    \centering
    \scalebox{1}{
        \begin{tabular}{|c|c|c|c|}
            \hline
            Pixel barrel layers & \multicolumn{3}{|c|}{Combinations of two pixel clusters for measuring $\Delta z$} \\
            \hline
            \hline
            1st, 2nd, 3rd layer & 1st-2nd layer & 1st-3rd layer & 2nd-3rd layer \\
            \hline
            1st, 2nd, 4th layer & 1st-2nd layer & 1st-4th layer & 2nd-4th layer \\
            \hline
            1st, 3rd, 4th layer & 1st-3rd layer & 1st-4th layer & 3rd-4th layer \\
            \hline
            2nd, 3rd, 4th layer & 2nd-3rd layer & 2nd-4th layer & 3rd-4th layer \\
            \hline
        \end{tabular} }
    \caption{The combination of pixel barrel layers for measuring $\Delta z$ in the region: $|\eta| <$ 0.8.}
    \label{tab:dz-combi}
\end{table}

\subsubsection{Reconstructing the pixel tracks within the isolation cone}

\begin{figure}[htbp]
    \centering
    \subfloat[]{\includegraphics[width=0.5\textwidth]{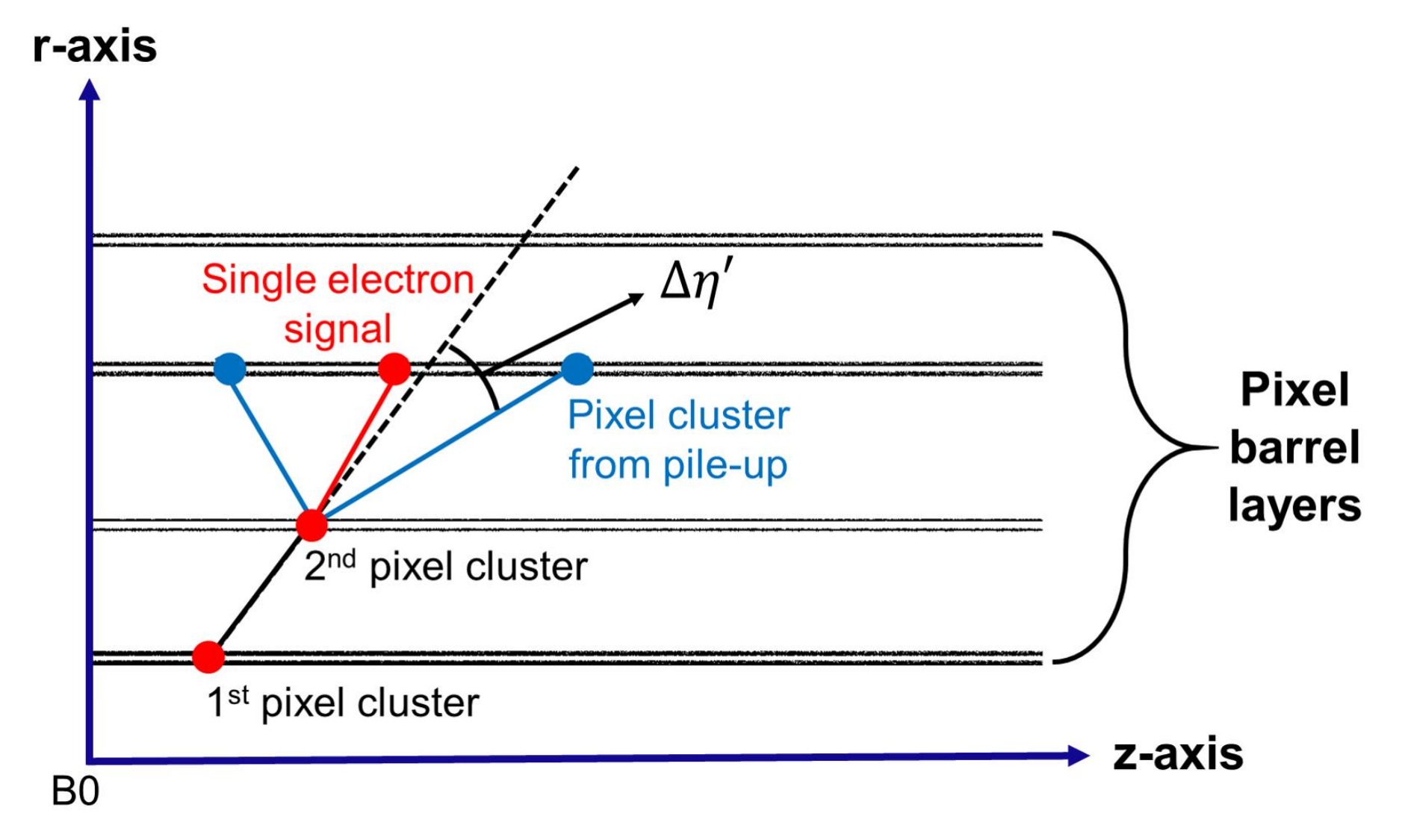}}
    \subfloat[]{\includegraphics[width=0.5\textwidth]{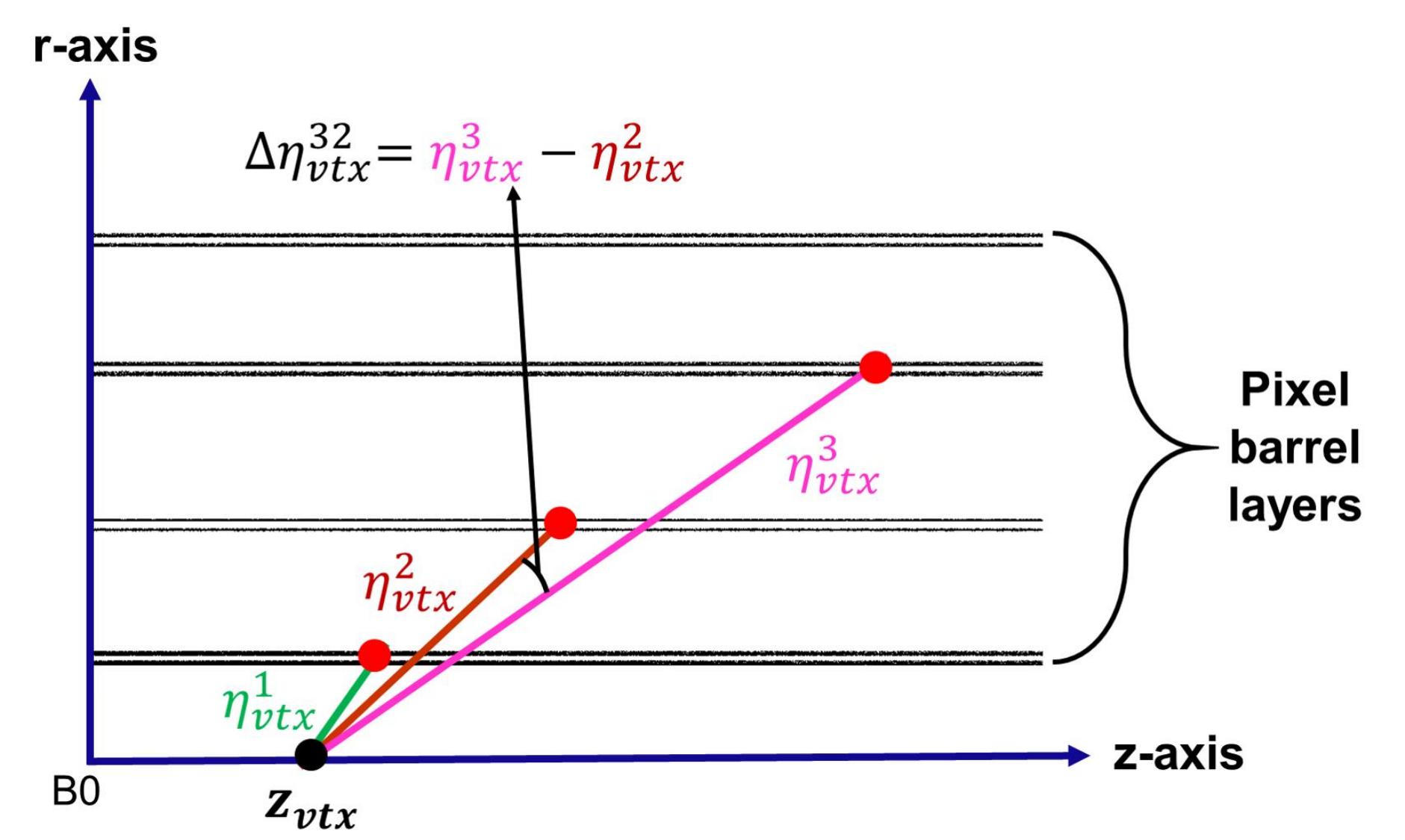}}
    \caption{Examples of reconstructing pixel track segment (a) using three pixel clusters and (b) using the L1 electron PiXTRK reconstructed vertex and pixel clusters. The example here is with layers 1, 2, and 3 starting from the innermost layer.}
    \label{fig:longitAngleDiff}
\end{figure}

%Once the first and second pixel clusters inside the isolation cone are selected,
A third pixel cluster must be found that combines the two first pixel clusters selected above.
%for to reconstruct the pixel track. 
To do so, the pseudorapidity angle difference among the three pixel clusters as shown in Fig.~\ref{fig:longitAngleDiff} (a) is measured. It can be expressed as (\ref{eq:deta1}):

\begin{equation} \label{eq:deta1}
    \Delta\eta_{\textrm{\scriptsize pix}} = \eta\left(L_{i},L_{j}\right) - \eta\left(L_{j},L_{k}\right) < 3\sigma
\end{equation}
    
where {\it i, j, k} $=$1…4 or 5 (if disks only) and {\it i} $\neq$ {\it j} $\neq$ {\it k}.

The pixel track segment is made by the connection between the leading electron track vertex and the selected pixel cluster (Fig.~\ref{fig:longitAngleDiff} (b)). The reconstructed electron vertex is used to calculate the pseudorapidity angle. The angle difference is calculated according to (\ref{eq:deta2}):

\begin{equation} \label{eq:deta2}
    ^{ji}\Delta\eta_{\textrm{\scriptsize vtx}} = \eta\left(z_{\textrm{\scriptsize vtx}},L_{i}\right) - \eta\left(z_{\textrm{\scriptsize vtx}},L_{j}\right) < 3\sigma
\end{equation}

where {\it i, j} $=$1…4 or 5 (if disks only) and {\it i} $\neq$ {\it j}.
Based on the measured pseudorapidity angle difference, the pixel clusters generated from pileup can be excluded from the pixel track reconstruction. 

The duplicated pixel tracks are removed by indexing each reconstructed track.

\subsubsection{Measuring the azimuthal $\phi$ curvature}

\begin{figure}[htbp]
    \centering
    \includegraphics[width=0.6\textwidth]{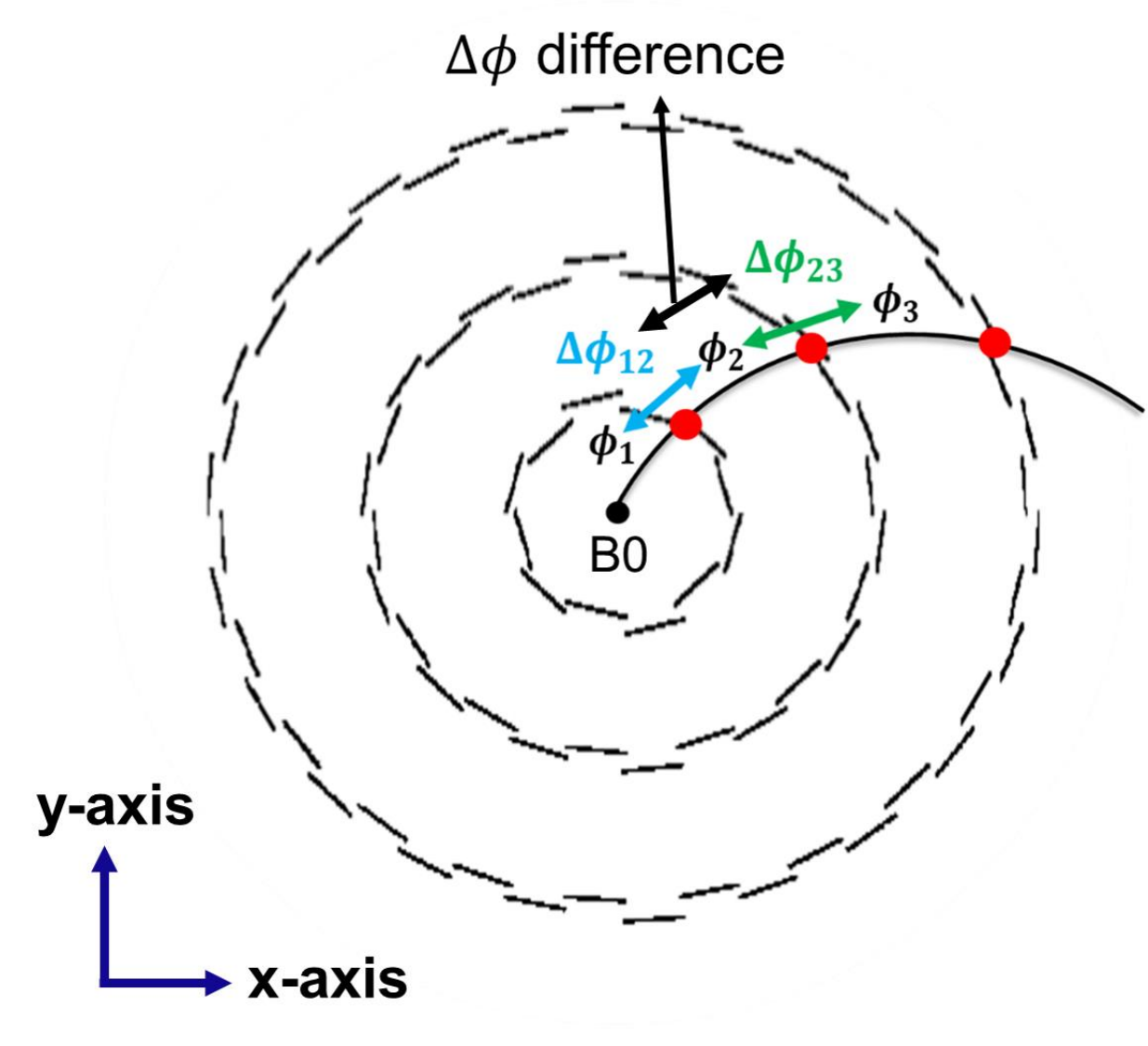} \\
    \caption{Schematic transverse view of the $\Delta\phi$ difference for the barrel region.}
    \label{fig:transAngleDiff}
\end{figure}

%The final step to reconstruct the pixel track in the cone is by measuring
The $\Delta\phi$ angle difference in the transverse plane (see  Fig.~\ref{fig:transAngleDiff}).
%The definition of the $\Delta\phi$ difference 
is expressed as follows:

%\begin{dmath}
\begin{equation}
    \Delta\phi\;\textrm{difference} = [\phi(\textrm{B0},L_{i})-\phi(L_{i},L_{j})] - [\phi(L_{i},L_{j})-\phi(L_{j},L_{k})] < 3\sigma
\end{equation}
   
%\end{dmath}

where {\it i, j, k} $=$1…4 or 5 (if disks only) and {\it i} $\neq$ {\it j} $\neq$ {\it k}.
The $\Delta\phi$ difference of the pixel track segments %reconstructed from the electron 
will be smaller when the corresponding track $p_{\textrm{\scriptsize T}}$ is higher.
%This indicates that the probability of reconstructing pixel track segments is increased if the pixel track isolation chooses the smallest $\Delta\phi$ difference value.

This is the last stage to reject fake tracks coming from combinatorial backgrounds.

%Once the pixel tracks within the cone around the leading electron are reconstructed, as explained in step 3, the next step consists in computing the $p_{\textrm{\scriptsize T}}$ of these tracks. This is performed as described in step 4.

%%%%%%%%%%%%%%%%%%%%%%%% step 4 %%%%%%%%%%%%%%%%%%%%%%%%%%%%%%%%%%%%%%
%%%%%%%%%%%%%%%%%%%%%%%% step 4 %%%%%%%%%%%%%%%%%%%%%%%%%%%%%%%%%%%%%%
%%%%%%%%%%%%%%%%%%%%%%%% step 4 %%%%%%%%%%%%%%%%%%%%%%%%%%%%%%%%%%%%%%
%%%%%%%%%%%%%%%%%%%%%%%% step 4 %%%%%%%%%%%%%%%%%%%%%%%%%%%%%%%%%%%%%%
%%%%%%%%%%%%%%%%%%%%%%%% step 4 %%%%%%%%%%%%%%%%%%%%%%%%%%%%%%%%%%%%%%

\subsection{Computation of the pixel-based tracks $p_{\textrm{\scriptsize T}}$}
\label{sec:computing_pixtrk}

% \item {\it Step 4: Computing pixel-based tracks $p_{\textrm{\scriptsize T}}$ and application to pixel track isolation}

\begin{figure}[htbp]
    \centering
    \subfloat{\includegraphics[width=0.8\textwidth]{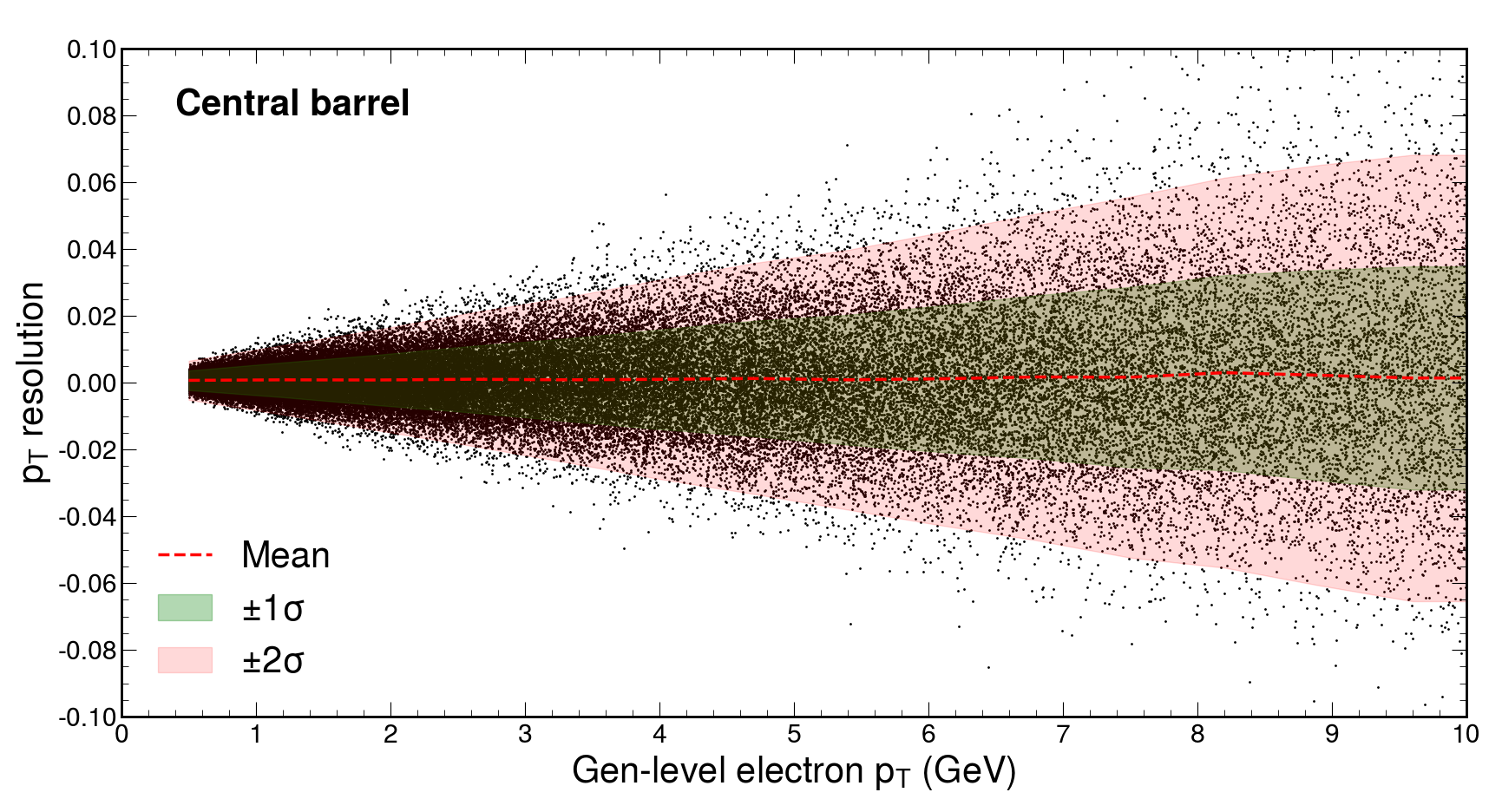}}

    \subfloat{\includegraphics[width=0.8\textwidth]{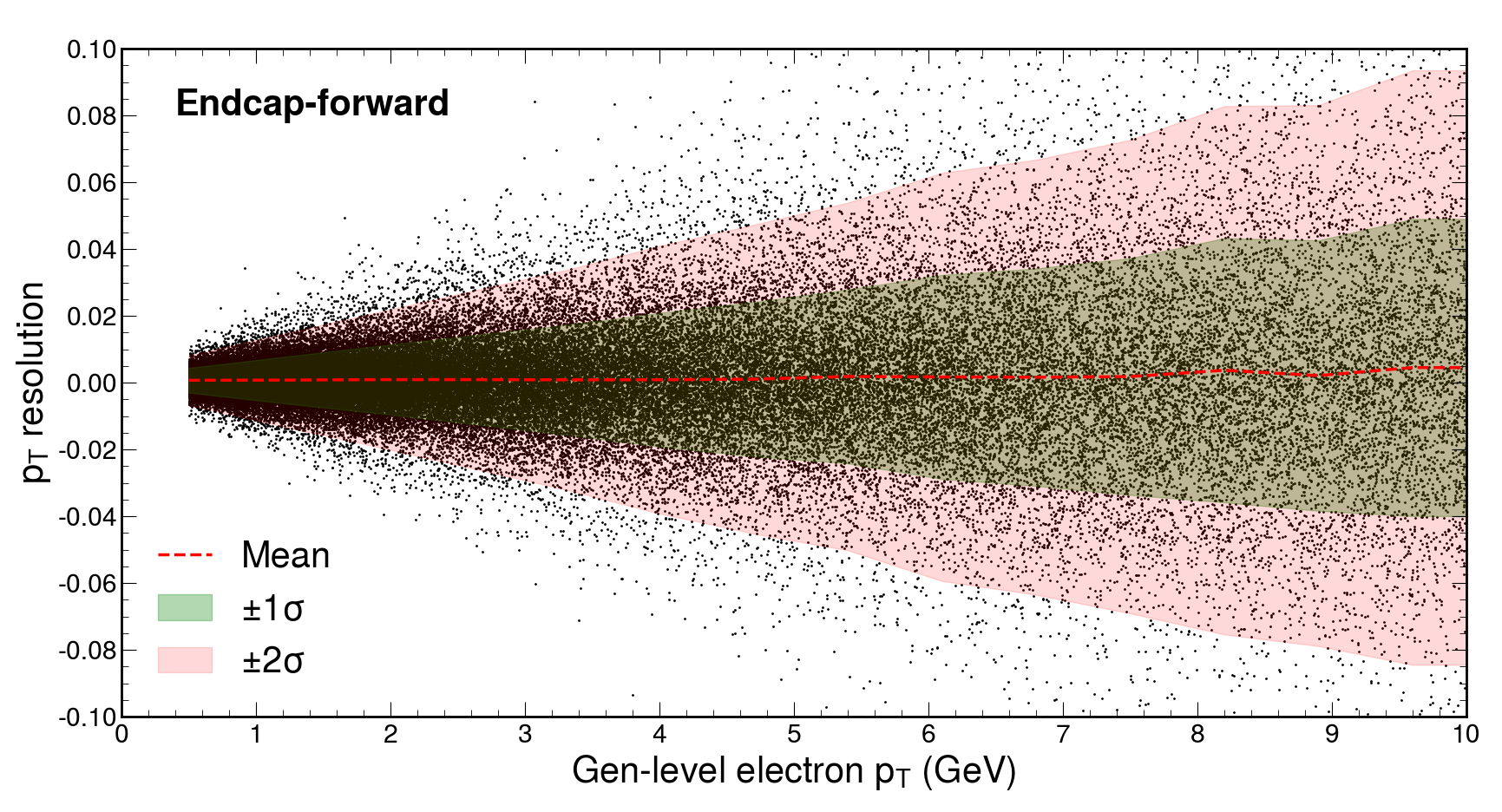}}
    \caption{The resolution of the reconstructed track $p_{\textrm{\scriptsize T}}$ as defined in Equation~\ref{eq:ptresolution}, in function of the gen-level track $p_{\textrm{\scriptsize T}}$ is shown on the top plot for tracks in the central barrel and on the bottom plot for tracks in the endcap region, based on a pure electron sample thus with no PU.}
    \label{fig:pt_resolution}
\end{figure}

After reconstructing the pixel-based tracks, additional information is provided by this detector, namely the $p_{\textrm{\scriptsize T}}$ of the tracks.
A new method is developed to compute this parameter. It is based on the well known formula:
%The $p_{\textrm{\scriptsize T}}$ computation is based on the well known formula:
\begin{equation}
    \centering
    p_{\textrm{\scriptsize T}} = 0.003 * B * R_{track}
\end{equation}

where $B$ is the magnitude of the magnetic field in which the tracker is merged, and $R_{track}$ (in cm) is the radius of the circle made with B0 and two of the pixel clusters relative to the reconstructed pixel track in the transverse plane (see  Appendix C for details).\\ 
The track $p_{\textrm{\scriptsize T}}$ is computed using a \texttt{DELPHES} sample of single electrons with $p_{\textrm{\scriptsize T}}$ ranging from 0 to 100 GeV and no pileup included. The $p_{\textrm{\scriptsize T}}$ resolution is defined by:

\begin{equation} \label{eq:ptresolution}
    \frac{(\textrm{reconstructed track}\: p_{\textrm{\scriptsize T}} - \textrm{gen-level track}\: p_{\textrm{\scriptsize T}})}{\textrm{gen-level track}\: p_{\textrm{\scriptsize T}}}
\end{equation}

where the gen-level quantity is provided by the simulated single electron tracks.\\
The results as a function of the gen-level $p_{\textrm{\scriptsize T}}$, are shown in Fig.~\ref{fig:pt_resolution}. Note that this is a pure electron sample i.e. with no PU.
%Two cases are shown, one ( 
For example, the top plot shows the resolution for tracks with the two clusters in the central barrel (Layers 1 and 4). 
%only (e.g. layers 1 and 4);
The bottom plot corresponds to tracks in the endcap-forward region with the two clusters in Disk 2 and Disk 5. 
The tracks with only  $p_{\textrm{\scriptsize T}}$ between 0.5 and 10 GeV %as we are interested here in 
are plotted, as rather low $p_{\textrm{\scriptsize T}}$ tracks are relevant for computing the track isolation. \\
In the central barrel, the resolution is smaller than 1\% up to 3 GeV $p_{\textrm{\scriptsize T}}$ and smaller than 3\% up to 10 GeV $p_{\textrm{\scriptsize T}}$. 
For most of the central tracks the $p_{\textrm{\scriptsize T}}$ resolution remains within 10\% even for tracks up to 30 GeV. For tracks in the endcap and forward regions, the $p_{\textrm{\scriptsize T}}$ resolution increases to 5\% up to 10 GeV and stays around at most 20\% for tracks larger than 10-15 GeV $p_{\textrm{\scriptsize T}}$.\\

%%%%%%%%%%%%%%%%%%%%%%%% step 5 %%%%%%%%%%%%%%%%%%%%%%%%%%%%%%%%%%%%%%
%%%%%%%%%%%%%%%%%%%%%%%% step 5 %%%%%%%%%%%%%%%%%%%%%%%%%%%%%%%%%%%%%%
%%%%%%%%%%%%%%%%%%%%%%%% step 5 %%%%%%%%%%%%%%%%%%%%%%%%%%%%%%%%%%%%%%
%%%%%%%%%%%%%%%%%%%%%%%% step 5 %%%%%%%%%%%%%%%%%%%%%%%%%%%%%%%%%%%%%%
%%%%%%%%%%%%%%%%%%%%%%%% step 5 %%%%%%%%%%%%%%%%%%%%%%%%%%%%%%%%%%%%%%

\subsection{Estimate of the electron track isolation}
\label{sec:estimate_traiso}

%\item {\it Step 5: Estimate of the electron track isolation}

 If the number of pixel tracks is zero or one in the isolation cone, the L1 electron track is isolated.
%the pixel track isolation algorithm.
%If the number of reconstructed pixel tracks in the isolation cone is larger than one, an isolation value can be calculated.
If the isolation cone contains \textit{n} >1 pixel tracks, with $p_{\textrm{\scriptsize T}}^{1}$ to $p_{\textrm{\scriptsize T}}^{n}$ in increasing order,
the pixel track isolation value is calculated as the ratio of two linear $p_{\textrm{\scriptsize T}}$ sum shown in Equation~\ref{eq:IsoValueDefine}.

\begin{equation} \label{eq:IsoValueDefine}
    \centering
    \textrm{Pixel Track Isolation} = \frac{\sum_{i=1}^{n-1} p_{\textrm{\scriptsize T}}^{i}}{\sum_{i=1}^{n} p_{\textrm{\scriptsize T}}^{i}}  
\end{equation}

The numerator is a linear sum of all the reconstructed pixel tracks $p_{\textrm{\scriptsize T}}$ except the one of the leading track, i.e. the electron track. 
The denominator is a linear sum of all the pixel tracks $p_{\textrm{\scriptsize T}}$. 
A minimum $p_{\textrm{\scriptsize T}}$ of 0.5 GeV is assumed for all the reconstructed tracks in the cone, except for the one of the electron track, which is at least 10 GeV.

In order to validate the procedure for determining the pixel track isolation, the isolation distributions of the electron (signal) events are compared with the background events. 
A sample of single electrons with 200 pileup is used for the signal events while a minimum bias sample is used for the background. 

Since the pixel track isolation algorithm depends on the number of tracks inside of the isolation cone, the simulation must well reproduce the content of the overall track including the soft tracks. 
The additional tracks from bremsstrahlung have thus to be included in the signal events of the \texttt{DELPHES} simulation, as \texttt{DELPHES} does not include this physics process.\footnote{A detailed procedure is developed, based on the bremsstrahlung as defined in a full LHC detector simulation, for modeling another single electron sample, without pileup, by properly adding to it the tracks due to the bremsstrahlung. 
The \texttt{DELPHES} sample corresponding to the single electron with pileup is then combined with the signal electron events without pileup modeled with the detailed bremsstrahlung simulation as just described.

A comparison of the isolation curves between the overall DELPHES modeled sample and the full detector simulation case shows a good agreement; this demonstrates that the additional tracks due to bremsstrahlung are well taken into account.}

\begin{figure}[hbtp]
    \centering
    \subfloat[Region 1]{\includegraphics[width=0.4\textwidth]{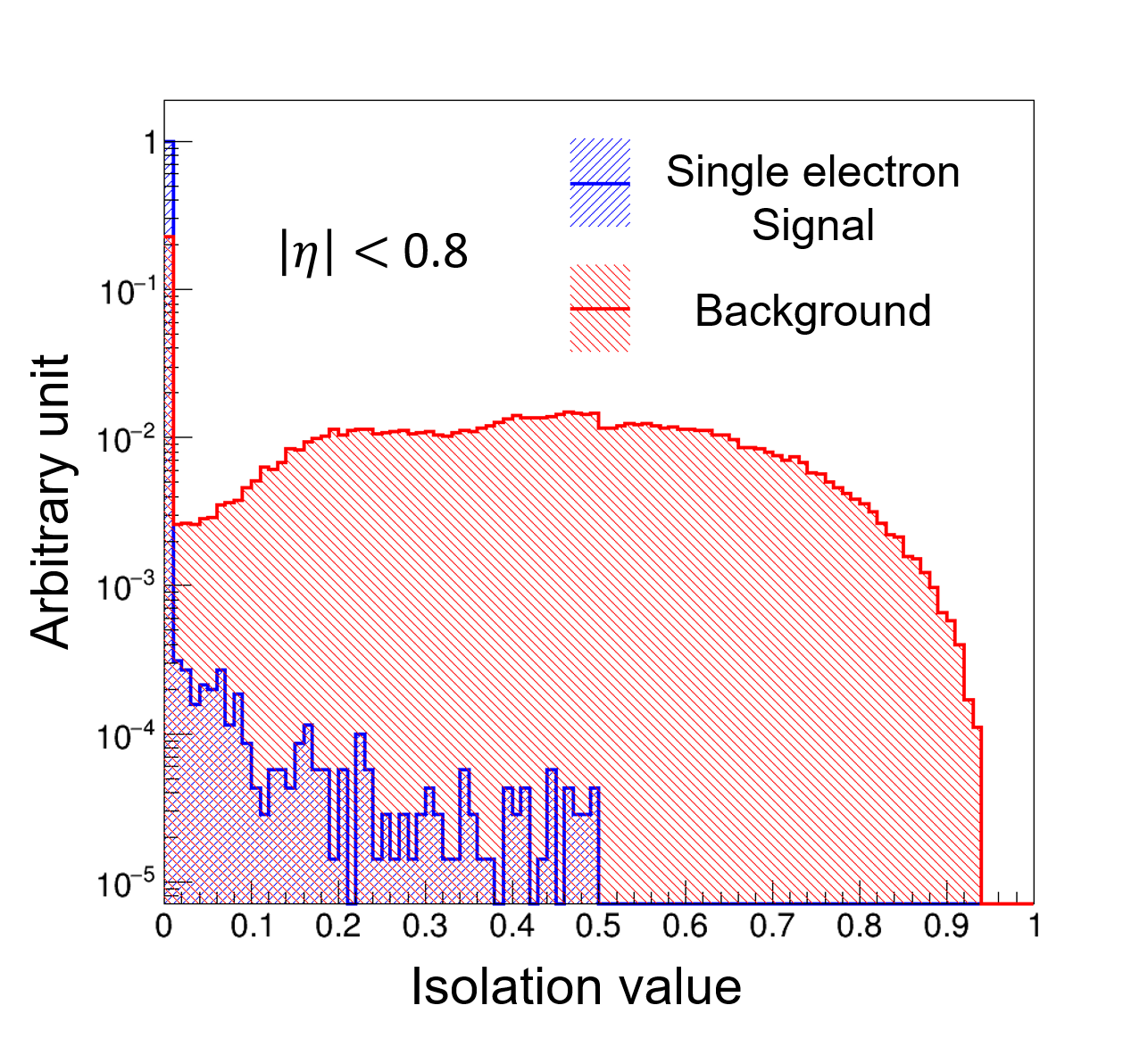}}
    \subfloat[Region 2]{\includegraphics[width=0.4\textwidth]{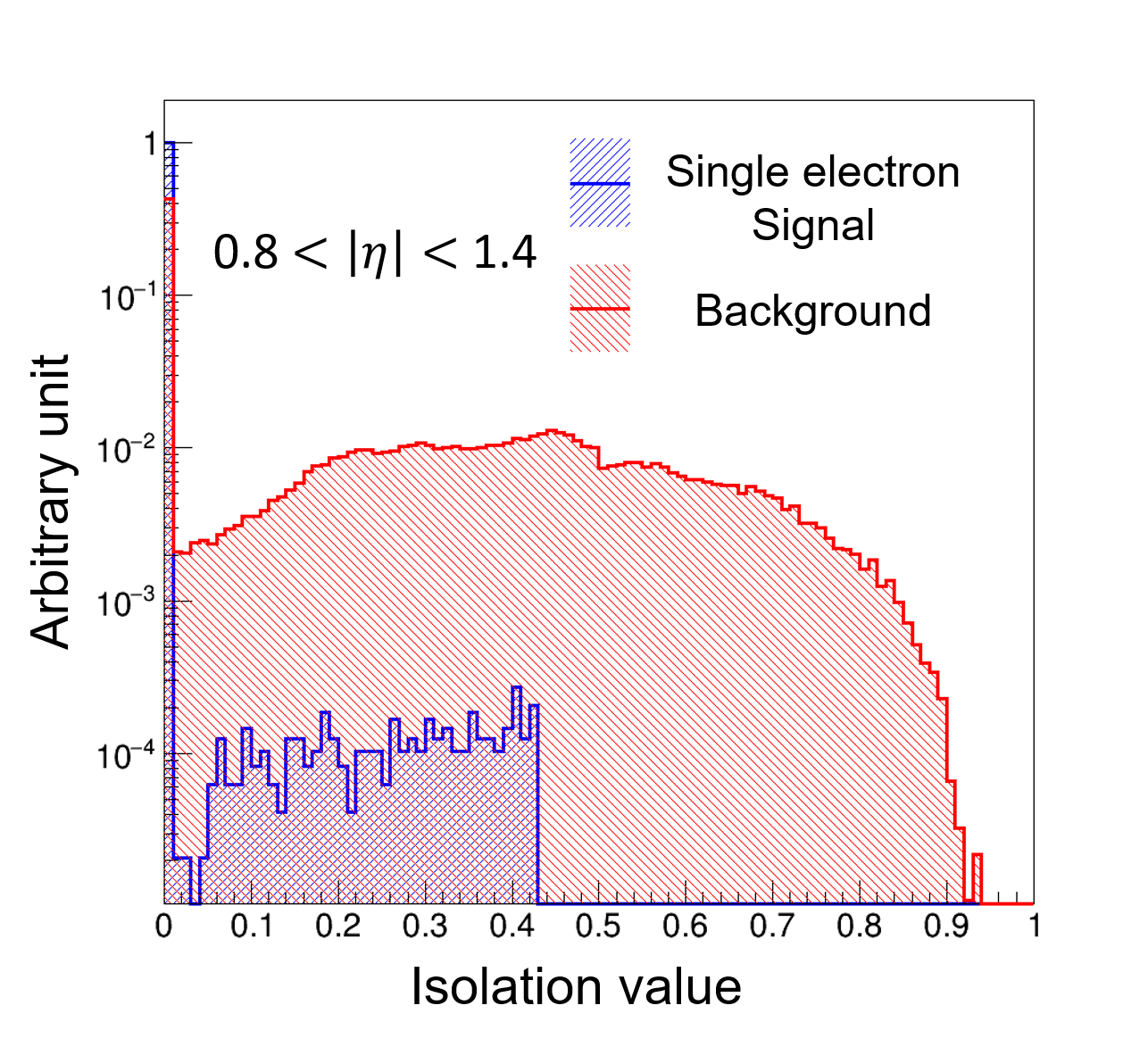}} 
    
    \subfloat[Region 3]{\includegraphics[width=0.4\textwidth]{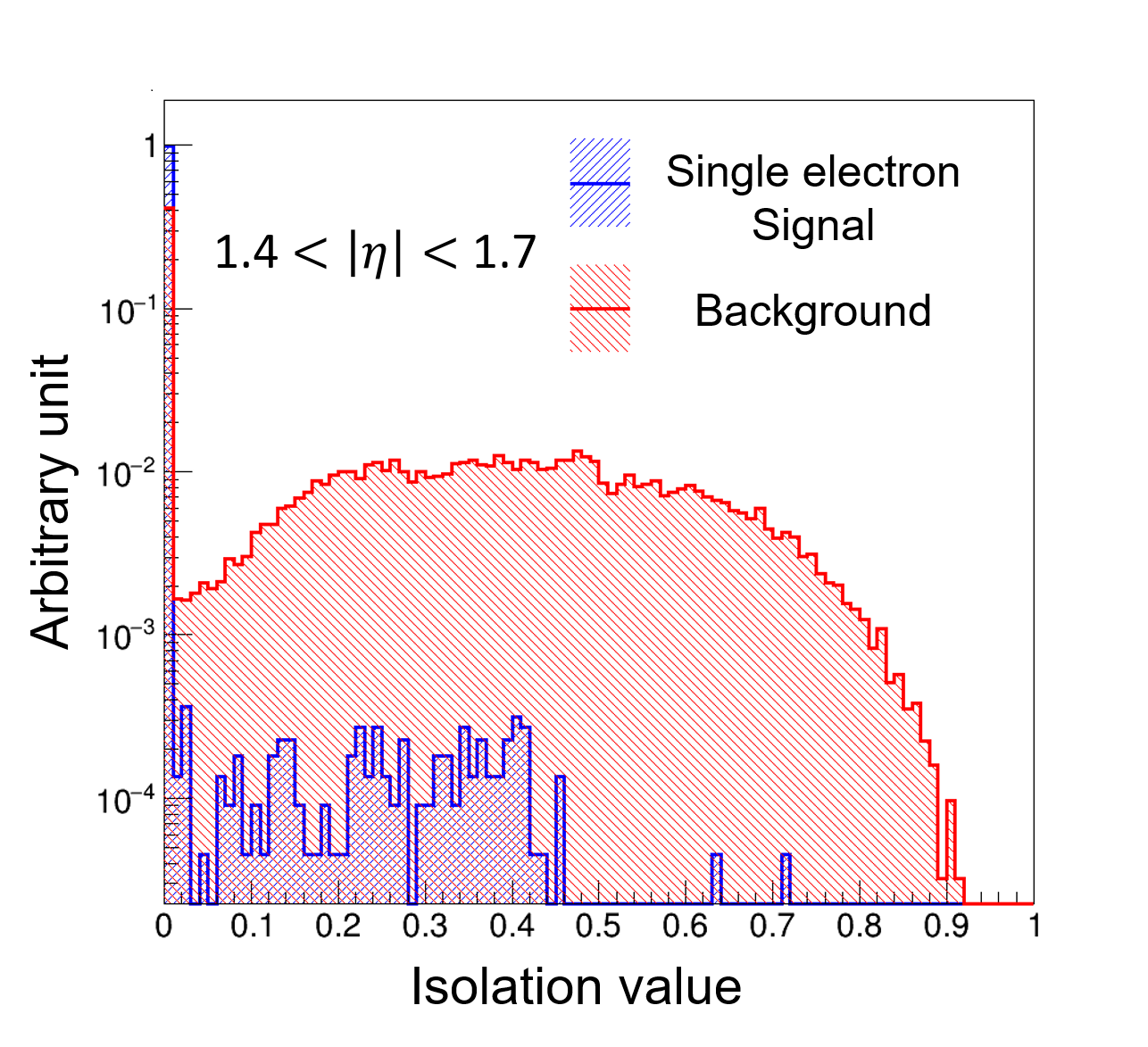}}
    \subfloat[Region 4]{\includegraphics[width=0.4\textwidth]{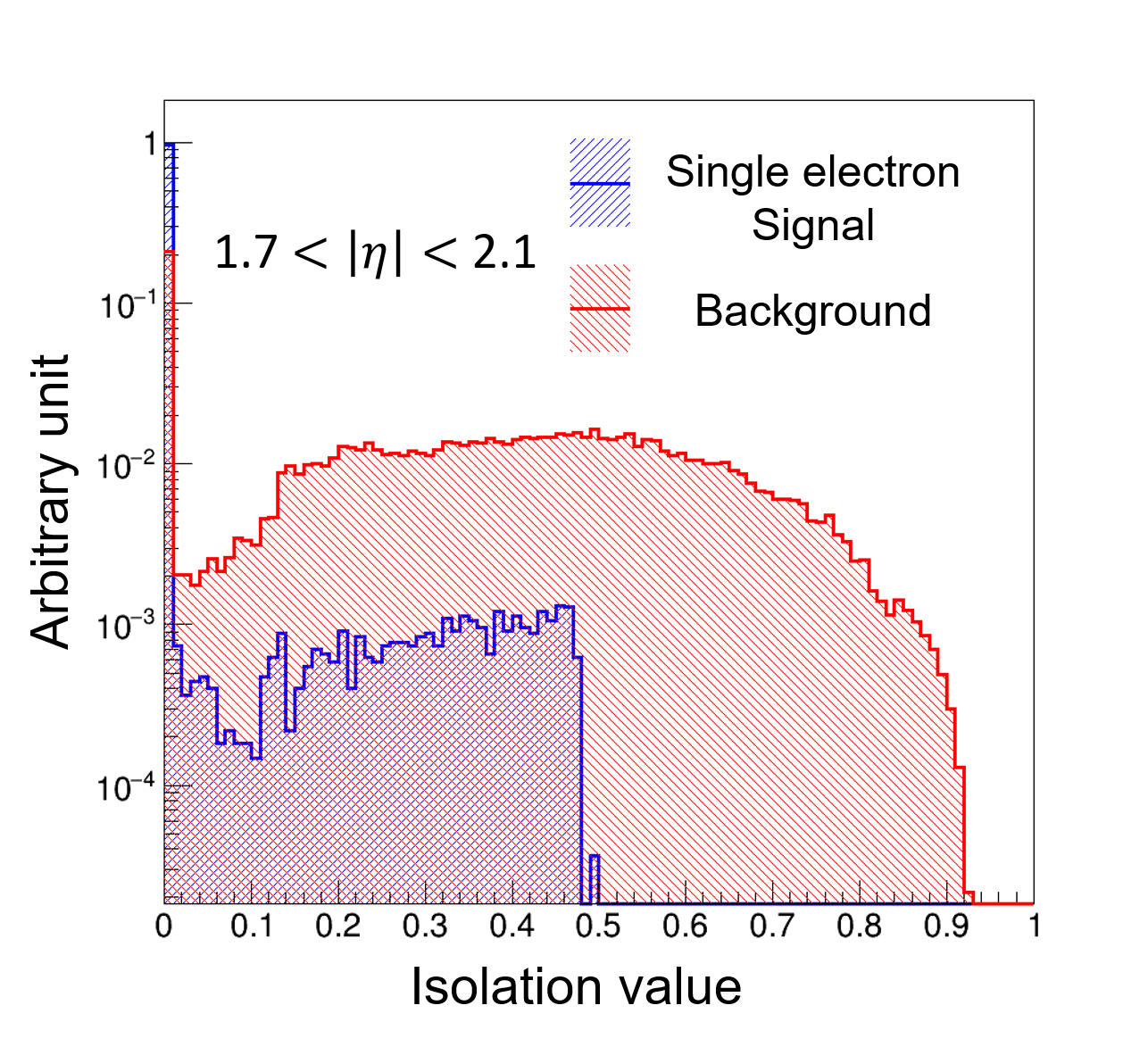}} 
    
    \subfloat[Region 5]{\includegraphics[width=0.4\textwidth]{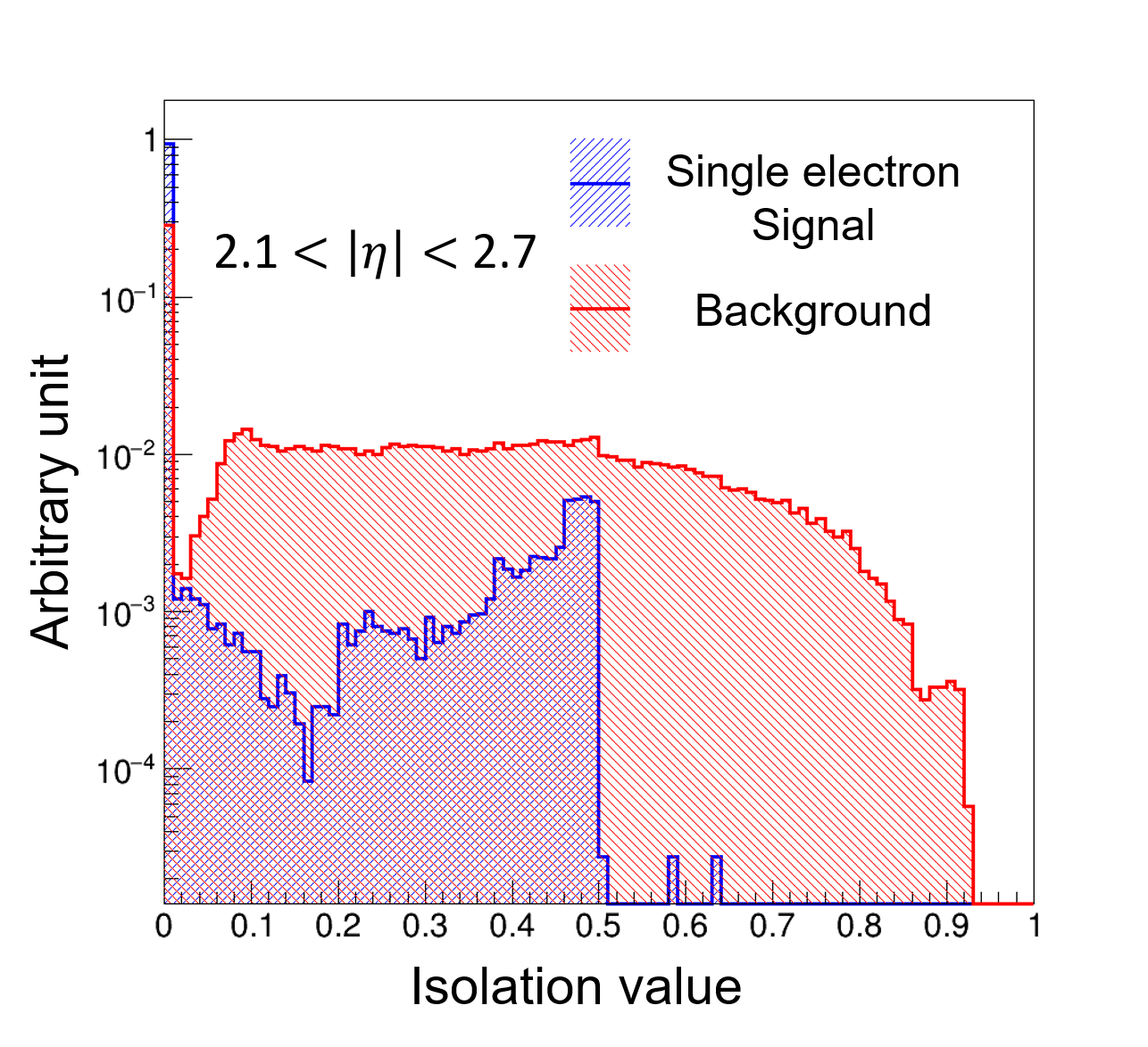}}
    \subfloat[Region 6]{\includegraphics[width=0.4\textwidth]{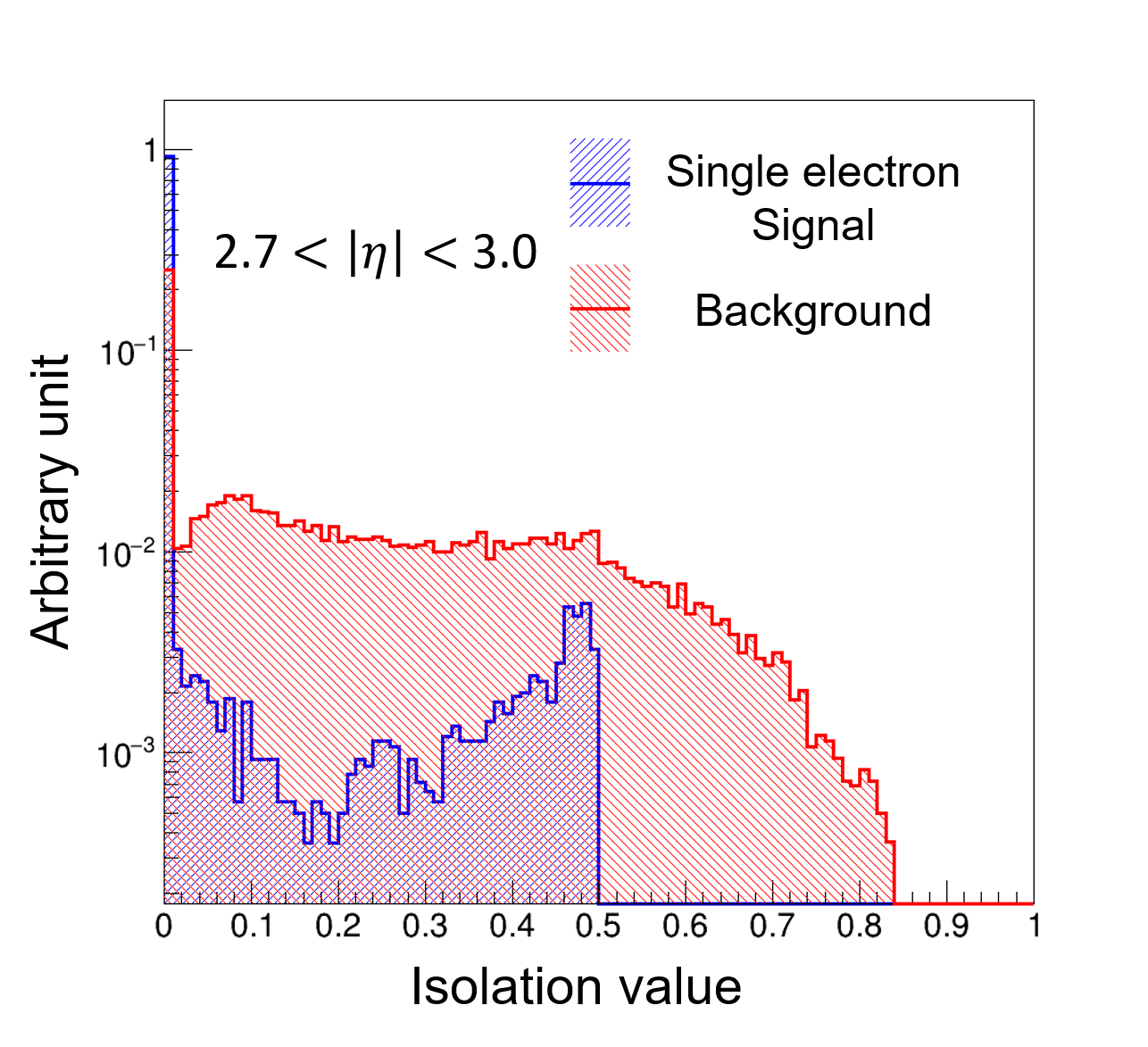}} 
    \caption{The pixel-based track isolation distributions of signal events (blue) and background events (red) in different $\eta$ regions.}
    \label{fig:IsovalueDist}
\end{figure}

Figure~\ref{fig:IsovalueDist} shows the isolation distributions of the signal events (in blue) and the overall background, i.e. minimum bias plus bremsstrahlung events, (in red), in each $\eta$ range, and for electrons with  $E_{\textrm{\scriptsize T}}$ larger than 20~GeV.

%In this Figure, the overall area of the histogram representing 
The distribution in a number of events as a function of the pixel track isolation value, for the signal (in blue) and for the background (in red), are normalized to one. 
%Thus the vertical coordinate is in arbitrary units in those plots.
The first bin corresponds to a pixel track isolation equal to zero, i.e.  %represents the case where there is
no additional track with at least 0.5 GeV in the isolation cone.

The pixel track isolation for the electron tracks depends on the $\eta$ range (Fig.~\ref{fig:IsovalueDist}). Additional tracks can be produced by bremsstrahlung due to the higher material budget in the endcap regions. 
These tracks contribute to a fair fraction of the single electron signal events in the forward region (2.1 $< |\eta| <$ 3.0) thus increasing the value of the pixel track isolation.
As a result, the background rejection is lowered in the endcaps.

Using these distributions, a cut value in this isolation parameter is determined for disentangling at best the signal from the background. The results are summarized in Table~\ref{tab:theIsoVal_op1}, 
in terms of the obtained signal efficiency and the corresponding overall background rejection, 
for each of the six regions in $\eta$ and the corresponding isolation cut value. 
Another option is also considered for the two largest $\eta$ regions, corresponding to $\eta > 2.1$, for an electron of 20 GeV and results are shown in Table~\ref{tab:theIsoVal_op2}.\\

\begin{table}[htbp]
    \centering
    \scalebox{1}{
        \begin{tabular}{|c|c|c|c|}
            \hline
            $\eta$ range & Isolation cut value & Signal efficiency & Background rejection \\ 
            \hline
            \hline
            $|\eta| < $ 0.8 & 0.10 & 99.8\% & 74.2\% \\
            \hline
            0.8 $ < |\eta| < $ 1.4 & 0.10 & 99.6\% & 54.7\% \\
            \hline
            1.4 $ < |\eta| < $ 1.7 & 0.17 & 99.6\% & 53.0\% \\
            \hline  
            1.7 $ < |\eta| < $ 2.1 & 0.28 & 98.0\% & 59.0\% \\
           \hline
           % 2.1 $ < |\eta| < $ 2.7 & 0.40 & 98.1\% & 29.8\% \\ 
           2.1 $ < |\eta| < $ 2.7 & 0.27 & 95.4\% & 46.3\% \\ 
           \hline
           % 2.7 $ < |\eta| < $ 3.0 & 0.43 & 98.1\% & 30.0\% \\
           2.7 $ < |\eta| < $ 3.0 & 0.21 & 94.8\% & 45.7\% \\
            \hline
        \end{tabular} }
    \caption{The signal efficiency and background rejection obtained with the pixel track isolation algorithm for each $\eta$ region, for an electron of 20 GeV.} 
    \label{tab:theIsoVal_op1}
\end{table}

\begin{table}[htbp]
    \centering
        \scalebox{1}{
        \begin{tabular}{|c|c|c|c|}
            \hline
            $\eta$ range & Isolation cut value & Signal efficiency & Background rejection \\ 
            \hline
            \hline
            2.1 $ < |\eta| < $ 2.7 & 0.44 & 98.0\% & 22.5\% \\
           \hline
            2.7 $ < |\eta| < $ 3.0 & 0.47 & 98.0\% & 17.6\% \\
            \hline
        \end{tabular} }
    \caption{The signal efficiency and background rejection obtained with the pixel track isolation algorithm for the two largest $\eta$ regions, corresponding to $|\eta| > 2.1$, for an electron of 20 GeV.} 
    \label{tab:theIsoVal_op2}
\end{table}

A signal efficiency of 99.8\% with a background rejection of 75\% is obtained in the central barrel region i.e. for $\eta$ up to 0.8. 
The signal efficiency remains at about the same value between 99.8 and 99.6\% with a background rejection between 55 and 59\% for $\eta$ between 0.8 and 2.1.
 
It then slightly drops to 95.8\% and 95.2\% while keeping the background rejection at around 45\% in the forward region i.e. $\eta$ between 2.1 and 3.0, if we choose to slightly decrease the performance in signal efficiency while we maintain a relatively high background rejection. 
This is the option 1. 
A second option (option 2) is based on maintaining a very high efficiency of about 98\% for the two more forward regions while the rejection rate is decreased to about 20\%, thus allowing a slightly higher L1 trigger rate.
%The results are discussed in the next section.  However, let's already stress here that they have to be taken with a grain of salt: it is worth reminding once more of the rather simplistic pixel detector design used in this study together with a less detailed simulation (DELPHES). 
Because of a simplified detector layout and simulation, the results reported here are underestimated with respect to what will be achievable with the sophisticated designs of the ATLAS and CMS pixel detectors for HL-LHC also serving as examples for future machines.

 %%%%%%%%%%%%%%%%%%%%%%%% section 5 Results and Performances %%%%%%%%%%%%%%%%%%%%%%%%%%%%%%%%%%%%%%
 %%%%%%%%%%%%%%%%%%%%%%%% section 5 Results and Performances %%%%%%%%%%%%%%%%%%%%%%%%%%%%%%%%%%%%%%
 %%%%%%%%%%%%%%%%%%%%%%%% section 5 Results and Performances %%%%%%%%%%%%%%%%%%%%%%%%%%%%%%%%%%%%%%
 %%%%%%%%%%%%%%%%%%%%%%%% section 5 Results and Performances %%%%%%%%%%%%%%%%%%%%%%%%%%%%%%%%%%%%%%
 %%%%%%%%%%%%%%%%%%%%%%%% section 5 Results and Performances %%%%%%%%%%%%%%%%%%%%%%%%%%%%%%%%%%%%%%

\section{Results and Performances}
\label{sec:result}
%This section summarizes the results of the performance studies presented in this paper, addressing the benefits of using the real-time information from the pixel detectors, with the electrons as a showcase. 
%
This section summarized the main results of the performance studies. It stresses the benefits of the L1 trigger performances by including the pixel information in the electron trigger as an example. Section 6 summarizes the two main categories of technological challenges to be overcome to make this option feasible within the HL-LHC scenario. This implies the Pixel Front-End ASIC and the real-time related algorithms to perform this triggering scheme, thanks also to the novel development in the processor technology. Section 7 concludes by showing the perspectives for a possible beyond baseline upgrade at HL-LHC and also for application to future colliders.

\subsection{Performance in real-time selection: efficiency and rate reduction}
\label{sec:perfo1}

The performance of the real-time selection is measured in terms of the real-time selection efficiency (also called Level-1 trigger efficiency) over the full $\eta$ range of the detector and of the corresponding trigger rate reduction.

The efficiency of the PiXTRK real-time track reconstruction algorithm without (blue) or with (green) including the pixel charged isolation (in green for option 1 and orange for option 2 in the endcaps) is measured as the trigger efficiency for electrons with a threshold of 35 GeV in $E_{\textrm{\scriptsize T}}$. 

\begin{figure}[hbtp]
    \centering
        \includegraphics[width=0.65\textwidth]{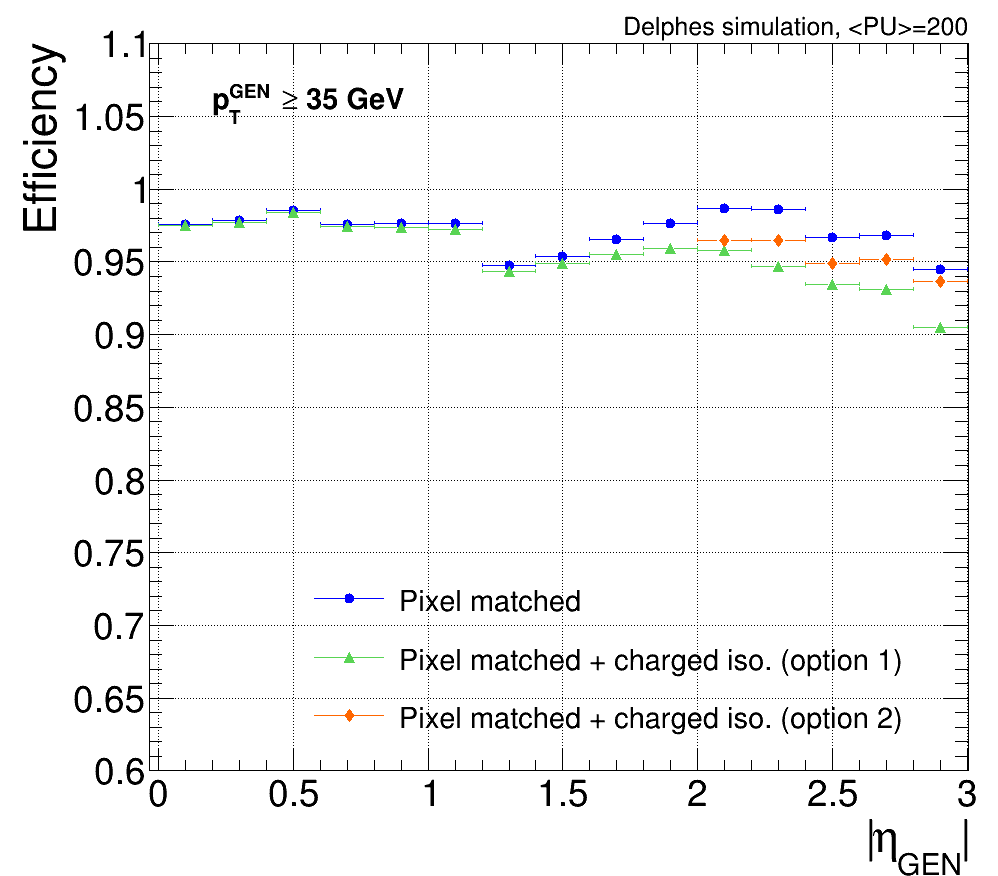} 
    \caption{Real-time trigger efficiency as a function of $\eta$ at the generator level, $|\eta_{GEN}|$, for electrons of at least 35 GeV in $E_{\textrm{\scriptsize T}}$, and 200 pileup at the HL-LHC, as achieved by the PiXTRK algorithm without (blue) and with (green) pixel track isolation if option 1 or with (orange) pixel track isolation if option 2, with \texttt{DELPHES} based simulation.}
    \label{fig:pixTrkEff}
\end{figure}

The efficiency is shown in Fig.~\ref{fig:pixTrkEff}, as a function of the $\eta$ at the generator level, $|\eta_{GEN}|$, of the electron candidates for the two presented options, on the overall covered $\eta$ range by the pixel detector.

Table~\ref{tab:pixTrkEfficiency} shows the average trigger efficiency for the different L1 trigger cases in four different regions, namely:$|\eta| <$ 1.0,  1. $< |\eta| <$ 1.5, 1.5 $< |\eta| <$ 2.5 and 2.5 $< |\eta| <$ 3.0 and taking into account the two considered options for the far-end or forward regions.\\

%% option 2 - average efficiency
\begin{table}[htbp]
    \centering
    \scalebox{0.85}{
        \begin{tabular}{|l|c|c|c|c|}
            \hline
            L1 trigger & $|\eta| < $ 1.0 & 1.0 $ < |\eta| < $ 1.5 & 1.5 $ < |\eta| < $ 2.5 & 2.5 $ < |\eta| < $ 3.0 \\
            \hline
            \hline
            Pixel matched & 97.8\% & 95.8\% & 97.7\% & 95.2\% \\
            \hline
            Pixel matched + track iso. (option 1) & \multirow{2}{*}{97.7\%}  & \multirow{2}{*}{95.5\%} & 95.2\% & 90.9\% \\
            \cline{1-1}\cline{4-5}
            Pixel matched + track iso. (option 2) &  &  & 96.1\% & 93.9\% \\
            \hline
        \end{tabular} }
    \caption{Average efficiency for the electron track reconstruction of the real-time PiXTRK algorithm, in the barrel ($|\eta| <$ 1), the transition region barrel-endcap (1.0 $< |\eta| <$ 1.5), the near-endcap region (1.5 $< |\eta| <$ 2.5) and the far-endcap or forward (2.5 $< |\eta| <$ 3.0) regions.}
    \label{tab:pixTrkEfficiency}
\end{table}

Another important parameter in the real-time selection performances of a detector is the impact on the trigger rate reduction. This is determined as a function of L1 e/$\gamma$ object $E_{\textrm{\scriptsize T}}$ threshold, over the overall $\eta$ coverage of the detector.

\begin{figure}[hbtp]
    \centering
    \subfloat[Rate reduction in the barrel ($|\eta| <$ 1.5)]{\includegraphics[width=0.5\textwidth]{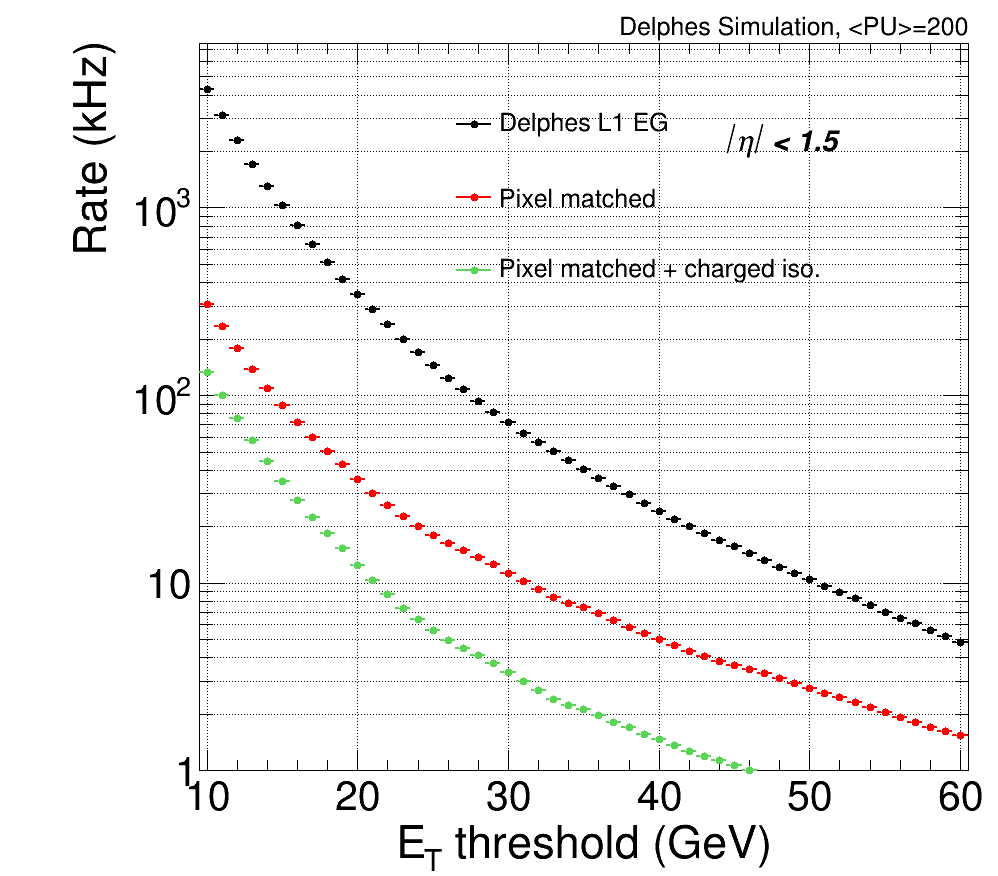}} %% option 2
    \subfloat[Rate reduction in endcap (1.5 $< |\eta| <$ 2.5)]{\includegraphics[width=0.5\textwidth]{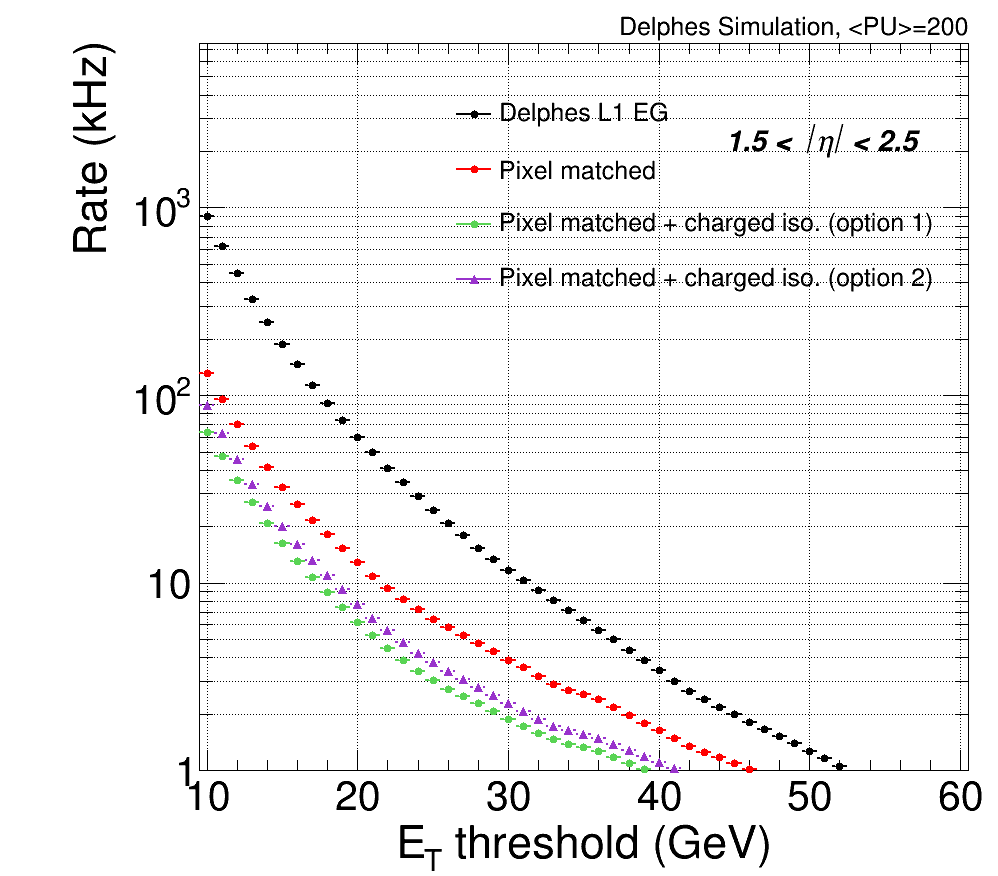}} %% option 2
    
    \subfloat[Rate reduction for overall region ($|\eta| <$ 2.5)]{\includegraphics[width=0.5\textwidth]{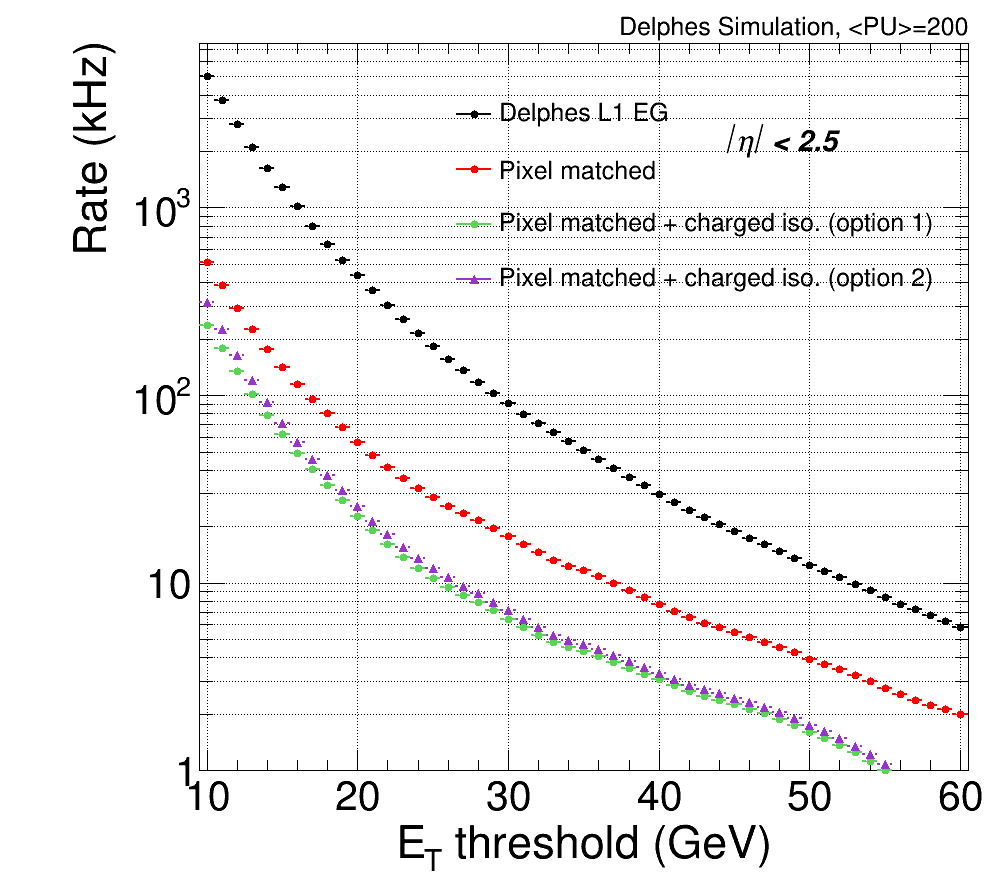}} %% option 2
    \subfloat[Rate reduction in forward (2.5 $< |\eta| <$ 3.0)]{\includegraphics[width=0.5\textwidth]{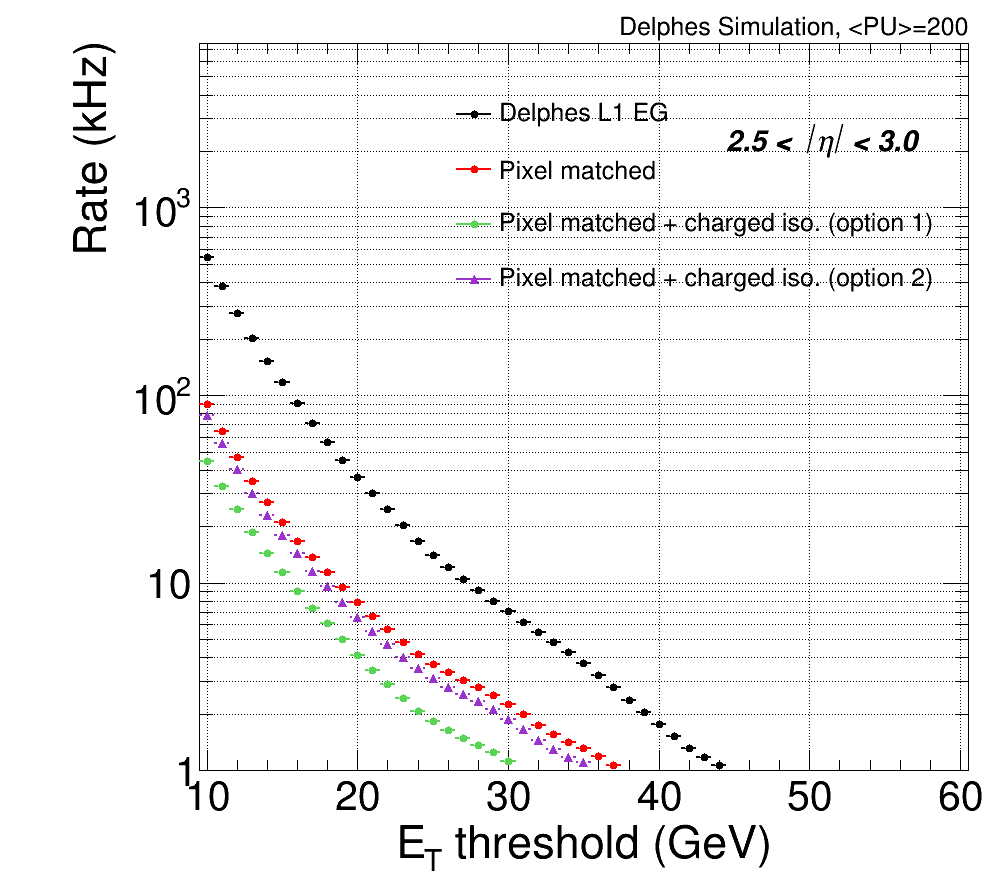}} %% option 2
    
    \caption{Real-time trigger rate as a function of $E_{\textrm{\scriptsize T}}$ with \texttt{DELPHES} based simulation for different $\eta$ regions and including for the endcap and forward regions the two considered options by choosing two different isolation cut values.
    }
    \label{fig:trigrate}
\end{figure}

The rate reduction is compared to what is achieved by the Level-1 calorimeter trigger (black), by the PiXTRK real-time reconstruction algorithm without (red), and with the pixel track isolation for the two considered options (green for option 1 and magenta for option 2). 
The corresponding curves, of the rate reduction as a function of the L1 e/$\gamma$ object $E_{\textrm{\scriptsize T}}$ threshold provided by the L1 EM calorimeter, for the different regions in $\eta$, namely: in the extended barrel ($|\eta| <$ 1.5), the near-endcap region (1.5 $< |\eta| <$ 2.5), the combined overall $|\eta| <$ 2.5 region and the far-endcap or forward region (2.5 $< |\eta| <$ 3.0) where tracking can only be made by the pixel detector, are shown on Fig.~\ref{fig:trigrate}.

The results in rate reduction are summarized as well, in Table~\ref{tab:trigrate}.
It gives the average rates obtained by applying PiXTRK algorithm without or with track isolation, for the electron reconstruction in the extended barrel ($|\eta| <$ 1.5), the endcap region (1.5 $< |\eta| <$ 2.5), the overall region covered by the outer tracker ($|\eta| <$ 2.5) and the far-endcap or forward region (2.5 $< |\eta| <$ 3.0) only covered by the pixel tracker. 
It is worth noting that, in the considered scenario the (2.5 $< |\eta| <$ 3.0 end-cap/forward region) is only covered by the Pixel detector, from the tracking system point of view, while the endcap calorimeter covers this region. 

\begin{table}[htbp]
    \centering
    \scalebox{0.75}{
    \begin{tabular}{|l|c|c|c|c|c|c|}
        \hline
        \multirow{2}{*}{L1 trigger} & \multicolumn{2}{c|}{$|\eta| < $ 1.5} & \multicolumn{2}{c|}{1.5 $ < |\eta| < $ 2.5} & \multicolumn{2}{c|}{2.5 $ < |\eta| < $ 3.0} \\
        \cline{2-7}
            & Rate & Rej. factor & Rate & Rej. factor & Rate & Rej. factor \\
        \hline\hline
        Calorimeter-only & 345 kHz & - & 60.0 kHz & - & 36.8 kHz & - \\
        \hline
        Pixel matched & 35.6 kHz & 9.7 & 12.8 kHz & 4.7 & 7.9 kHz & 4.6 \\
        \hline
        Pixel matched + track iso. (option 1) & \multirow{2}{*}{12.4 kHz} & \multirow{2}{*}{27.8} & 6.2 kHz & 9.7 & 4.1 kHz & 8.9 \\
        \cline{1-1}\cline{4-7}
        Pixel matched + track iso. (option 2) & & & 7.7 kHz & 7.8 & 6.5 kHz & 5.6 \\
        \hline
   \end{tabular}
    }
    \caption{Rates and rejection factor at $E_{\textrm{\scriptsize T}} =$ 20 GeV threshold obtained by PiXTRK for the electron reconstruction in the barrel ($|\eta| <$ 1.5), endcap (1.5 $< |\eta| <$ 2.5), and forward (2.5 $< |\eta| <$ 3.0) regions.}
    \label{tab:trigrate}
\end{table}

\subsection{Real-time processing of the pixel detector information: further potential}
\label{sec:benefits}

Beyond the improvements in the real-time selection (or Level-1 trigger), this section summarizes the additional benefits the pixel information provides if processed in real-time. Because of its location and design, the innermost part of tracking systems, made of extremely fine granularity pixels presents the challenge of the huge data rate (thus the need for seeded handling of its information) but in counterpart offers three major virtues: closest to the beam-pipe and thus to the interaction point, low material budget, very forward expandability.

%To conclude this section let's stress and summarize these benefits. Because of its location and design the pixelated section (extremely fine granularity) of the overall tracking system in the design of experiments presents three major virtues: closest to the beam-pipe and thus the interaction point, low material budget, the most extendable to very large $\eta$ range.

This device is thus unique for determining with a very high resolution of the primary vertex position of the events, tagging b- or c-quarks produced in the interaction (high resolution secondary vertex position), handling the pileup, discriminating tracks from bremsstrahlung, reconstructing relatively low $p_{\textrm{\scriptsize T}}$ tracks, being the tracking device linked to the endcap and forward parts of the calorimeters and the muon detectors because covering a tracking region that cannot be addressed by the outer tracker. All this is performed at the High-Level Trigger and off-line. But having these possibilities in real-time will be more and more needed by physics and achievable thanks to the advances in AI-based algorithms and new processors.\\

This study does not address all of them, but using the electron case, a fair amount of the potential of this device for real-time processing is shown.\\

By being closest to the interaction point and with the finest granularity, the vertexing capability of this device is unique. Section ~\ref{sec:vertex} describes a detailed study of how to exploit it in real-time with the electrons as a showcase. Table~\ref{tab:pv_resolution_at_sec5} here below, summarizes the results of this study with the performances in the resolution of the primary vertex position as determined, with the Pixel detector information for single electrons plus 200 pileup events.\\

\begin{table}[hbtp]
    \centering
    \begin{tabular}{|c|c|}
        \hline
        $\eta$ Range & Vertex resolution \\
        \hline
        $|\eta| < $ 0.8 & 19.7 $\mu$m \\
        \hline
        0.8 $ < |\eta| < $ 1.4 & 28.2 $\mu$m \\
        \hline
        1.4 $ < |\eta| < $ 1.7 & 63.3 $\mu$m \\
        \hline
        1.7 $ < |\eta| < $ 2.1 & 58.1 $\mu$m \\
        \hline
        2.1 $ < |\eta| < $ 2.7 & 244 $\mu$m \\
        \hline
        2.7 $ < |\eta| < $ 3.0 & 379 $\mu$m \\
        \hline
    \end{tabular}
    \caption{Resolution on the primary vertex position for the electron plus 200 pileup events, in the six considered regions corresponding to different $\eta$ ranges. }
    \label{tab:pv_resolution_at_sec5}
\end{table}

The z-resolution in the central region ($\eta$<1.4) is below 30 $\mu$m. It remains around 60 $\mu$m up to $\eta$ of 2.1. In the forward region, it reaches 380 $\mu$m. More generally, the vertex position resolution obtained with the pixel detector is better than an order of magnitude compared to the one obtained with the outer tracker only, as reported in the performance study, including the pixel information in a real-time b-tagging trigger ~\cite{Moon_2016}.\\

The low material budget and again the fine granularity of this device provide other advantages of interest for the real-time processing of its information. 
%Among these advantages, the capability to consider tracks with very low $p_{\textrm{\scriptsize T}}$ to measure them with a rather good resolution.
Among these advantages, the capability to consider tracks with very low $p_{\textrm{\scriptsize T}}$ means that they can be measured with fairly good resolution.
This is addressed in detail in the study of the track isolation (Section ~\ref{sec:TrkIso}).
Decreasing the $p_{\textrm{\scriptsize T}}$ cut threshold for the tracks in the isolation cone from 2 GeV to 0.5 GeV diminishes the background rejection from 65\% to 10\% but allows keeping 99\% instead of 92\% of the signal events. 

To optimize the real-time selection, one has to make a compromise between the efficiency and the rate reduction.
It is worth noting that keeping the trigger efficiency very high to the expense of a decrease in rate reduction gives rather similar results in terms of rate reduction for $|\eta|$ up to 2.5. Instead, for $\eta > 2.5$, the charged isolation does not have a real impact on rate reduction but this trigger rate is not dramatically high, and indeed still affordable.

%The performances obtained for this even simplified design can be compared for instance with the ones of the L1 tracking trigger system as developed by CMS for the HL-LHC, based on a novel outer tracker design~\cite{CMS-phase2L1TDR} with its unique triggering at 40 MHz capability. It indicates the possible additional benefits of including the pixel detector within the real-time trigger architecture.

Besides the benefits of the real-time selection, lowering the $p_{\textrm{\scriptsize T}}$ cut threshold of smaller values, has an important impact on increasing the Physics reach on a number of important topics such as Heavy Flavour Physics, rare $\tau$ lepton decays, etc.\\

%Besides its capability to extend much forward the tracking capability of an experiment, this detector also allows considering tracks with a lower $p_{\textrm{\scriptsize T}}$, namely 0.5 GeV instead of 2 GeV. This is interesting for a number of reasons and especially here for an improved estimate of the charged isolation

The expandability of the pixel detector to very large values in $\eta$ is another asset of this device. It makes it unique as a tracking device in the forward regions, to be coupled with the calorimeters and the muon detectors.
%Even more Important are the Physics cases that benefit from the capability of lowering this threshold to smaller values such as low $p_{\textrm{\scriptsize T}}$ B-Physics, rare $\tau$ lepton decays, etc. 
Although the simplified pixel layout considered here extends only to $\eta$ of 3, because of the present coverage of the endcap calorimeters for HL-LHC, this study already indicates the tracking capability and performances of the pixel detectors in the forward region as covered for the first phase (at least) of the HL-LHC. 
It must be noted that the present microvertex being built for HL-LHC covers up to $\eta$ of 4. 
It should be pointed out that, because of the Physics needs at the HL-LHC, an extension of the calorimetry and of the muon detection down to $\eta$ of 4, might be part of the second stage of the HL-LHC upgrades of the experiments for Run 5, expected to start in 2035~\cite{HL-LHC-Schedule}. 
Thus, the results of this study can be extended with a still good efficiency and rejection rate for the electron case to $\eta$ of 4 and extrapolated as well to other physics objects such as muons or jets.

The extension of the current pixel detectors of both ATLAS and CMS down to $\eta$ of 4, will thus be of unique value for reconstructing forward electrons. 
Following the example of the electrons, muons as well as $\tau$ leptons, forward jets, or tagging of b-quarks to the forward regions (boosted objects) will be feasible, all this in real-time triggering. 
This will be the object of other studies on jet real-time reconstruction performances and b-tagging performances with the pixel detectors.

It is important for the physics potential at the HL-LHC, and even more when considering future machines with higher c.m. energy. to extend the tracker to larger $\eta$. 
In that later case, preliminary designs for FCC-hh~\cite{FCC} for instance, extend the usable $\eta$ range of the pixel trackers to at least $ |\eta|$ of 4 or even 6, because of the Physics requirements.

Besides, the HL-LHC will be a unique playground for learning how to handle these physics cases even more important at higher energy machines.

\section{Pixel information in real-time: Main Technological challenges}
\label{sec:future}
The goal is to include the microvertex detector information in the real-time selection stage at LHC,  i.e. at 40 MHz input rate and within a trigger latency of order 10 $\mu$s. Therefore, the two challenging parameters are the bandwidth and the latency achievable by the level-1 electron pixel-based trigger. 
To be feasible, its bandwidth should not exceed a few percent of the total Level-1 trigger bandwidth (currently 750 kHz at HL-LHC, and possibly increased up to at least 1 MHz ~\cite{atlasTDR-TDAQ-Add}. 
Its latency should fit within typically 10 $\mu$s, for the total corresponding L1 processing. This mainly adverts to very demanding challenges on 1) the hardware and detector aspects, including in particular the R\&D on new pixel technology and the associated Front-End Electronics (FEE), and 2) the software computing aspect with the implementation of fast (real-time)  and highly performing algorithms and processor units. 

%The first challenge is primarily on the Front-End pixel ASIC, sitting at the forefront of the signal processing.
%The second challenge is on an intelligent processing of the Front-End information delivered by the FEE, able to provide within the short trigger latency, relevant and beneficial information, further improving this first selection stage. 
%This second challenge is related to real-time algorithms and fast new processors. 
A preliminary estimate of the bandwidth and the latency requested by the proposed L1 electron pixel detector is given in the next subsection. Several R\&D activities are ongoing for confronting the environmental conditions and the Physics demands at Run 5 at HL-LHC and the future machines. They are briefly listed in the last two subsections. They open promising ways to address the technical feasibility of the proposed trigger, even in such demanding environmental conditions. The technical feasibility of the L1 pixel trigger will be the object of another paper.  

 \subsection{Bandwidth and Latency of a real-time Level-1 pixel trigger: preliminary estimate}
The bandwidth and the latency are the two key parameters for ensuring the feasibility of such a trigger that must handle a large flow of information at 40 MHz. 
The readout system must have sufficient bandwidth to allow the selected pixel information to be delivered to the trigger system.
The considered readout scenario is based on a seeded trigger. 
The RoI size is defined by a region in $\phi$ of $\pm$ 0.1 radian and no constraint in $\eta$, i.e. $|\eta|$=3.0. Two regions in $\phi$ need to be read out, since the electron charge is not determined by the calorimeter. 
For this defined RoI size, a total of (2x0.2 rad)/2$\pi$ = 6.4\% of the pixel area in each barrel layer will need to be read out. As can be seen in Fig.~\ref{fig:trigrate}, the rate of em clusters is 400 kHz, for 200 pileups, tracks above 20 GeV $p_T$, and the considered detector $\eta$ range. 
Thus the required bandwidth equivalent to the full readout of the pixel detector corresponds to 6.4\% of 400 kHz i.e. 26 kHz. 
This dedicated electron pixel-based L1 trigger therefore represents 3.4\% of the total L1 readout bandwidth i.e. quite affordable within the total L1 trigger bandwidth.

\begin{figure}[htbp]
        \centering
        \includegraphics[width=1.0\textwidth]{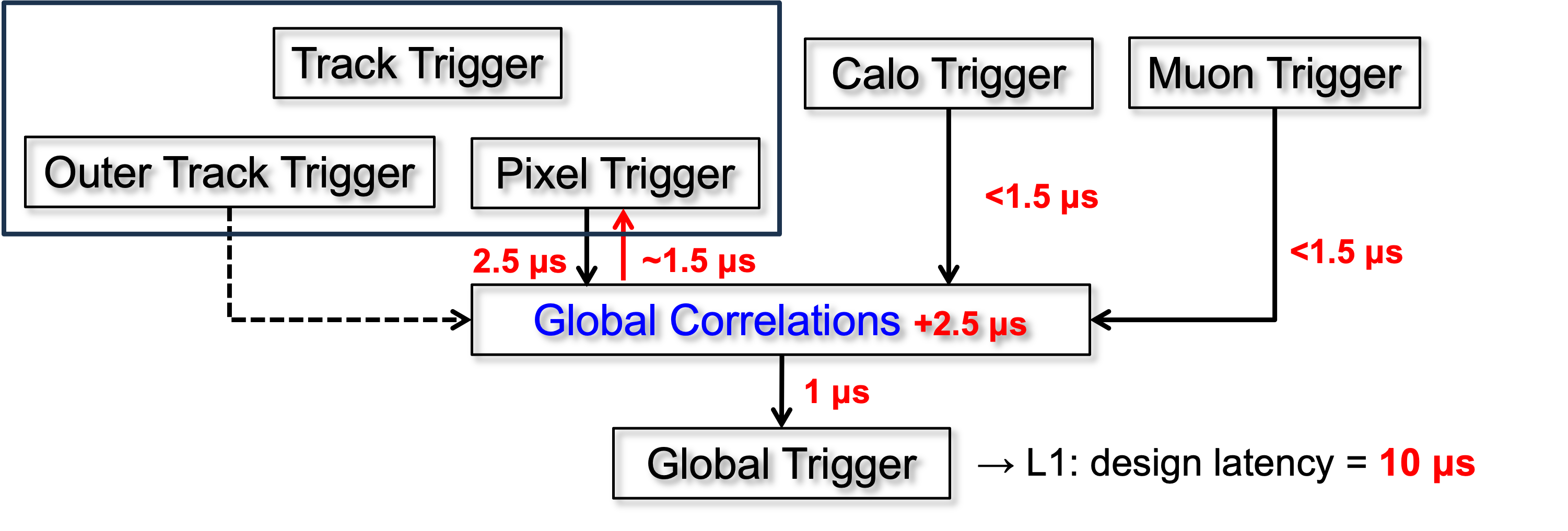}
        \caption{Generic Schema of real-time standalone pixel-based electron trigger}
        \label{fig:L1PixelTrigLatency}
    \end{figure}

A rough estimate of the latency needed to process such a trigger is sketched in Fig.~\ref{fig:L1PixelTrigLatency}, assuming a generic experimental case and only tackling the standalone proposed electron pixel-based trigger. 
It includes first the time to send the L0 trigger signal from the calorimeter to the Pixel detector FEE-ASIC, equipped with a "fast trigger" signal. 
This fast L0 trigger is indeed available in the RD53 FEE ASIC, as designed for ATLAS pixel detector. It can be used to define the clusters in the FEE ASIC, that are hit by this fast trigger and thus the corresponding RoI regions in each concerned ASIC. 
The geographical cluster information can be then processed through the FEE ASIC digital part and sent to a global correlator unit as schematized in Fig.~\ref{fig:L1PixelTrigLatency} within 2.5 $\mu$s. This time corresponds to this preliminary first stage of processing the information. The corresponding data (digitized hit cluster addresses, i.e. geographical location) are sent to the global trigger unit, taking an additional 1 $\mu$s. 
This overall flow chart very rough based estimate gives a total of less than 10 $\mu$s. 
In order to perform a more realistic estimate of this latency a benchmarking platform will be set up and the overall corresponding study reported in another paper.

The next two subsections give a preview of some of the ongoing R\&D activities to upgrade/replace in part or totally the pixel detectors in construction or to build a new generation of pixel detectors.

\subsection{Pixel detector and Front-End hardware new developments and challenges}

The ability to achieve real-time triggering using pixel detector information will benefit greatly from the development of new silicon pixels and associated FEE electronics.

In view of Run 5 and beyond at the HL-LHC, an active R\&D is underway in all 4 LHC experiments. 
Let's just cite some of them among the most promising. For instance, ATLAS is pursuing the R\&D on 3D pixel Si sensors that will be installed on the first innermost barrel layer already for Run 4, together with planar single-sided Si sensors for the other 4 pixel layers in the barrel as well as all the pixel layers that will equip the intermediate $\eta$ region and the forward/backward disks. 

For Run 5 and beyond LHCb is actively carrying on the upgrade of the Silicon Vertex Detector (VELO) with the aim to include timing~\cite{VELO-II-upgrade}. Two Si pixel technologies under consideration for this 4D device are new 3D pixels (e.g. diamond pixels) or new LGAD sensors. 
The aim is an excellent space resolution together with a timing precision better than 50 ps per hit leading to a track-time stamp resolution of about 20 ps. Besides the associated FEE is aimed to be made in 28nm CMOS technology. 

ALICE is also carrying on active R\&D aiming to ultra thin MAPS (Monolithic Active Pixel sensors)~\cite{NewMAPS}.

A 4D tracking device especially in the vertex region is essential for confronting the high luminosity. 
The time-stamping of the tracks will allow overcoming very high pileups at HL-LHC and even more at the future colliders such as a 100 TeV hadron collider. 
Besides, time stamping requests are being developed for all the detectors such as the calorimeters or RICH (in LHCb), etc. Thus the promoters of the LGAD-based detectors are developing new ultra Fast Si detectors for 4D tracking in this technology~\cite{UFast-Timing}. 
This is a new family of particle detectors merging excellent position and timing resolution with GHz counting capabilities, very low material budget, radiation hardness, fine granularity, low power, and affordability based on LGAD technology. 
Implementing a 4D tracking sensor technology for the microvertex detector either in the innermost layers or in all the devices will indeed strengthen the potential of a real-time pixel-based trigger.

The Front-End ASIC, PSI46V2 ~\cite{Barbaro-PSI_2006}, of the first Pixel detector designed and built by the PSI group in CMS already addressed the implementation of the pixel information in the L1 trigger via the double column concept. 
A cluster multiplicity counter, i.e. a fast trigger signal was made available to the L1 trigger system. 
It included two thresholds that could be set by this mechanism: one setting the minimum number of hits within a double column, whereas the other tuning the number of hit double columns above which a trigger signal is issued. 

Both ATLAS and CMS are developing an FE ASIC within the international R\&D collaboration RD53~\cite{Garcia-Sciveres:2113263}.
ATLAS pioneered some of the key features of this device, with the development of a new FE ASIC for the upgrade of the ATLAS Microvertex at the LHC Phase-1~\cite{GARCIASCIVERES2011S155, Garcia_Sciveres_2018}. 
This FE chip (FEI4) includes together with advanced processing of the pixel hits, two main features, namely: i) the logic to gather several pixel hits within a single ``cluster'' ({\it clusterization}) and ii) a two-trigger-signal scheme by adding to the usual 40 MHz L1 trigger, a second fast real-time trigger signal (labeled as L0). 
The L0 trigger allows a prompt extraction of the clusters of pixel hits that correspond to the 40 MHz trigger and their transmission to the next level of signal processing. 
These two features are essential assets to perform real-time signal processing at the front-end level. 

%The experience gained with this first generation of new ``intelligent'' FE ASICs working in the harsh HL-LHC conditions will be instrumental to build the updated version able to fully address the challenges of a real-time L1 pixel-based trigger.

The experience gained with this first generation of new ``intelligent'' FE ASICs operating in harsh HL-LHC conditions will be instrumental in the development of the updated version which will be able to fully address the challenges of a real-time L1 pixel-based trigger.

Besides new means of high-speed, high-rate data transmission are being explored within the new photonics-based technology~\cite{HSpeed-Rate-PhoTransmission}.

\subsection{The use of AI and new Processor tools}
%Over the last years the progress in building fast and sophisticated algorithms %is impressive.
%This is due to the development of AI tools (e.g. Neural networks tools etc..) and of new processing units.  It allows performing complicated algorithms in record times and with high data rates. This is more and more experienced in the HEP domain for instance in the L1 triggers of LHC experiments.

Over the last 5 years, the use of AI-based tools in several aspects of the signal and data processing chain is making impressive advances in the HEP domain and especially the LHC experiments. 
This goes together with the increase in performances of the new processors (e.g. new FPGAs). 

The application of AI (e.g. Neural Networks etc.) software tools allows performing sophisticated algorithms in record times and with high data rates. 
This is more and more developed in different triggering levels (even at the L1 level) or DAQ stages, coupled with high performance FPGA or GPU units in the LHC experiments.

Besides the use of AI at the software level, new interesting developments have recently occurred based on embedding AI in the hardware processing of the detector signal.
The developed hls4ml tool \cite{hls4ml} is an open-source software-hardware codesign workflow to interpret and translate machine learning algorithms for implementation with both FPGA and ASIC technologies. 
It allows near-sensor real-time processing.
It is already used in the design and fabrication of a reconfigurable neural network ASIC for detector front-end data compression at HL-LHC \cite{MLhardware-appli}. 
Preliminary applications to high granularity devices (pixel detectors or high granularity calorimeters) are considered or under study.

The study reported in this paper is done with simple algorithms and software tools (e.g. LUTs) easily performed with the present software and processor tools (current FPGAs). 
It stresses the improvements in the trigger performances by including the pixel information. 
A benchmarking platform using the new hardware-based developments (i.e. new Front-End ASIC design and AI-based tools) applied to a pixel detector demonstrator is the subject of an R\&D, to be reported in another paper. 

%The advances in new processing units and AI tools will further improve these algorithms and thus their performances in time and precision.

\section{Perspectives and concluding remarks}
\label{sec:remark}
A second phase of the HL-LHC is foreseen with the aim to reach an even higher instantaneous luminosity (7.5$\times$10\textsuperscript{34}cm\textsuperscript{-2}s\textsuperscript{-1}) and thus get 4 ab\textsuperscript{-1} total integrated luminosity by the end of HL-LHC~\cite{atlasTDR-TDAQ}.
Besides, following the updated HL-LHC schedule~\cite{HL-LHC-Schedule}, a possible slight increase of the LHC collider c.m. energy could be feasible at Run 5 (after 2035). 
This is part of the worldwide developments on Higher Field Magnets for future hadron colliders.
Furthermore, various detector upgrade stages are part of the routine life of the experiments at long-life machines. This will apply to the HL-LHC which will last for more than 10 years.

%ATLAS has defined an evolution scenario for handling the Run 5 challenges~\cite{atlasTDR-TDAQ}. 
%It keeps the strategy as briefly reminded in this paper but with an increase in data rate transmission through the DAQ up to 2 or 4 MHz, and an extension of the trigger latency to 30 or 35~$\mu$s. Within this scheme as in the first part of the HL-LHC run, the full tracker system information will be used, including all or part of the pixel information. CMS has developed a Track Trigger detector and associated Front End Electronics for the outer Tracker part (all but the Microvertex) able to work at a 40 MHz input rate. The Pixel information is used at the High-Level Trigger stage.

ATLAS first defined an evolution scenario for handling the Run 5 challenges~\cite{atlasTDR-TDAQ}.
It keeps the 3 stages strategy as reminded in this paper but with an increased input rate at L1 (up to 2 or 4 MHz) and an extension of the L1 trigger latency to 30 or 35 $\mu$s.
Within this scheme as in the first part of the HL-LHC, the pixel information will be partly included in the overall track trigger of the ATLAS experiment part of the L1 trigger,
working with 2 up to 4 MHz reduced input rate, after the L0-trigger which works with the full 40 MHz input rate.
This is now superseded by the amendment to the TDR~\cite{atlasTDR-TDAQ-Add}. 
The objective is to keep, over the whole HL-LHC, the same trigger scheme and to overcome the machine's increased performances in a possible second stage of HL-LHC, thanks to a revised Event Filter, as described in this amendment.
This Event Filter relies on the advances in the industrial world, on CPUs, GPUs, and FPGAs as well as on the AI field.
The ongoing impressive progress in these areas will allow for the building of a fancy and highly performing heterogeneous system associated with sophisticated and quite efficient algorithms on the software side.
CMS has developed a Track Trigger detector and associated FEE for the outer Tracker part (all but the Microvertex) able to work at a 40 MHz input rate.
The Pixel information is used at the High-Level Trigger stage.

Moreover, let us stress the interest in developing a similar first-level trigger for the pixel-based timing detectors, under active R\&D for HL-LHC. 
The accurate timing information at the first-level trigger will be of utmost importance (e.g. increase of pileups). This is a piece of major information to incorporate at the first-level trigger in the forthcoming trigger system upgrades.

The performance-based study presented in this paper goes beyond the ATLAS and CMS HL-LHC track trigger present scenarios by showing the benefits of using the Microvertex information, in the trigger architecture, at the real-time level. 
It shows how including the information from this essential detector, will improve the overall trigger real-time selection performances. This will be highly beneficial for increasing the Physics potential of these detectors. 
It will allow extracting most of the HL-LHC era Physics and get ready for the next generation(s) of detectors to be running at future high-energy machines.
The setting up of a dedicated benchmark platform including the hardware devices and software tools will be the subject of another paper.

%%%%%%%%%%%%%%%%%% Acknowledgments %%%%%%%%%%%%%%%%%% 
\acknowledgments

This work was supported by the National Research Foundation of Korea (NRF) grant funded by the Korean government (MSIT) (Grants No. 2018R1A6A1A06024970, No. 2020R1A2C1012322 and Contract NRF-2008-00460), the computing resources of Global Science experimental Data hub Center (GSDC) in Korea Institute of Science and Technology Information (KISTI). The research leading to these results has also received funding from the EU Community Marie Curie International Incoming Fellowship (IIF), FP7-PEOPLE-2011-IIF, Contract No. 302103, TauKitforNewPhysics, and from the People ITN Programme Marie Curie Actions FP7-PEOPLE-2012-ITN, INFIERI, under REA grant agreement No. 317446. One of us (ASN) is indebted to LPC at FNAL for hospitality and support as a visiting scientist in 2011, when launching this work.

A few pioneering Front-End ASIC designs for pixels have inspired this study. The PSI46 device with the double columns layout, developed by the PSI group in CMS, led S. Kwan (FNAL) and ASN (CNRS) to initiate a collaborative effort for studying the different stages of integrating an L1 trigger using the pixel information. The development of the FEI4 for the Internal Barrel Layer (IBL) of ATLAS by M. Garcia-Sciveres (LBL) and T. Hemperek (Bonn University) and collaborators, is an essential step for an intelligent signal processing of the Pixel information at LHC. This is pursued now by the RD53 International Collaboration. This work is very much indebted to the R\&D on these essential advanced Integrated Circuits.

Thanks to the CMS Collaboration for the CMS simulation frameworks used in the dedicated CMS studies that preceded this work.
Thanks to the DELPHES authors for the general DELPHES simulation package for LHC experiments, on which this work is based.

A number of people contributed to the various stages of these studies, among whom Petra Merkel (FNAL), David Christian (FNAL), Michael Wang (FNAL), Geumbong Yu (SNU), and several young PhD students or postdocs supported by the INFIERI EU program among whom: Benedetta Nodari,  Alvin Sashala Naik, Anton Bogachev and Sergei Lapin. This work greatly benefited from discussions and exchanges with renowned experts including Wesley Smith (Wisconsin University). We thank him for his support.
Finally, we are indebted to Ian Tomalin (RAL) from the CMS collaboration and Yoshinobu Unno (KEK) from the ATLAS collaboration, for their critical and expert reading of the paper, and their valuable comments and inputs.

\appendix

\section{The DELPHES simulation}

The DELPHES simulation used as a framework for these performance studies is detailed in this Appendix.
%As pointed out in Section~\ref{sec:intro}, including the pixel detector within the overall L1 trigger CMS upgrade for HL-LHC has not been endorsed by the CMS collaboration~\cite{CMS-phase2L1TDR}. %^Therefore, the work reported here is performed at a “generic level” and in a more general approach, but also keeping the overall CMS-Phase2 upgrade as an example case. 

Three types of simulated MC samples are generated using \texttt{DELPHES} in this study: (i) 5 million single electron gun events without pileup to measure 3$\sigma$ boundaries of signal windows; (ii) 1 million single electron gun events with 200 PU for measuring the L1 trigger efficiency; (iii) 10 million of minimum bias events with 200 PU for estimating the L1 trigger rate; (ii) and (iii) samples are used for the pixel-based charged isolation algorithm.

The calorimeter only information (no tracking included) allows the defining of the so-called L1 electron/photon (L1 e/$\gamma$) objects. 
%Indeed without tracks associated or not to these objects, it is not possible to discriminate between an electron and a photon. 
A more precise L1 e/$\gamma$ definition is provided through the L1 trigger tower objects. 
%The Phase-2 CMS calorimetry is taken as a showcase.
The barrel calorimeter granularity is defined by:
%granularity and the pixel clusters are considered. 
%barrel calorimeter made of crystals has a granularity of 
(0.0174, 0.0174) in ($\eta, \phi$) ~\cite{CMS-barrelECAL}.
%improved from the current Phase-1 towers with a granularity of (0.087, 0.087) will provide better energy and position~\cite{CMS-barrelECAL}.
A high granularity endcap calorimeter is promoted to stress the essential role of the pixel detector tracking at large  $\eta$. We thus consider a calorimeter based on silicon sensors, providing a position resolution better than 1~mm because of the small cell sizes; the angular uncertainty in $\theta$ is of 7~mrad for a $p_{\textrm{\scriptsize T}}$ shower of 25~GeV without corrections and tuning. An example is the HGCAL endcap calorimeter for the CMS upgrade~\cite{CMS-HGCAL}.

As for the Phase-2 upgrades of ATLAS and CMS ~\cite{ATLAS-phase2tracker, CMS-phase2tracker}, the microvertex detector simulation considers small pitch silicon pixel sensors of 100-150~$\mu$m thickness, with pixel size of 50$\times$50~$\mu$m\textsuperscript{2} for the barrel part, and 25$\times$100~$\mu$m\textsuperscript{2} for the endcap and forward parts.
Pixel clusters are built from contiguous pixels. 
A typical threshold of 1200 electrons is set for the pixel readout.
%in a Time over Threshold based readout (e.g.~\cite{CMS-phase2tracker}).

\texttt{DELPHES} provides a simplified description of the two L1 key parameters, i.e. the trigger towers for the calorimetry, and the L1 tracks for the tracking system. These are the physical/detector entities we refer to at Level-1.
%they are described in detail in a full detector simulation such as CMSSW for CMS.
The trigger tower objects are produced to match the granularity of both the barrel and the endcap calorimeter defined here above.
%the Phase-2 CMS calorimeter. 
Pixel clusters are simulated from the tracks in \texttt{DELPHES} simulation by smearing position resolution to be consistent with a full detector simulation. 
%for instance with that of CMSSW.

\begin{figure}[hbtp]
   \centering
   \ \subfloat{
      \includegraphics[width=0.27\textwidth]{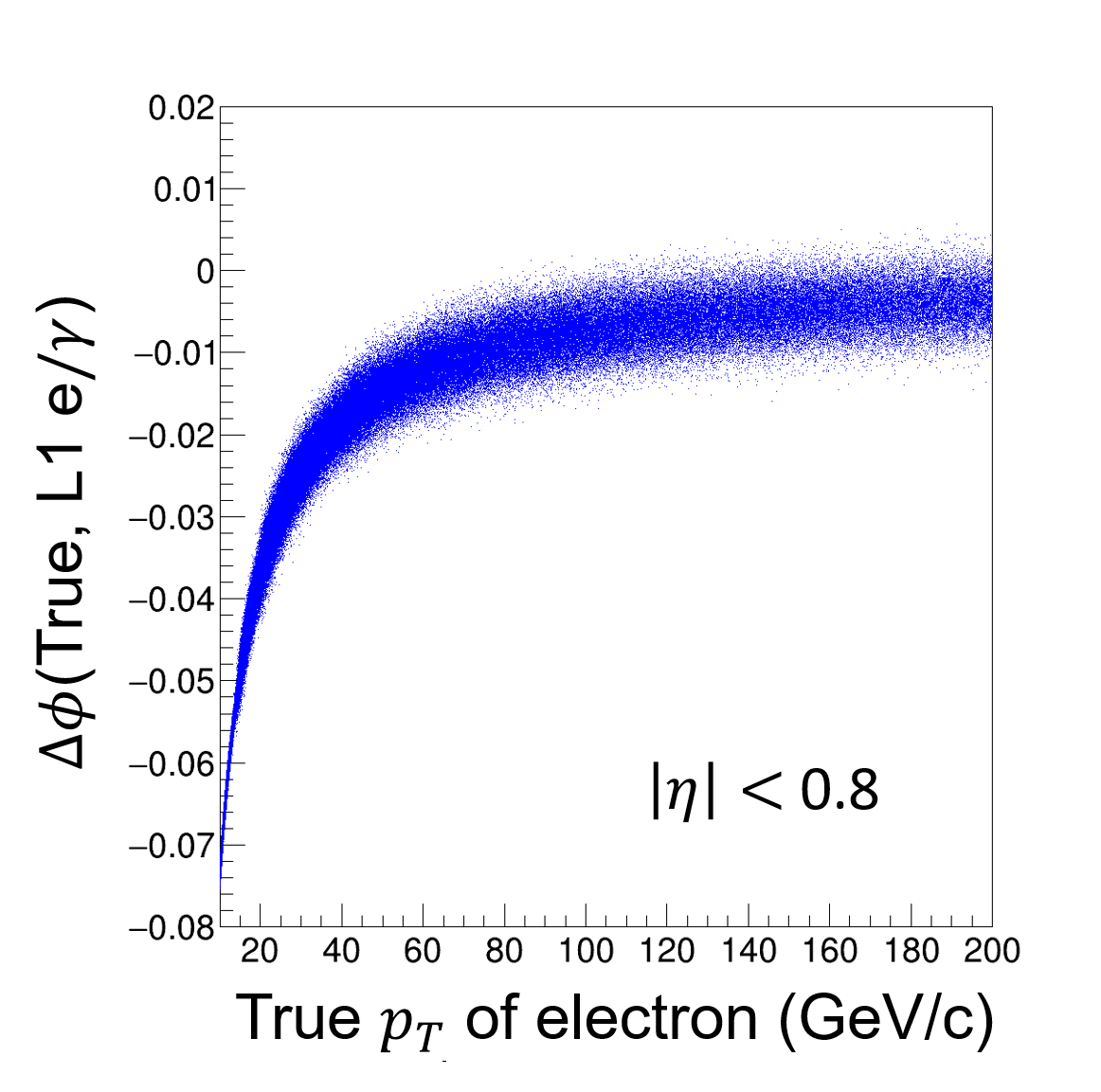}
       \includegraphics[width=0.27\textwidth]{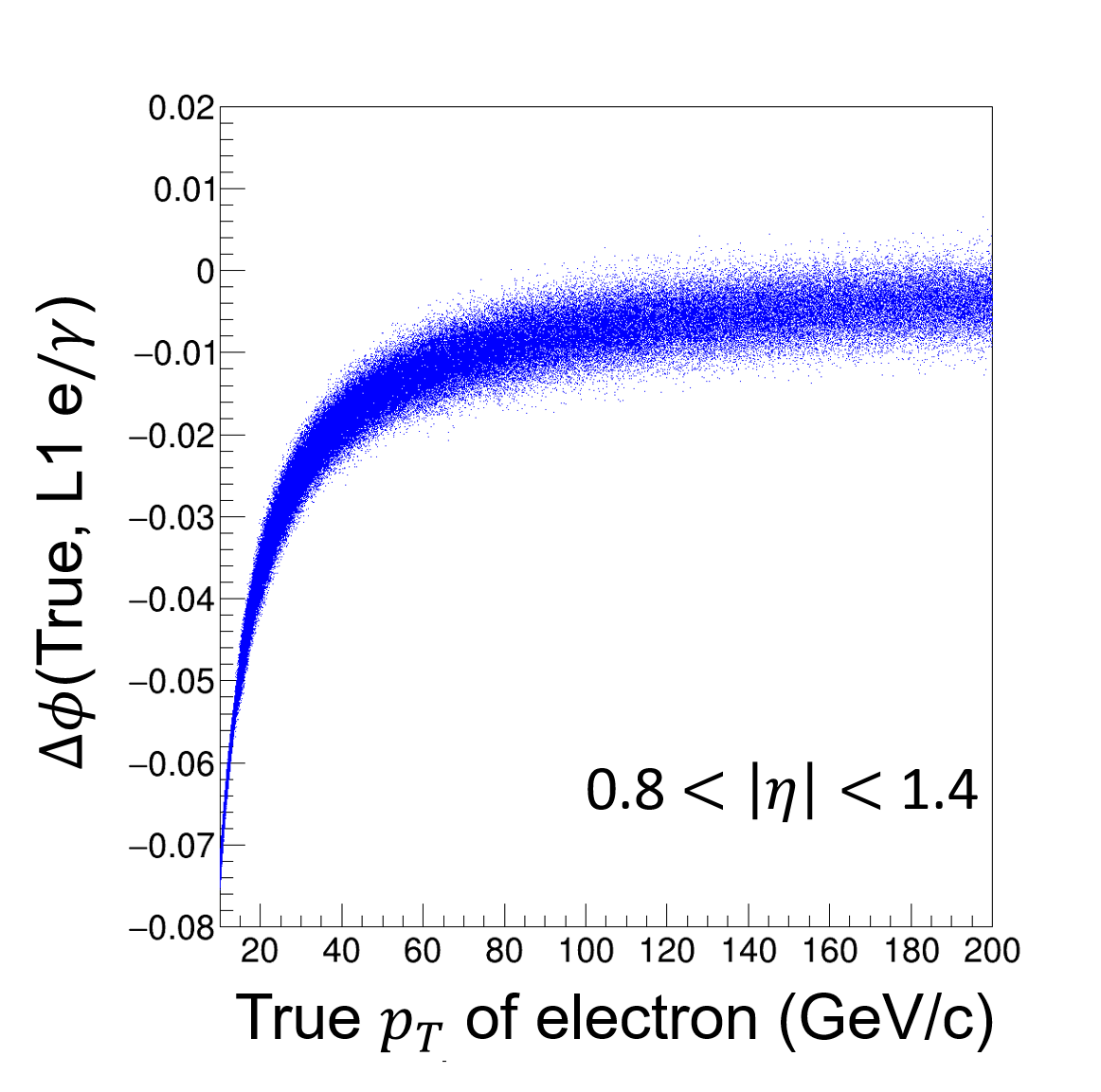}
        \includegraphics[width=0.27\textwidth]{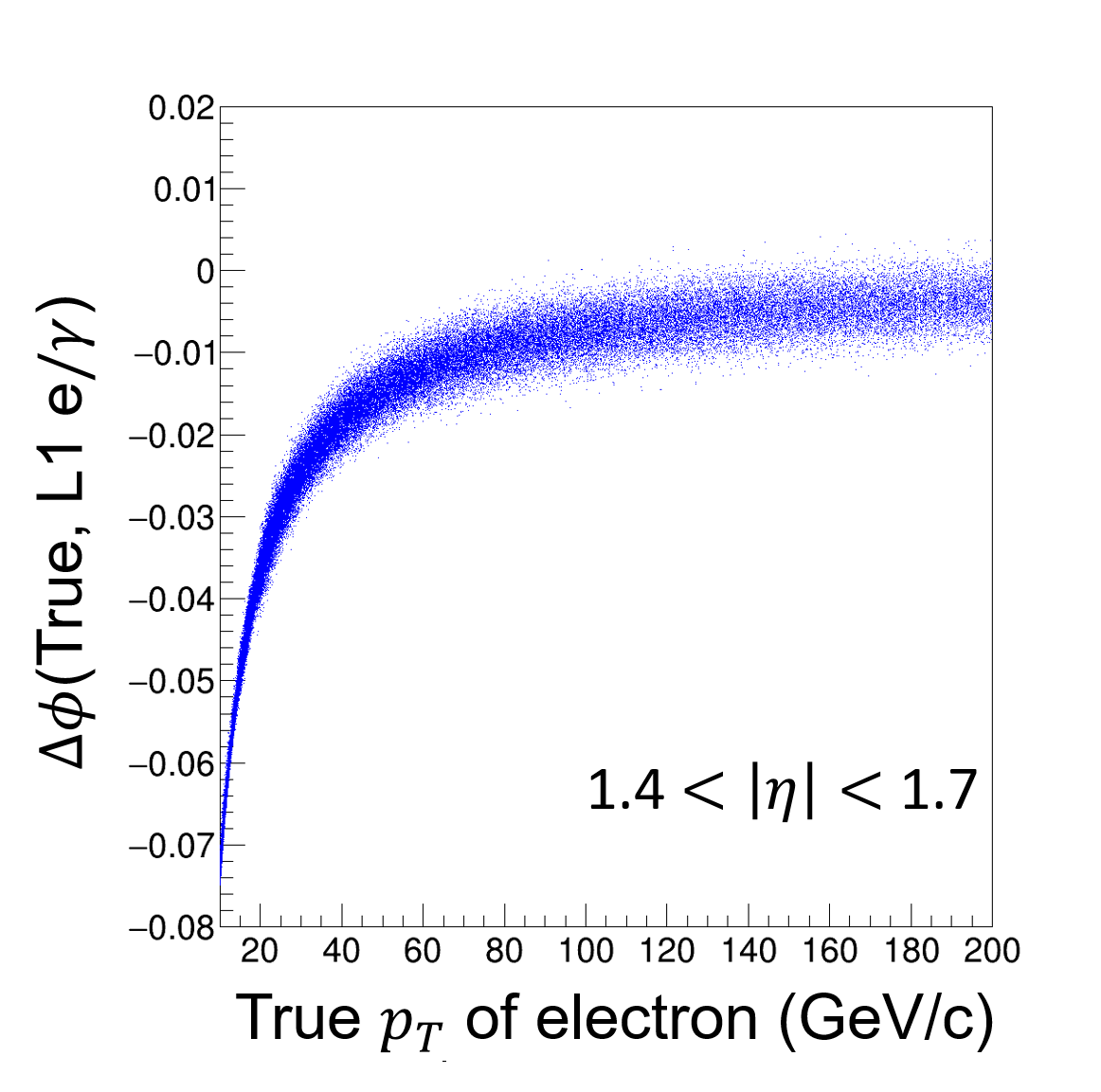}
   }
    
    \subfloat{
       \includegraphics[width=0.27\textwidth]{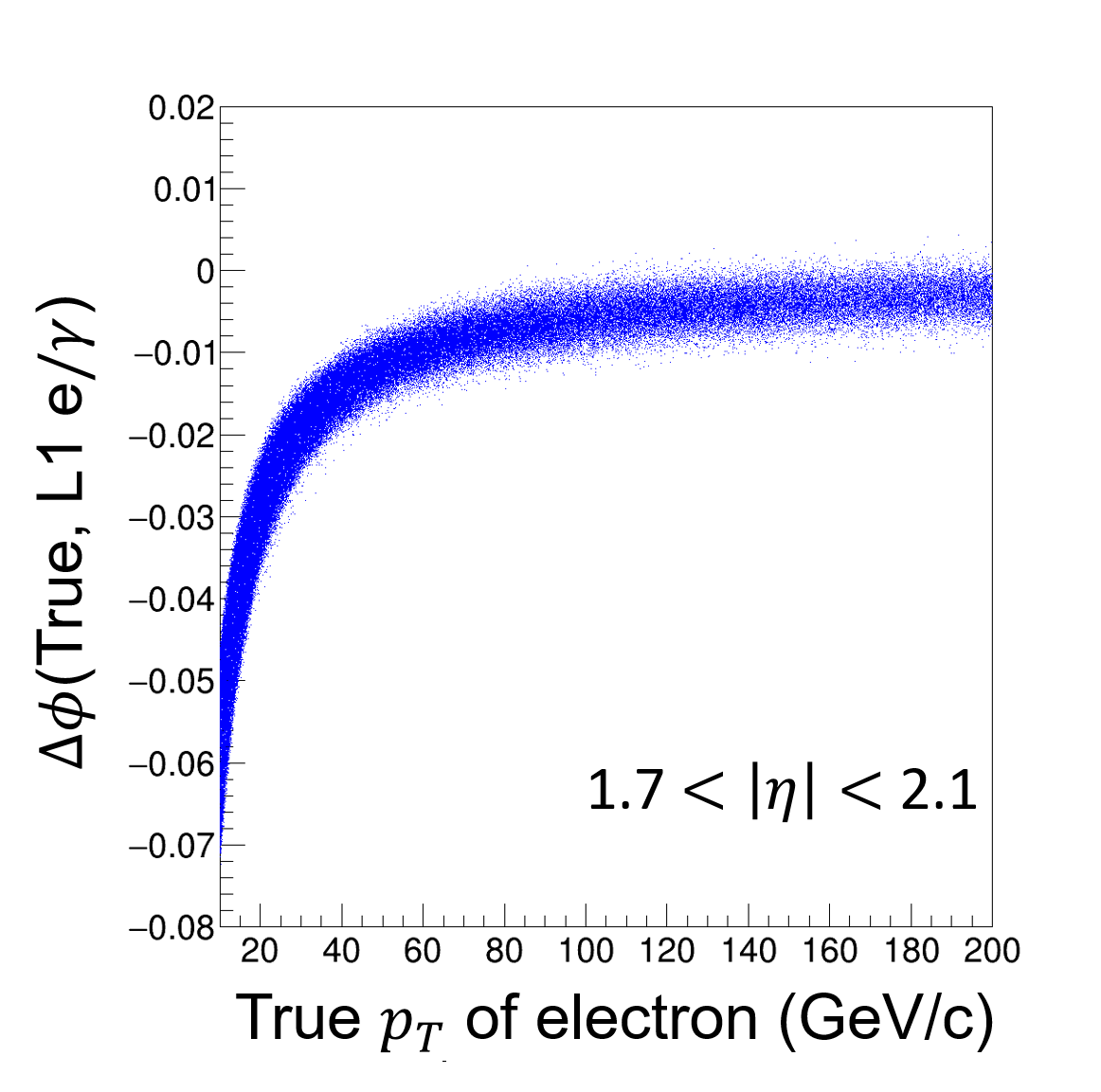}
        \includegraphics[width=0.27\textwidth]{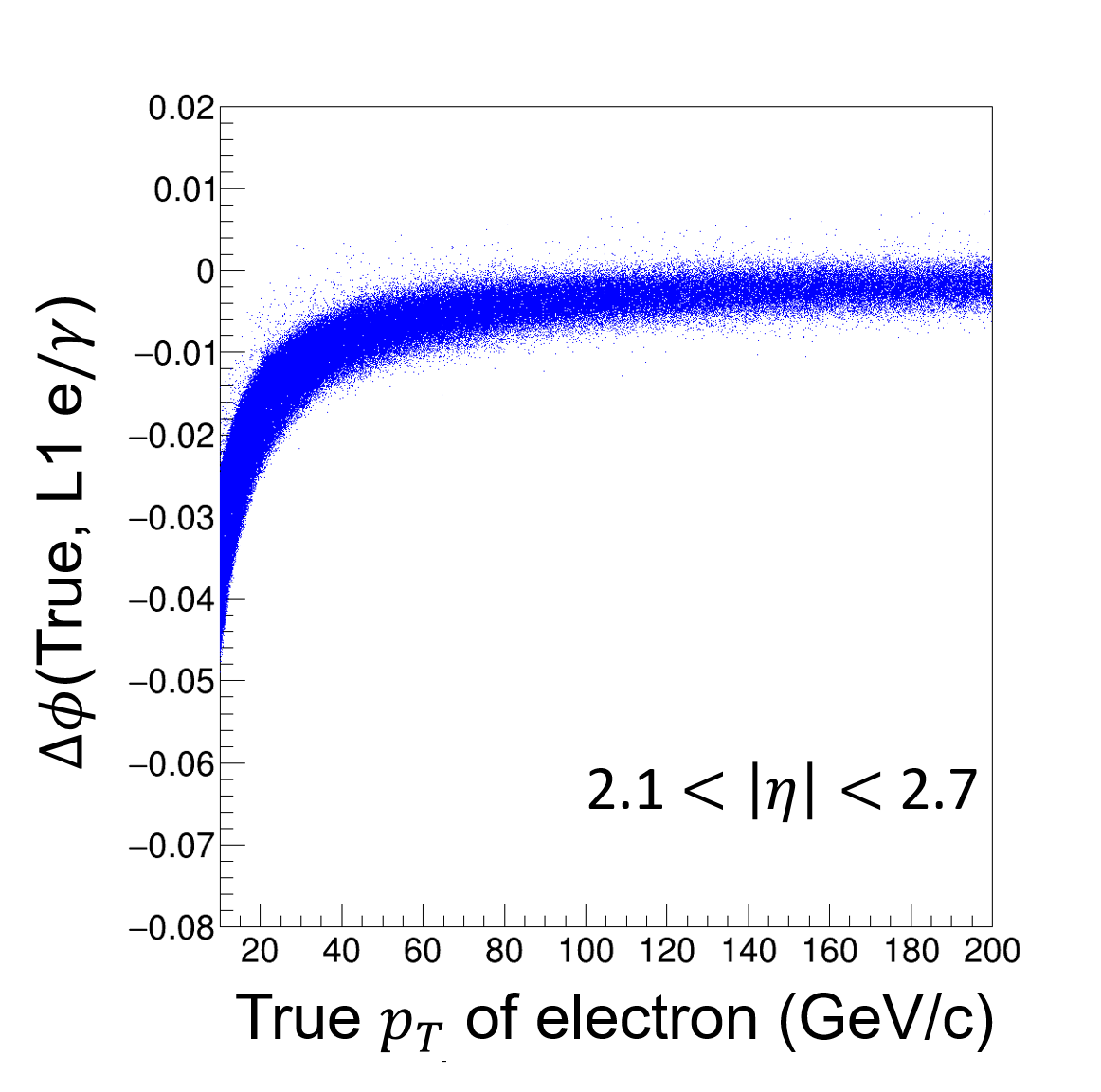}
        \includegraphics[width=0.27\textwidth]{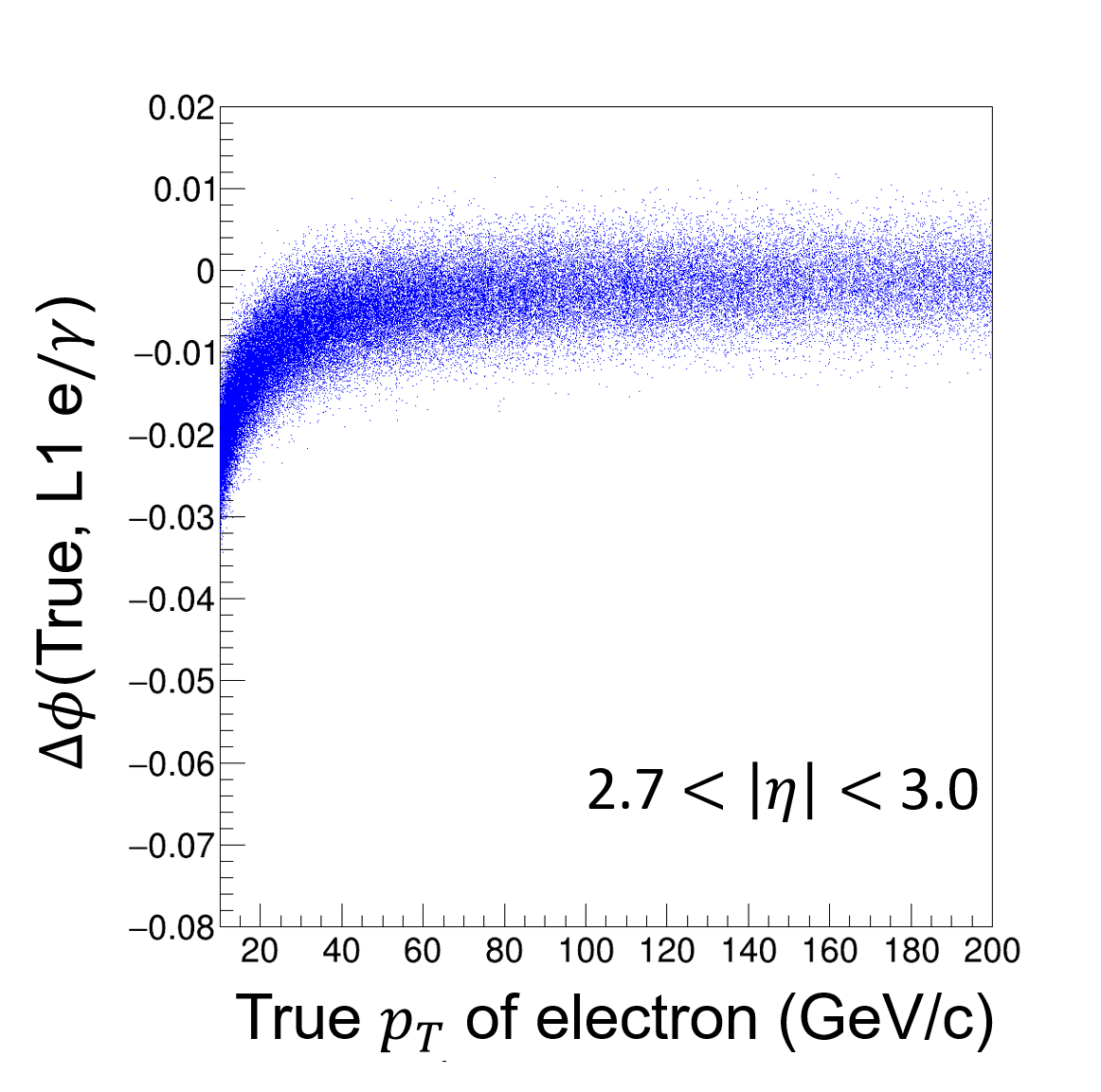}
   }
    \caption{$\Delta\phi$ distributions showing the difference in $\phi$ between the ``true'' $\phi$ i.e. from the gen-level electron and the $\phi$ from the L1 e/$\gamma$ object, as a function of the ``true'' gen-level $p_{\textrm{\scriptsize T}}$, for the considered $\eta$ regions. The corresponding $\Delta\phi$ distributions for the positrons are symmetrical w.r.t. $\Delta\phi$ = 0 axis, to the ones of the electrons.}
    \label{fig:resolution1}
\end{figure}

\begin{figure}[hbtp]
    \centering
   \subfloat{
        \includegraphics[width=0.27\textwidth]{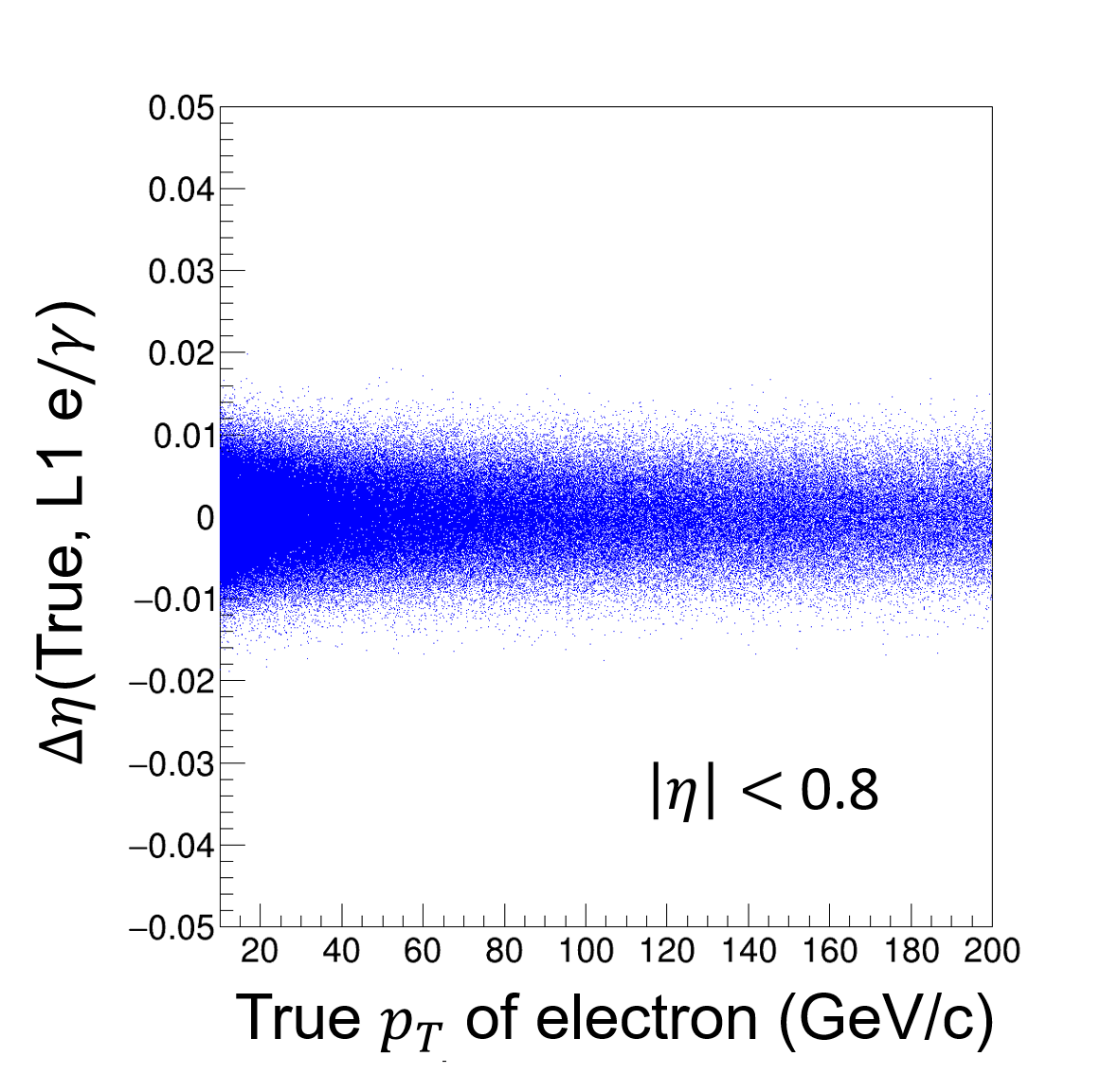}
        \includegraphics[width=0.27\textwidth]{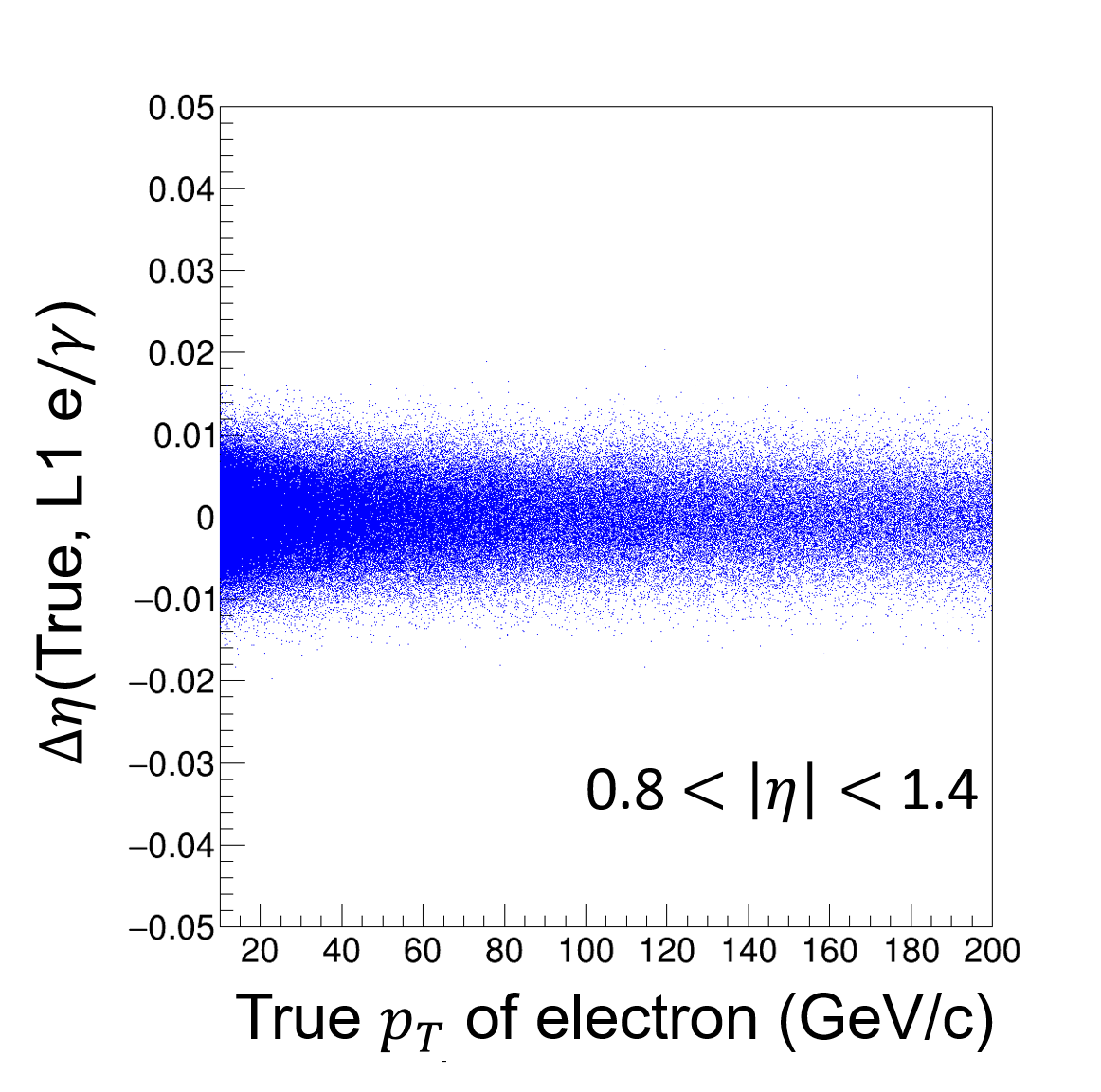}
        \includegraphics[width=0.27\textwidth]{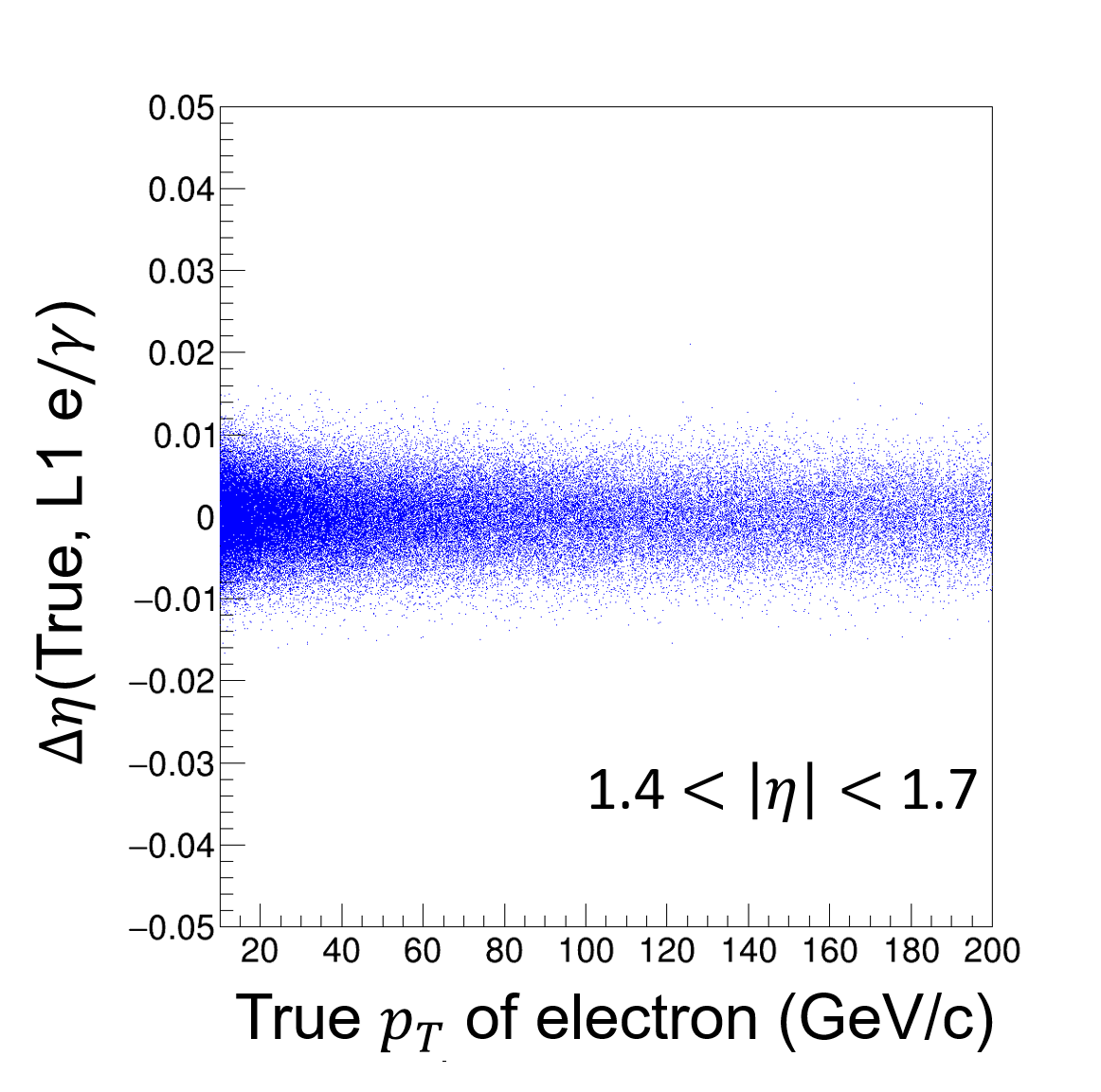}
    }
    
    \subfloat{
        \includegraphics[width=0.27\textwidth]{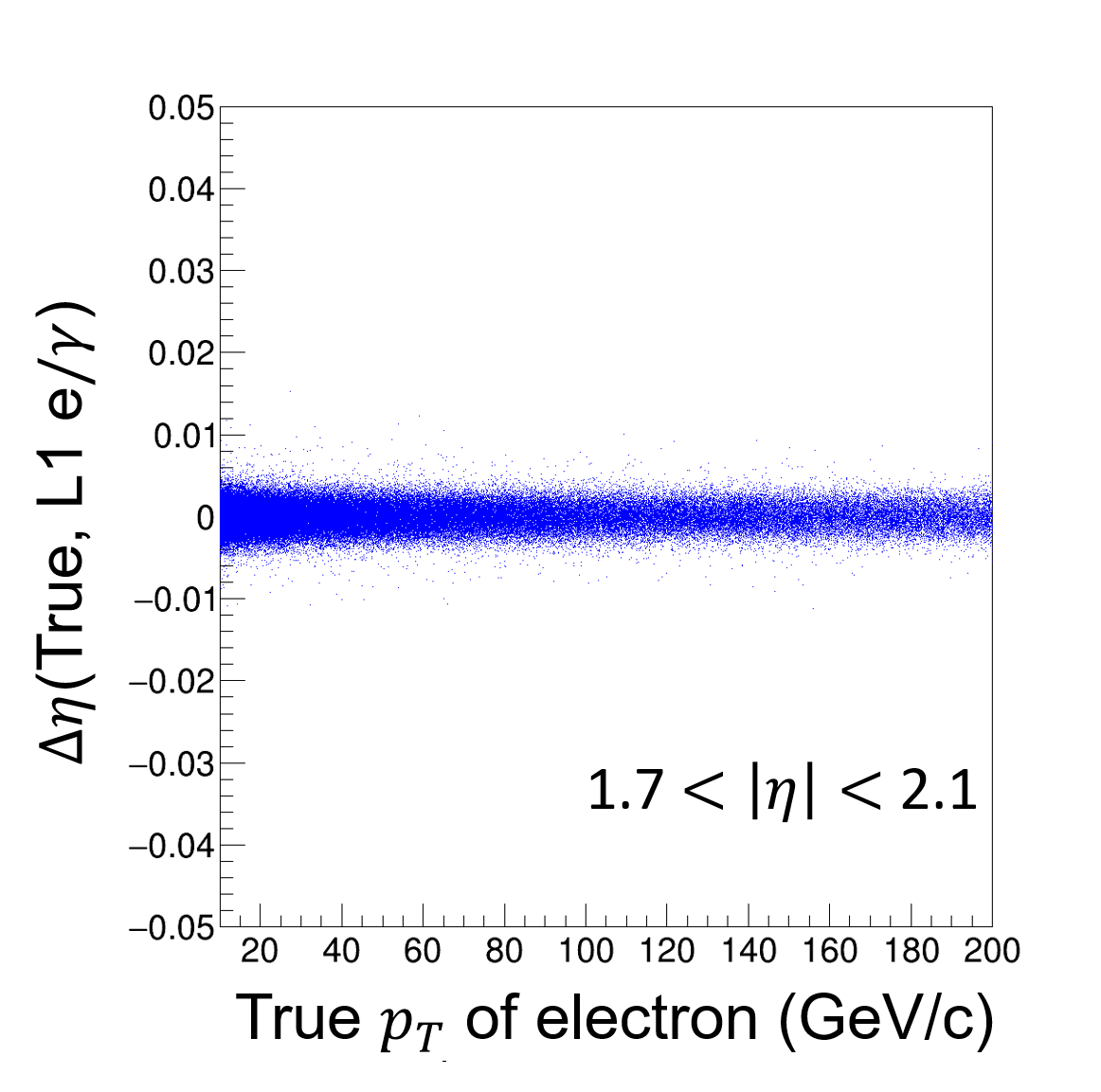}
        \includegraphics[width=0.27\textwidth]{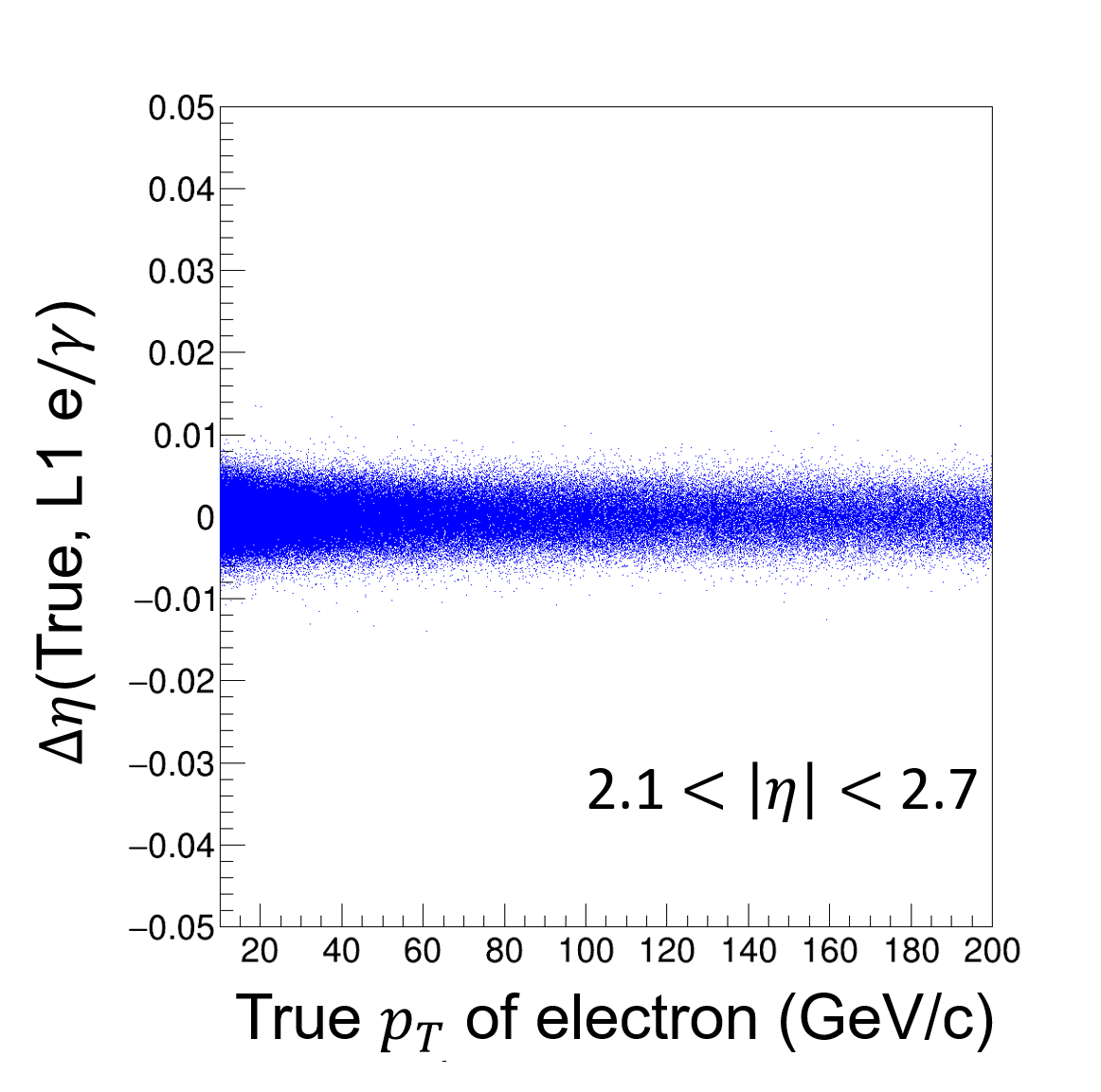}
        \includegraphics[width=0.27\textwidth]{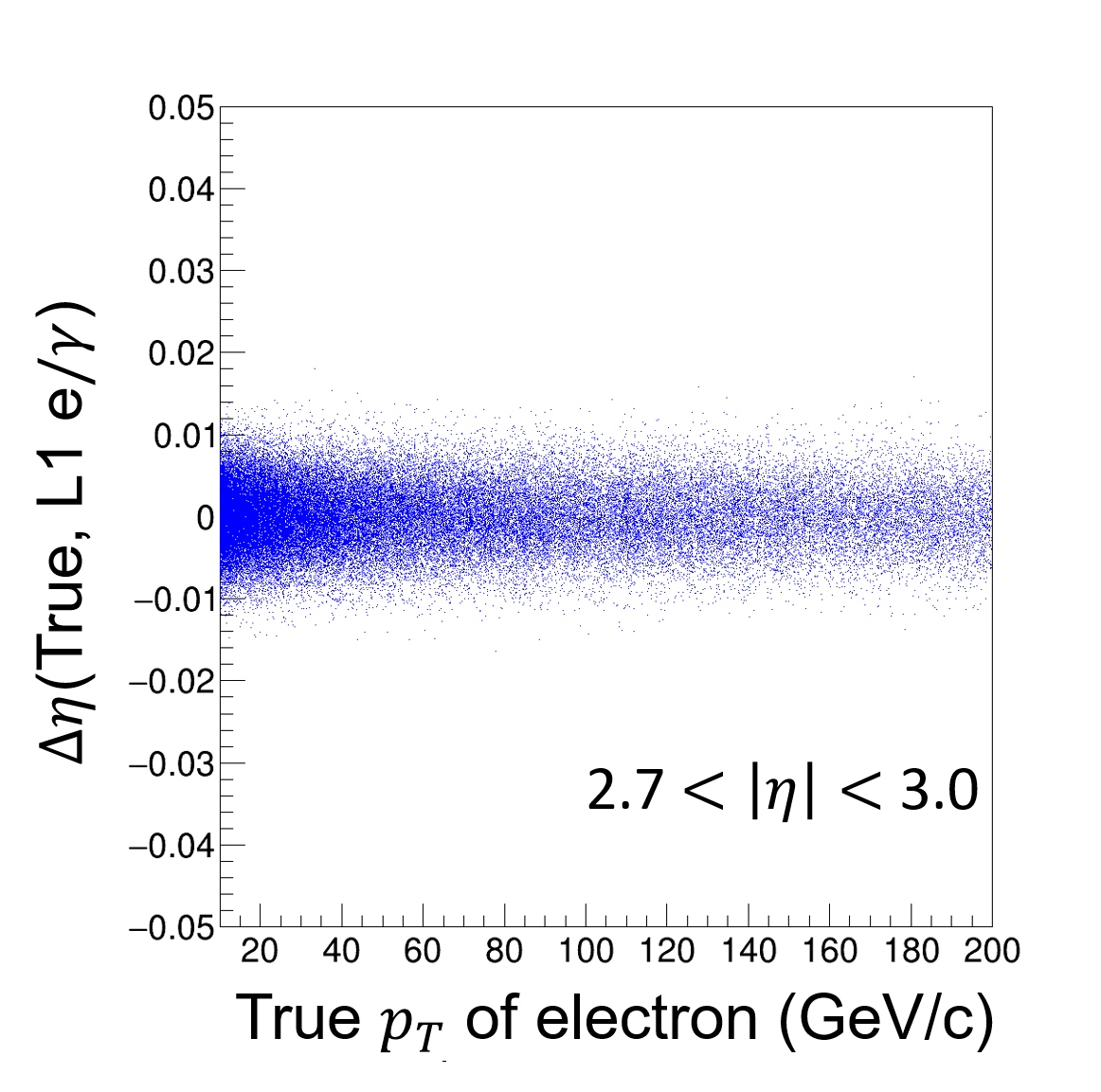}
    }
    \caption{$\Delta\eta$ distributions representing the difference in $\eta$ between the ``true'' $\eta$ i.e. from the gen-level electron and the one from the L1 e/$\gamma$ object, as a function of the ``true'' gen-level $p_{\textrm{\scriptsize T}}$, for the considered $\eta$ regions. Positrons have the same $\Delta\eta$ distributions. }
    %of gen-level electron and L1 e/$\gamma$ object for the different $\eta$ regions.}
    \label{fig:resolution2}
\end{figure}

The resolutions of the $\phi$ and $\eta$ position parameters of the produced \texttt{DELPHES} L1 e/$\gamma$ objects are shown in Fig.~\ref{fig:resolution1} and Fig.~\ref{fig:resolution2}.
The angular resolutions $\Delta\phi = \phi_{\textrm{\scriptsize gen}} - \phi_{\textrm{\scriptsize tower}}$ and $\Delta\eta = \eta_{\textrm{\scriptsize gen}} - \eta_{\textrm{\scriptsize tower}}$ are defined as a function of the generator level (gen-level) electron $p_{\textrm{\scriptsize T}}$, while the vertex correction is applied to the L1 e/$\gamma$ objects. The parameters $\phi_{\textrm{\scriptsize gen}}$ and $\eta_{\textrm{\scriptsize gen}}$ are the ``true'' generated $\phi$ and $\eta$ position parameters. The parameters $\phi_{\textrm{\scriptsize tower}}$ and $\eta_{\textrm{\scriptsize tower}}$ are the position parameters in $\phi$ and $\eta$ of the single tower in the barrel EM calorimeter, respectively the elementary tower element in the endcap EM calorimeter.
For measuring the position resolution of L1 e/$\gamma$ objects, a single electron sample is used. It is worth noting that because of the highest granularity of the considered end cap calorimeter, the region with $\eta$ above 1.7 shows a higher resolution.

\clearpage

\section{\texorpdfstring{$\Delta\phi$}{Lg}-LUT for real-time pixel-based track reconstruction algorithm}
 %In step 2, $\Delta\phi$ resolution is dominated by the calorimeter granularity and distance to %beam axis, whereas in Step 3, the pixel high granularity is the dominant factor.

\subsection{Step 2 cases}

\begin{table}[!ht]
    \centering
    \scalebox{0.5}{
    \begin{tabular}{|c|c|c|c|c|c|c|c|c|c|c|c|c|c|c|c|c|c|c|c|c|}
    \hline
    L1 e/$\gamma$ $E_{\textrm{\scriptsize T}}$ (GeV) & 10 & 11 & 12 & 13 & 14 & 15 & 16 & 17 & 18 & 19 & 20 & 22 & 24 & 26 & 28 & 30 & 35 & 40 & 45 & 50 \\
    \hline
    3$\sigma$ upper boundary & 0.089 & 0.082 & 0.077 & 0.072 & 0.069 & 0.065 & 0.063 & 0.060 & 0.058 & 0.056 & 0.054 & 0.050 & 0.047 & 0.045 & 0.043 & 0.041 & 0.038 & 0.035 & 0.033 & 0.031 \\ 
    \hline
    3$\sigma$ lower boundary & 0.048 & 0.043 & 0.038 & 0.034 & 0.030 & 0.027 & 0.025 & 0.022 & 0.020 & 0.018 & 0.016 & 0.013 & 0.011 & 0.009 & 0.007 & 0.005 & 0.002 & -0.001 & -0.003 & -0.004 \\
    \hline
    \end{tabular}
    }
    \caption{Step 2: $\Delta\phi$-LUT for $\eta$ Region 1, 2, 3, i.e. $\eta$<1.7}
    \label{tab:phiLUTr123}
\end{table}

\begin{table}[!ht]
    \centering
    \scalebox{0.5}{
    \begin{tabular}{|c|c|c|c|c|c|c|c|c|c|c|c|c|c|c|c|c|c|c|c|c|}
    \hline
    L1 e/$\gamma$ $E_{\textrm{\scriptsize T}}$ (GeV) & 10 & 11 & 12 & 13 & 14 & 15 & 16 & 17 & 18 & 19 & 20 & 22 & 24 & 26 & 28 & 30 & 35 & 40 & 45 & 50 \\
    \hline
    3$\sigma$ upper boundary & 0.071 & 0.067 & 0.063 & 0.060 & 0.057 & 0.054 & 0.052 & 0.050 & 0.048 & 0.046 & 0.045 & 0.042 & 0.040 & 0.039 & 0.037 & 0.036 & 0.033 & 0.031 & 0.029 & 0.028 \\ 
    \hline
    3$\sigma$ lower boundary & 0.033 & 0.029 & 0.025 & 0.022 & 0.020 & 0.017 & 0.015 & 0.014 & 0.012 & 0.010 & 0.009 & 0.007 & 0.005 & 0.003 & 0.002 & 0.000 & -0.002 & -0.004 & -0.006 & -0.007 \\
    \hline
    \end{tabular}
    }
    \caption{Step 2: $\Delta\phi$-LUT for $\eta$ Region 4, i.e. 1.7<$\eta$<2.1.}
    \label{tab:phiLUTr4}
\end{table}

\begin{table}[!ht]
    \centering
    \scalebox{0.48}{
    \begin{tabular}{|c|c|c|c|c|c|c|c|c|c|c|c|c|c|c|c|c|c|c|c|c|}
    \hline
    L1 e/$\gamma$ $E_{\textrm{\scriptsize T}}$ (GeV) & 10 & 11 & 12 & 13 & 14 & 15 & 16 & 17 & 18 & 19 & 20 & 22 & 24 & 26 & 28 & 30 & 35 & 40 & 45 & 50 \\
    \hline
    3$\sigma$ upper boundary & 0.048 & 0.046 & 0.043 & 0.042 & 0.040 & 0.039 & 0.037 & 0.036 & 0.035 & 0.034 & 0.033 & 0.032 & 0.031 & 0.030 & 0.029 & 0.028 & 0.027 & 0.026 & 0.025 & 0.024 \\ 
    \hline
    3$\sigma$ lower boundary & 0.010 & 0.008 & 0.006 & 0.005 & 0.004 & 0.003 & 0.002 & 0.001 & 0.000 & -0.001 & -0.001 & -0.003 & -0.004 & -0.005 & -0.006 & -0.006 & -0.008 & -0.009 & -0.010 & -0.011 \\
    \hline
    \end{tabular}
    }
    \caption{Step 2: $\Delta\phi$-LUT for $\eta$ Region 5, i.e. 2.1<$\eta$<2.7.}
    \label{tab:phiLUTr5}
\end{table}

\begin{table}[!ht]
    \centering
    \scalebox{0.47}{
    \begin{tabular}{|c|c|c|c|c|c|c|c|c|c|c|c|c|c|c|c|c|c|c|c|c|}
    \hline
    L1 e/$\gamma$ $E_{\textrm{\scriptsize T}}$ (GeV) & 10 & 11 & 12 & 13 & 14 & 15 & 16 & 17 & 18 & 19 & 20 & 22 & 24 & 26 & 28 & 30 & 35 & 40 & 45 & 50 \\
    \hline
    3$\sigma$ upper boundary & 0.037 & 0.035 & 0.034 & 0.032 & 0.031 & 0.031 & 0.030 & 0.029 & 0.028 & 0.028 & 0.027 & 0.026 & 0.026 & 0.025 & 0.025 & 0.024 & 0.023 & 0.022 & 0.022 & 0.021 \\ 
    \hline
    3$\sigma$ lower boundary & 0.000 & -0.001 & -0.002 & -0.003 & -0.004 & -0.004 & -0.005 & -0.006 & -0.006 & -0.007 & -0.007 & -0.008 & -0.009 & -0.010 & -0.010 & -0.011 & -0.011 & -0.012 & -0.013 & -0.013 \\
    \hline
    \end{tabular}
    }
    \caption{Step 2: $\Delta\phi$-LUT for $\eta$ Region 6, i.e. 2.7<$\eta$<3.0.}
    \label{tab:phiLUTr6}
\end{table}

\subsection{Step 3 cases}

\begin{table}[!ht]
    \centering
    \begin{tabular}{|c|c|c|c|c|c|c|c|c|}
    \hline
    L1 e/$\gamma$ $E_{\textrm{\scriptsize T}}$ (GeV) & 10 & 11 & 13 & 15 & 19 & 24 & 35 & 50 \\
    \hline
    3$\sigma$ upper boundary & 0.006 & 0.005 & 0.005 & 0.004 & 0.004 & 0.003 & 0.003 & 0.002 \\
    \hline
    3$\sigma$ lower boundary & 0.002 & 0.002 & 0.001 & 0.001 & 0.000 & 0.000 & -0.001 & -0.001 \\
    \hline
    \end{tabular}
    \caption{Step 3: $\Delta\phi$-LUT for $\eta$<0.8, with pixel track reconstruction using the beam origin B0, and the two innermost barrel layers 1 and 2, thus the triplet (B0L1L2).}
%\end{table}

%\begin{table}[!ht]
    \centering
    \begin{tabular}{|c|c|c|c|c|}
    \hline
    L1 e/$\gamma$ $E_{\textrm{\scriptsize T}}$ (GeV) & 10 & 13 & 19 & 35 \\
    \hline
    3$\sigma$ upper boundary & 0.007 & 0.006 & 0.005 & 0.004 \\
    \hline
    3$\sigma$ lower boundary & 0.001 & 0.000 & -0.001 & -0.002 \\
    \hline
    \end{tabular}
    \caption{Step 3: $\Delta\phi$-LUT for 1.7<$\eta$<2.1, with the pixel track reconstruction using the beam origin B0, the innermost barrel layer and nearest to barrel disk, thus triplet (B0L1D1).}
%\end{table}
%\begin{table}[!ht]
    \centering
    \begin{tabular}{|c|c|c|c|}
    \hline
    L1 e/$\gamma$ $E_{\textrm{\scriptsize T}}$ (GeV) & 10 & 12 & 19 \\
    \hline
    3$\sigma$ upper boundary & 0.006 & 0.005 & 0.004 \\
    \hline
    3$\sigma$ lower boundary & 0.000 & -0.001 & -0.002 \\
    \hline
    \end{tabular}
    \caption{Step 3: $\Delta\phi$-LUT for 2.1<$\eta$<2.7, with the pixel track reconstruction using the beam origin B0, the second and third nearest to barrel disks, thus triplet (B0D2D3).}
\end{table}

\clearpage

\section{Method for the determination of the \texorpdfstring{$p_{\textrm{\scriptsize T}}$}{Lg} of the track based on the pixel clusters}

\begin{figure}[hbtp]
    \centering
    \subfloat[Before rotation]{\includegraphics[width=0.4\textwidth]{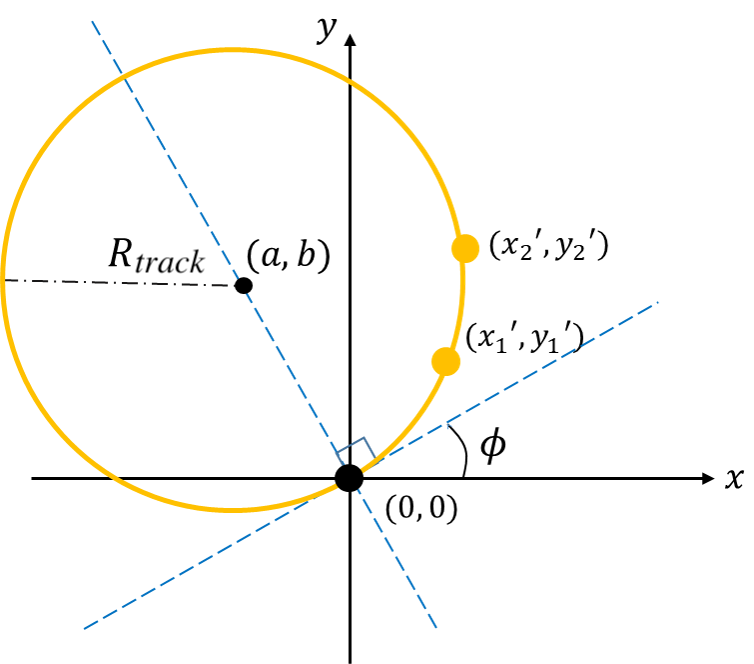}} \hspace{0.1\textwidth}
    \subfloat[After rotation]{\includegraphics[width=0.4\textwidth]{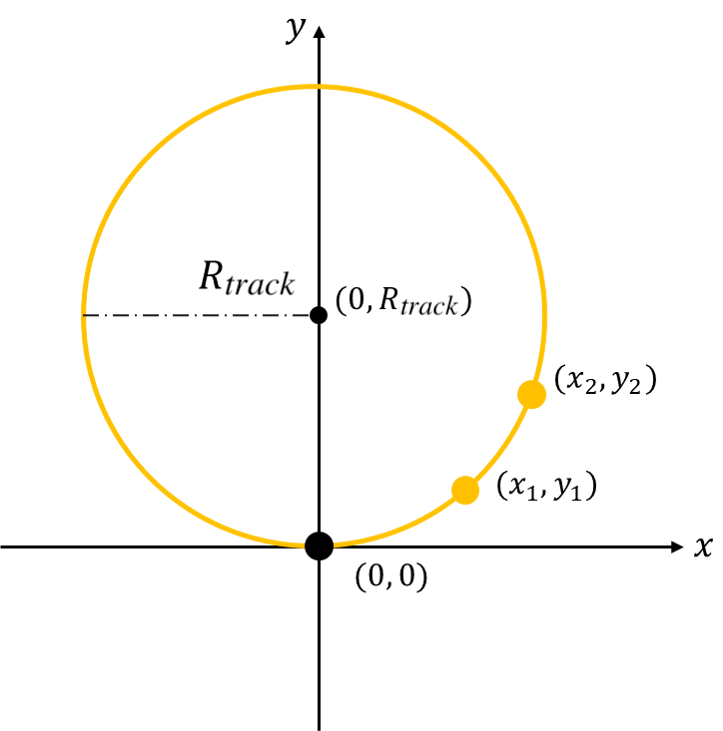}}
    \caption{Schematic of the reconstructed circle in the transverse plane before rotation (a) and after rotation (b).}
    \label{fig:reconstruction-pT}
\end{figure}

We reconstruct in the transverse plane the circle passing by the B0 coordinate in that plane and two of the pixel clusters of the considered track.
The reconstructed circle is rotated by an azimuthal angle $\phi$ with respect to the electron candidate (Fig.~\ref{fig:reconstruction-pT}).

The rotated coordinates can be expressed with the following equations:
\begin{equation}
    \begin{split}
    x_{i} = x_{i}^{'}\,cos(-\phi) - y_{i}^{'}\,sin(-\phi) \\ 
    y_{i} = x_{i}^{'}\,sin(-\phi) + y_{i}^{'}\,cos(-\phi)
    \end{split}
\end{equation}

where $i =$ 1 or 2. The equation of the circle in the new rotated frame is:

\begin{equation} \label{eq:circle}
    x^{2} + (y-R_{track})^{2} = R_{track}^{2}
\end{equation}

By substituting two rotated coordinates $x_{i}$ and $y_{i}$ into Equation~\ref{eq:circle}, we obtain:

\begin{equation} \label{eq:middle1}
    \begin{split}
    x_{1}^{2} + (y_{1}-R_{track})^{2} = R_{track}^{2} \\ 
    x_{2}^{2} + (y_{2}-R_{track})^{2} = R_{track}^{2}
    \end{split}
\end{equation}

Replacing $x_{i}^{2} + y_{i}^{2}$ by $R_{i}^{2}$ after expanding Equation~\ref{eq:middle1}, we get:

\begin{equation} \label{eq:middle2}
    \begin{split}
    R_{1}^{2} - 2y_{1}R_{track} = 0 \\ 
    R_{2}^{2} - 2y_{2}R_{track} = 0
    \end{split}
\end{equation}

And thus the radius of the circle $R_{track}$ is:

\begin{equation}
    R_{track} = \frac{R_{1}^{2} - R_{2}^{2}}{2(y_{1}-y_{2})} \; \textrm{or} \; \frac{R_{1}^{2} + R_{2}^{2}}{2(y_{1}+y_{2})}
\end{equation}

%%%%%%%%%%%%%%%%%% bibliography %%%%%%%%%%%%%%%%%% 

%\bibliographystyle{plain}
%\bibliography{pixel}

\end{document}